\documentclass[11pt, tightenlines, pra]{revtex4}

\makeatletter
\def\l@subsubsection#1#2{}

\makeatother
 
\usepackage{amsmath, amssymb, amsthm, amsfonts, latexsym, graphicx}
\usepackage{float}
\usepackage{verbatim} 
\usepackage[colorlinks]{hyperref}
\hypersetup{citecolor=blue}
\usepackage{url}
\usepackage{complexity}
\usepackage{hypcap}
\usepackage{dsfont}
\usepackage{wrapfig}
\usepackage{mathrsfs}
\usepackage{mathtools}

\DeclarePairedDelimiter{\abs}{\lvert}{\rvert}
\usepackage[caption=false]{subfig}
\input{Qcircuit}
\begin{document}
\title{An Invitation to Distributed Quantum Neural Networks}
\author{Lirandë Pira}
\email{lirande.pira@student.uts.edu.au}
\affiliation{University of Technology Sydney,
		Centre for Quantum Software and Information,
		Ultimo NSW 2007, Australia}
\author{Chris Ferrie}
\affiliation{University of Technology Sydney,
		Centre for Quantum Software and Information,
		Ultimo NSW 2007, Australia}

\begin{abstract}
Deep neural networks have established themselves as one of the most promising machine learning techniques. Training such models at large scales is often parallelized, giving rise to the concept of distributed deep learning. Distributed techniques are often employed in training large models or large datasets either out of necessity or simply for speed. Quantum machine learning, on the other hand, is the interplay between machine learning and quantum computing. It seeks to understand the advantages of employing quantum devices in developing new learning algorithms as well as improving the existing ones. A set of architectures that are heavily explored in quantum machine learning are quantum neural networks. In this review, we consider ideas from distributed deep learning as they apply to quantum neural networks. We find that the distribution of quantum datasets shares more similarities with its classical counterpart than does the distribution of quantum models, though the unique aspects of quantum data introduces new vulnerabilities to both approaches. We review the current state of the art in distributed quantum neural networks, including recent numerical experiments and the concept of \textit{circuit cutting}.
\end{abstract}
\date{\today}
\maketitle

{
  \hypersetup{hidelinks}
  \tableofcontents
}

\section{Introduction}
\label{sec:introduction}
By now we have sufficient evidence that classical computers \textit{can} learn. Decades of research in artificial intelligence (AI) \citep{russell_artificial_2010, mitchell_machine_1997, bishop_pattern_2006} and specifically deep learning (DL) \citep{lecun_deep_2015, goodfellow_deep_2016, schmidhuber_deep_2015}, have yielded powerful learning algorithms that are now employed in everyday tasks across several industries. With the rise of quantum computers, the natural question that arises is whether quantum computers, too, can learn. Quantum computing is the paradigm of computation which employs concepts from quantum mechanics \citep{nielsen_quantum_2011}. Attempting to answer this question requires a thorough exploration of quantum computing (QC) and machine learning (ML). With several directions in its agenda, the emerging field of quantum machine learning (QML) \citep{biamonte_quantum_2017, schuld_supervised_2018, wittek_quantum_2014, ciliberto_quantum_2018, cerezo_challenges_2022}, explores the intersection of quantum computing and machine learning. This interaction can have various objectives depending on whether the data or the environment is either classical or quantum \citep{aimeur_machine_2006, dunjko_machine_2018, schuld_supervised_2018}. The direction we consider here is to consider whether quantum computers can be used to provide benefits in training neural networks specifically. This question, too, has been asked and explored with various objectives in mind \citep{benedetti_parameterized_2019}. Present day quantum computers are known as noisy intermediate-scale quantum devices (NISQ) \citep{preskill_quantum_2018}. They are overshadowed by high error-rates and a small number of qubits, which hinders their capabilities. However, there is a growing debate over what algorithms these devices can be used for. In this paper, we overview the concept of distributed quantum neural networks and suggest that this might underpin the first \textit{real} application of quantum computers in the NISQ era. 

Drawing inspiration from artificial neural networks (ANNs), quantum neural networks (QNNs) have emerged as a new class of promising  quantum algorithms. While there are many approaches to training quantum neural networks, until recently they have all been inherently \textit{sequential}, aimed at training a quantum neural network on a single quantum computer. Yet, training a classical neural network on a single core is not always feasible in large scale classical machine learning. When working with large datasets or sophisticated models, training is often \textit{distributed} \citep{verbraeken_survey_2020}. Dubbed distributed deep learning (DDL), these techniques are employed either because of the size of the dataset or the model itself are too large to be processed on a single core. Employing multiple cores or even multiple machines overcomes this problem and typically leads to faster training time. 
Distributed deep learning brings together high performance computing communication protocols and the thriving field of deep neural networks \citep{bennun_demystifying_2019, chahal_hitchiker_2020, mayer_scalable_2019, langer_distributed_2020}. One work that is often cited as a large scale success story is Ref. \citep{goyal_accurate_2018}, which trains the ImageNet dataset \citep{deng_imagenet_2009} across $256$ graphical processing units (GPUs) in $1$ hour. In a single node fashion, training of the ImageNet would normally take several days. 

The limitations motivating DDL are even more pronounced in the quantum setting and an emerging set of techniques are being developed to mirror the classical paradigm. In this paper, we extend the ideas of distributed deep learning to quantum neural networks by reviewing and consolidating the existing literature. Our aim is to make more concrete the current set of vaguely similar ideas directing the research toward a more unified and directed goal of distributed QNNs. We define a distributed QNN as a quantum machine learning algorithm employing multiple quantum computers (quantum processing units (QPUs), by analogy), which we refer to as nodes. We identify some common themes in the distribution of QNNs and discuss the implications.

The rest of this paper is organised as follows. The next three Sections~\ref{section:deep}, \ref{sec:qcomp}, and \ref{sec:qml} give a primer on the ingredients required to understand distributed QNNs. Notably, Section~\ref{section:deep} introduces deep learning concepts and expands on some of the well-known classical distributed deep learning frameworks. Section~\ref{sec:qcomp} introduces quantum computing along with distributed quantum computing concepts.
Section~\ref{sec:qml} overviews quantum machine learning with a focus on quantum neural networks.
Section~\ref{section:data} gives a more detailed overview of data parallelism considered through the quantum lens while emphasizing two data encoding types and their distributed forms: basis encoding in Sec.~\ref{subsec:basis} and amplitude encoding in Sec.~\ref{subsec:amplitude}. Section~\ref{section:model} achieves the same for model parallelism while briefly commenting on vertical splitting of quantum circuits, and expanding more on some of the recent works in the so called ``circuit cutting'' schemes. In Section~\ref{section:discussion} we discuss the relevance of these works in the NISQ era and provide an overview of software that facilitates distributed deep learning as well as the quantum approaches.

\section{Distributed Deep learning}\label{section:deep}

\subsection{A Brief History of Deep Neural Networks}
Neural networks are the machinery behind the current most prevalent machine learning method --- deep learning \citep{lecun_deep_2015}. Fueled by the availability of big data and the increase in processing power, this disruptive technology provides an ecosystem for creating self-learning agents able to find abstractions that are oftentimes not visible to other types of ML algorithms.

\subsubsection{Structure of neural networks}

The building block of a neural network is the neuron. The artificial neuron --- very much inspired by the human biological neuron --- has a classical input-output structure. The first architectural model was proposed in Ref.~\citep{mcculloch_logical_1943} in 1943, known as the \textit{MP neuron}. The input values $x$ of a neuron, each of which has a corresponding weight coefficient $w$ – the parameter that determines how \textit{important} the input is to the output. The goal of a neuron is to connect with other neurons. A neural network has an input layer, so-called \textit{hidden layers}, and an output layer. The \textit{depth} of the network is determined by the number of hidden layers. The reason deep learning architectures are preferred to \textit{shallow} ones, lies on the ability of hidden layers to reach higher levels of abstraction, thus discovering more intricate patterns in datasets. The ability to extract information typically increases with the number of hidden layers, and as long as new and useful information is extracted from the data source, the number of layers is tuned accordingly.

Training begins by calculating the input sum of the weighted parameters (and the bias \textit{b}), thus:
\begin{align}
z = \sum_{i=1}^{n}w_ix_i + b. \label{eq:1}
\end{align}
The output can be noted as $y = f(z)$. Function \textit{f} is known as the \textit{activation function}, and it is highly \textit{non-linear}. The process of training the weights goes through two main processes: the first one is computing gradients using the \textit{backpropagation} algorithm \citep{rumelhart_learning_1986, hinton_fast_2006}, and secondly, an optimization procedure generally using \textit{gradient descent} methods \citep{kingma_adam_2015, ruder_overview_2016}. From Eq.~\eqref{eq:1}, the cost function (i.e., mean squared error) can be defined as:
\begin{align}
C(w) = \frac {1} {n} \sum_{i=1}^{n} (y'^{(i)}- y^i)^2 \label{eq:cost}
\end{align}
where $n$ is the number of samples, $y'$ is the predicted value and $y$ the actual value.

In its simplest form, given the one-directional transmission of information in a neural network, is called feedforward neural network. In the stacked layers of feedforward neural network architectures, it is, in fact, the output of a layer that defines the input of the following. When a feedforward neural network has no hidden layers, it is called a \textit{perceptron} \citep{rosenblatt_perceptron_1957}.   Besides feedforward neural networks, there exists another class of neural networks called hopfield neural networks \citep{hopfield_neural_1982}, that represent a class of \textit{recurrent} and fully interconnected networks.

Several stacked layers of a neural network introduce deep neural networks (DNNs) make such an architecture a \textit{deep} architecture. Even though deep learning is a much older paradigm, the last decades have brought the invention of many widely applied deep learning architectures \citep{goodfellow_deep_2016} based on feedforward and recurrent networks, notably convolutional neural networks (CNNs), several architectures of recurrent neural networks (RNNs), --- such as long-short term memory (LSTM) --- generative adversarial networks (GAN), deep Boltzmann machines (DBMs), variational autoencoders (VAEs) and others. Each of the available architectures might a better fit for different problems. CNNs for instance, work particularly well with images and are applied to problems in computer vision \citep{szeliski_computer_2010, krizhevsky_imagenet_2012}. Computer vision problems are machine learning applications that train the computer program to identify images. Along with CNNs, RNNs are usually go-to candidates for natural language processing (NLP) problems \citep{yin_comparative_2017}. NLP represents a set of problems that usually require identification of natural human language.

\subsubsection{Scaling DNNs}
It is evident that there are many problems for which neural networks are good candidates as a solution, including classifying objects, image recognition, forecasting, medical diagnosis and more. Inspired from the idea that classical approaches of neural networks and deep learning are a machine learning success story, these techniques have begun their journey in the quantum world as well. The quantum approaches and their achievements are further explored here in Section~\ref{sec:qml}.

\begin{figure}
    \centering
    \includegraphics[width=\textwidth]{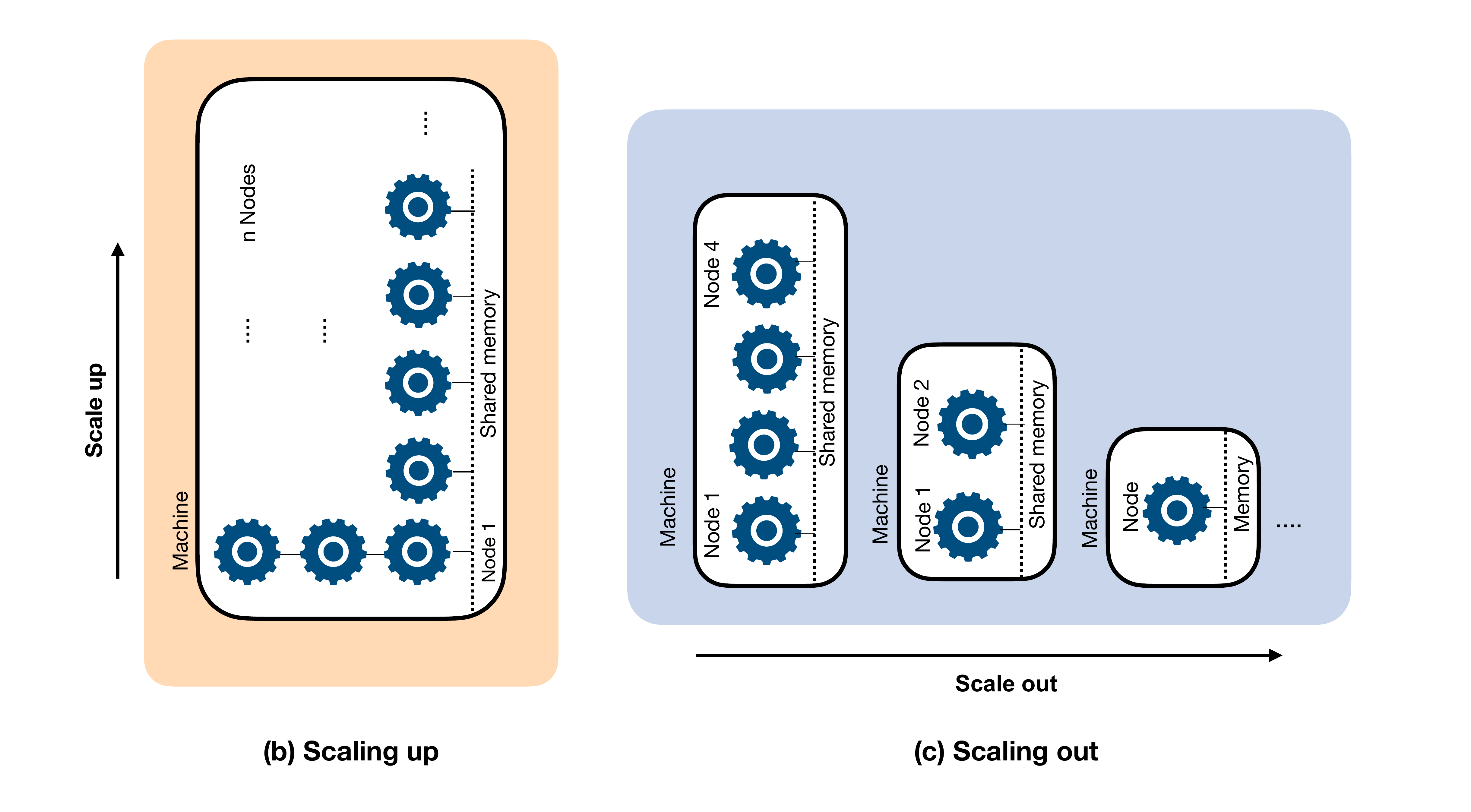}
    \caption{Computational architecture scaling. $(a)$ is an illustration of the \textit{scaling up} method of computation. In this approach, the processing power is increased as more core processing units are added to a single device. Whereas in $(b)$, the computational capacity is \textit{scaling out}, which represents connecting distinct smaller devices each with an individual number of processing units to achieve higher processing capabilities. The later is the distributed approach we assume for scalability here.}
    \label{fig:diagram}
\end{figure}

Oftentimes it suffices to have a single machine to perform tasks. But, processing a task requires computational power. More complex tasks require more computational power in which case the processing system needs to be scaled in terms of resources. For smaller scales of processing it remains convenient to add resources to the same processing machine. This approach is known as \textit{scaling up}, or vertical scaling (Fig.~\ref{fig:diagram}\textcolor{red}{a}). In reality, any processing machine can be scaled up, however the cost of production becomes exponential the higher we need to scale. A more pragmatic solution is often given by \textit{scaling out}, also known as horizontal scaling (Fig.~\ref{fig:diagram}\textcolor{red}{b}). In simple terms, this means having the required number of resources in different machines, rather than in a single machine. At large scales, this solution is more cost efficient. This outlines the need for distributed systems. 

It is often implied that a distributed system is running a single process (task) at a time. In other words, all the participating devices are working towards one single output. Albeit, reliable distribution of resources and processes has its own challenges. Having resources distributed to form a cluster requires communication an synchronization protocols. Relevant in this context, one issue that is prevalent directly in the training of neural networks is the communication overhead.

Another reason that motivates distributed training of deep learning architectures, is the fact that either the dataset or the model could get prohibitively large. It is because of these two elements that can be paralleled, that there exist two techniques of distribution: data parallelism and model parallelism (Fig.~\ref{fig:data-model-parallelism}). In these scenarios either the dataset or the model are split across nodes, respectively. Parallel or distributed processing often has different connotations. Parallel processing can be used in terms of multi-core processing in a single device; while distributed processing refers to the processing taking place in different nodes. The goal in either processing remains the same: to output the result coming from all the devices as if it were coming from one. Which is why oftentimes the two terms are used interchangeably. The data and model-parallel distribution architectures can be used in either context. In our theoretical assumptions, we assume that distribution takes place in different devices, which we will refer to as nodes. While we refer to the collection of nodes in a distributed architecture as cluster.

\subsection{Distributed Deep Learning}
\begin{figure}
    \centering
    \includegraphics[width=\textwidth]{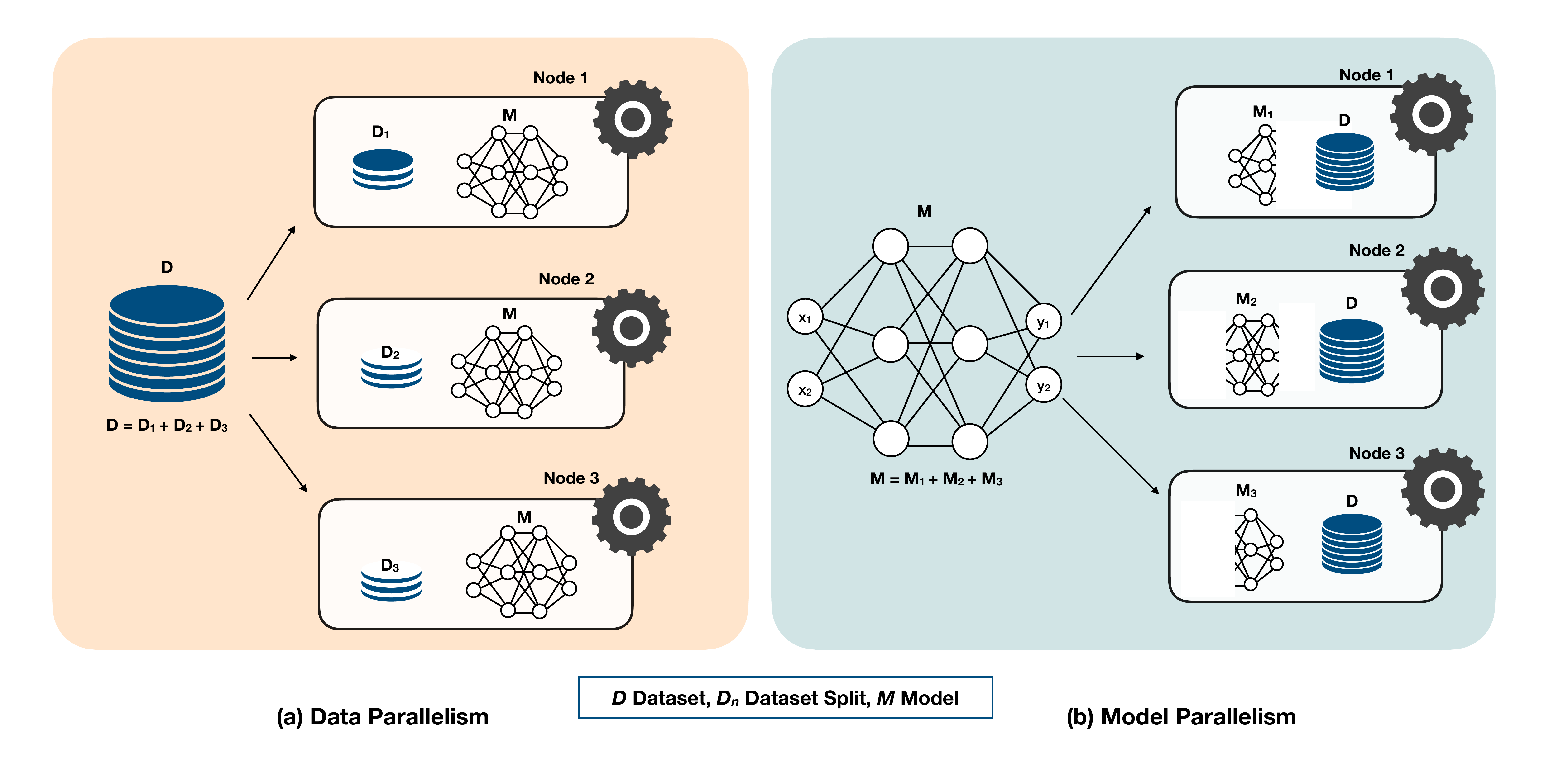}
    \caption{Architecture overview of data parallelism and model parallelism approaches in distributed neural network training. $a)$ The dataset $D$ is split in three equal parts ${(D_1, D_2, D_3)}$ across $n$ available devices (here three nodes), where each device holds an identical copy of the entire model $M$. $b)$ The model $M$ is split across $n$ devices (here three nodes), while each device holds a copy of the entire dataset $D$. In both scenarios, parameters are subsequently synchronised among the devices either asynchronously or synchronously. Gradients are exchanged using one of the parameter exchange protocols such as the MPI.}
    \label{fig:data-model-parallelism}
\end{figure}

When training a deep learning architecture, there are two elements that could become prohibitively large: the dataset or the model \citep{bennun_demystifying_2019}. Either the working dataset or the model may be too large to fit into a single available device. Inspired by techniques from parallel computing, the solution to overcoming this limitation is in distributing the largest elements. The first to consider is data parallelism. In this scenario, the dataset is split across the available nodes, while each node holds an entire copy of the model. The second approach, model parallelism, assumes the model is split across the nodes, while each node holds an entire copy of the dataset. Distribution of resources across several nodes takes several forms (Fig.~\ref{fig:data-model-parallelism}). The data and the model approach are inherently linked to other parameters to consider when building a distributed architecture. In data parallelism the dataset is distributed across different nodes, while each of the nodes hold an entire copy of the model. Model parallelism has the same logic, with the model distributed across the nodes, while each node contains an entire copy of the dataset.

\subsubsection{Data and model parallelism}
Data parallelism techniques used to train neural networks are very often focused on training CNNs \citep{zinkevich_parallelized_2010, niu_hogwild_2011, dean_large_2012}. For more, see Table 1 in Ref. \citep{mayer_scalable_2019} for a categorization based on the proposed architectures in the data-parallel approach. Model parallelism on the other hand has been explored in several works such as Refs. \citep{coates_deep_2013, dean_large_2012}. DistBelief \citep{dean_large_2012} is a framework that allows the training of a model in a parameter-sever architecture. Data parallelism is more used and explored in the DDL scheme, in part because it allows better cluster utilisation \citep{langer_distributed_2020}. In some works, there exists the so called domain parallelism approach which can be sub-categorized as a data parallel approach \citep{gholami_integrated_2017}. In domain parallelism, the data points themselves are split across different processors. Furthermore, beyond data and model parallel approaches, there are other approaches to classification of distributed protocols. One such notable architecture is pipeline parallelism \citep{huang_gpipe_2019, narayanan_pipedream_2019} which involves pipelining the network layers in different nodes. It can also be inferred that there exist hybrid approaches to distribution, which make use of distributing both the model as well as the dataset \citep{gholami_integrated_2017, xing_strategies_2015, jia_beyond_2019}. The DistBelief architecture mentioned earlier, is one such hybrid architecture. A study in Ref. \citep{krizhevsky_imagenet_2012} on parallelising the training of CNNs, it proposes to split the two different type of layers constituting the architecture of modern CNNs, in two different ways. Notably, to convolutional layers which contain the majority of computation, one can apply data parallelism. While for fully-connected layers which contain a small amount of computation, model parallelism may be more suited. In this work however, we focus on the primarily distinctions between data and model distribution in the quantum setting.

\subsubsection{Centralised and decentralised architecture}
\begin{figure}
    \centering
    \includegraphics[width=\textwidth]{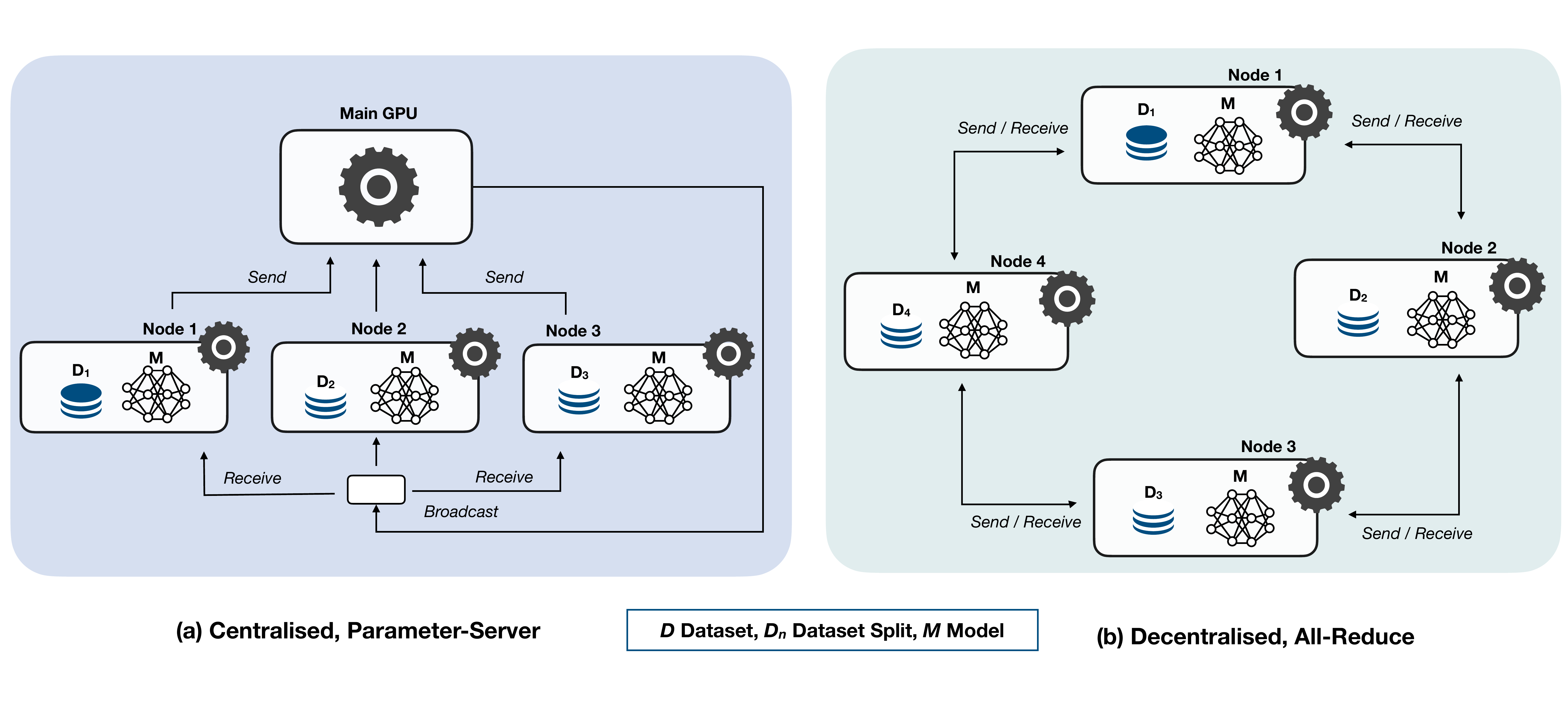}
    \caption{Centralised and decentralised exchange of gradients in two distributed setting architectures. In $a)$ we see the \textit{main node} and three secondary nodes sending gradients to the main node as well as receiving gradients broadcasted from the central node. This architecture is known as the parameter server scheme. In $b)$, four nodes each sending and receiving gradients in an all-reduce scheme without the need for a central node to orchestrate the communication. In both scenarios the dataset $D$ has been cut into $n$ equal splits, while the model $M$ remains intact in every node.}
    \label{fig:cent-decent}
\end{figure}

When designing concrete architectures based on either distribution, there are a number of choices one can make. First and foremost, the distributed architecture can be centralised or decentralised, as in Fig. \ref{fig:cent-decent}. In a centralised architectures, there is one appointed node that collects and broadcasts the information. In the jargon of DDL, this analogy is known as the parameter server architecture \citep{li_scaling_2014, gupta_model_2017}. In contrast, a decentralised architecture does not employ a parameter node that orchestrates communication \citep{sergeev_horovod_2018, daily_gossipgrad_2018}. It instead, employs communication techniques such as the all-reduce algorithm. In this scenario, each of the nodes has the same role of calculating, sending, and receiving gradients. It remains an open question as to whether the centralized or the decentralized approach is more suited to which scenarios. Evidently, that depends on several factors, and there may not be an architecture to fit all use-cases. The obvious drawback for the parameter server method is that the main nodes can quickly become communication bottlenecks, potentially leading to failure. On the other hand, in a decentralized architecture the communication cost increases with the number of nodes. This can lead to increased network maintenance complexity. There are works which evaluate the two approaches under certain conditions. For instance, Ref. \citep{lian_can_2017} concludes that there exists a regime in which decentralised algorithms outperform centralised ones in the distributed setting, in the scenario when the communication in the network remains low. As quantum technology evolves, it is likely that higher-level functions will continue to be performed by centralized classical devices, while low-level computations are distributed among several QPU nodes. 

\subsubsection{Synchronous and asynchronous scheduling}
Another distinctive feature of the topology of choice is the way in which parameters are exchanged --- a problem known as \textit{scheduling}. In the scenario of the deep learning distribution, the parameters that need to be exchanged are the calculated gradients. The scheduling can take the form of synchronous or asynchronous scheduling. In the former, the nodes \textit{wait} on each other for the exchange of the gradients and gradients are exchanged only when all the working nodes have finished the respective calculations. Given that some nodes may be faster than the others, this technique facilitates a uniform exchange. The result is broadcasted at the same time to all the nodes, once all the nodes have finished calculation \citep{iandola_firecaffe_2016, coates_deep_2013}. Asynchronous communication on the other hand, implies that the gradients are exchanged as soon as respective nodes have finished their designated calculations. When speaking of good cluster utilization, it is the asynchronous communication that comes to the picture. In asynchronous communication neither of the nodes waits for the progress of the other nodes. The faster nodes are not hindered by the slower ones. The result is broadcasted to the the nodes that have finished the communication without the barrier of waiting on the slower workers \citep{niu_hogwild_2011, dean_large_2012, keuper_asynchronous_2015}. There are evident advantages and disadvantages with either of the techniques further discussed in Ref. \citep{chahal_hitchiker_2020}. Beyond the canonical approaches, there exist more relaxed scheduling strategies such as the stale synchronous \citep{gupta_model_2017, ho_more_2013} and the non-deterministic communication methods \citep{bennun_demystifying_2019}. In the context of quantum computation, new limitations arise in communicating quantum information. However, classical co-processors will likely be employed in any use of QPUs, and will be relied heavily upon in such hybrid scenarios to optimize QNNs.

\subsubsection{Communication protocols}
When it comes to the exchange protocols used to facilitate the communication, this is where techniques from high performance computing (HPC) come in. One of the most used methods is the all-reduce algorithm that takes on various forms depending on the architecture \citep{thakur_optimization_2005}. Several out-of-the-box software packages provide access to distributed training. As such, gradients are exchanged using certain communication protocols. For instance, in Horovod \citep{sergeev_horovod_2018}, the training is supported in the ring-allreduce architecture to facilitate data parallel training approach \citep{patarasuk_bandwidth_2009}. Horovod uses message passing interface protocol (MPI) for sending and receiving the gradients \citep{walker_mpi_1996} among the nodes. As we will see shortly, quantum information cannot be copied, so the naive application of many communication protocols does not apply to the communication of quantum information. Generalizations exist, but require many advances in quantum technology infrastructure.

\section{Essential Quantum Computing}\label{sec:qcomp}
\subsection{Fundamental concepts}
Quantum computing is based on the principles of quantum mechanics. The idea of using the postulates of quantum mechanics to build a \textit{new kind of computer} was first introduced in the 1980s in two seminal studies by Benioff and Feynman \citep{benioff_computer_1980, feynman_simulating_1982}. Feynman's proposal of quantum computation is backed by the idea that our quantum universe can only be simulated by quantum computers –-- per contrast to classical computers. Another negative argument that we must move to quantum from classical is the end of Moore's law \citep{prati_quantum_2017, markov_limits_2014}, which famously extrapolated the trends of computing and predicted that computing power will double every two years. To achieve this, transistors have been shrinking in size at a comparable rate. However, things can only shrink so much before they are the size of an individual atom --- at which point, control over them would effectively render them as components of a quantum computer.

There are, of course, positive arguments for quantum computers as well, which often begin with the promise of exponential speed-ups for some quantum algorithms \citep{nielsen_quantum_2011}. By now, there are dozens of quantum algorithms that can provide speed-ups over their classical counterparts \citep{montanaro_quantum_2016}. Training of neural networks is one. But, before jumping straight into QNNs, we first overview the basic quantum information concepts required.

\subsubsection{Quantum Bits}

In quantum computers, the information is processed via the means of its building blocks called qubits. Unlike bits, qubits have the ability to be in \textit{superposition} and \textit{entanglement}. The parallel of a qubit in the classical world of computing is a bit. A bit has two states $0$ and $1$, whereas a qubit has two states $\ket{0}$ and $\ket{1}$, and many other states as well. The two states $\ket{0}$ and $\ket{1}$ technically form a \textit{basis} in a  two-dimensional complex vector space (the $\ket{\cdot}$ symbol denotes its vector nature). This ability of a qubit to be in a continuum of its two basis states is called \textit{superposition}. Superposition simply represents a \textit{linear combination} of classical states:
 
 \begin{align}
    \ket{\psi} = \alpha \ket{0} + \beta \ket{1}\label{eq:superposition}.
 \end{align}
 Coefficients $\alpha$ and $\beta$ are complex numbers $(\alpha, \beta \in \mathbb{C})$ and are often called \textit{amplitudes}.
 
 Multiple qubits are represented as superpositions in a higher-dimensional vector space. For $n$ qubits, the basis states consist of all binary strings of length $n$: $\ket{b} = |b_1 b_1\cdots b_n\rangle $. Since there are $2^n$ such basis vectors, the entire space has dimension $2^n$ and an arbitrary state of quantum information can be written as
 \begin{equation}\label{eq:general state}
     \ket{\psi} = \sum_{b=1}^{2^n} \alpha_b |b\rangle,
 \end{equation}
where the amplitudes must satisfy a \textit{normalization} condition,
\begin{equation}
    \|\ket{\psi}\|^2 = \sum_{b=1}^{2^n} |\alpha_b|^2 =1.
\end{equation}

\subsubsection{Superposition and Entanglement}

A state $\ket{\psi}$ from Eq.~\eqref{eq:general state} may be simply one of the basis states. In this case, there is no superposition and the information could be represented by the bits labeling it. Often, a quantum computation is assumed to start in the so-called \textit{zero state} $\ket{00\cdots 0}$. 

Two or more interacting qubits exhibiting properties of correlation can be \textit{entangled}, which is easiest to introduce by example. The prototypical entangled state is the so-called \textit{Bell state}: $\ket{\Psi^+}=\frac {1} {\sqrt{2}} (\ket{00} + \ket{11})$. The state is entangled because it cannot be written as two individual single-qubit states. For a system of many qubits, most states are entangled. The easiest way to interpret entangled states is as a superposition of correlated classical states.

Understanding the entire nature of superposition and entanglement is an open research question. But, suffice it to say, at least \textit{some} of each is necessary to achieve novelty in a computation --- otherwise a classical computer could straightforwardly replicate it. Since most quantum computations are assumed to begin in the \textit{un}entangled state $\ket{00\cdots 0}$, entanglement must be \textit{built up} as the computation proceeds. 

\subsubsection{Quantum Gates and Circuits }

The high-level ideas of computation remains the same in the quantum setting as in the classical setting. Similar to classical computers that use gates, quantum computers manipulate qubits via \textit{quantum} gates. Gates map quantum states into other quantum states. In digital logic, the NAND gate is \textit{universal} --- any other logical function can be implemented using only this gate. Similarly any quantum gate can be decomposed into a sequence of one- and two-qubit gates drawn from a small finite set of universal gates. As such, is both sufficient and convenient to distinguish between gates that act on a single qubit and gates that act on two qubits.  

\begin{figure}
    \centering
    \includegraphics[width=0.50\textwidth]{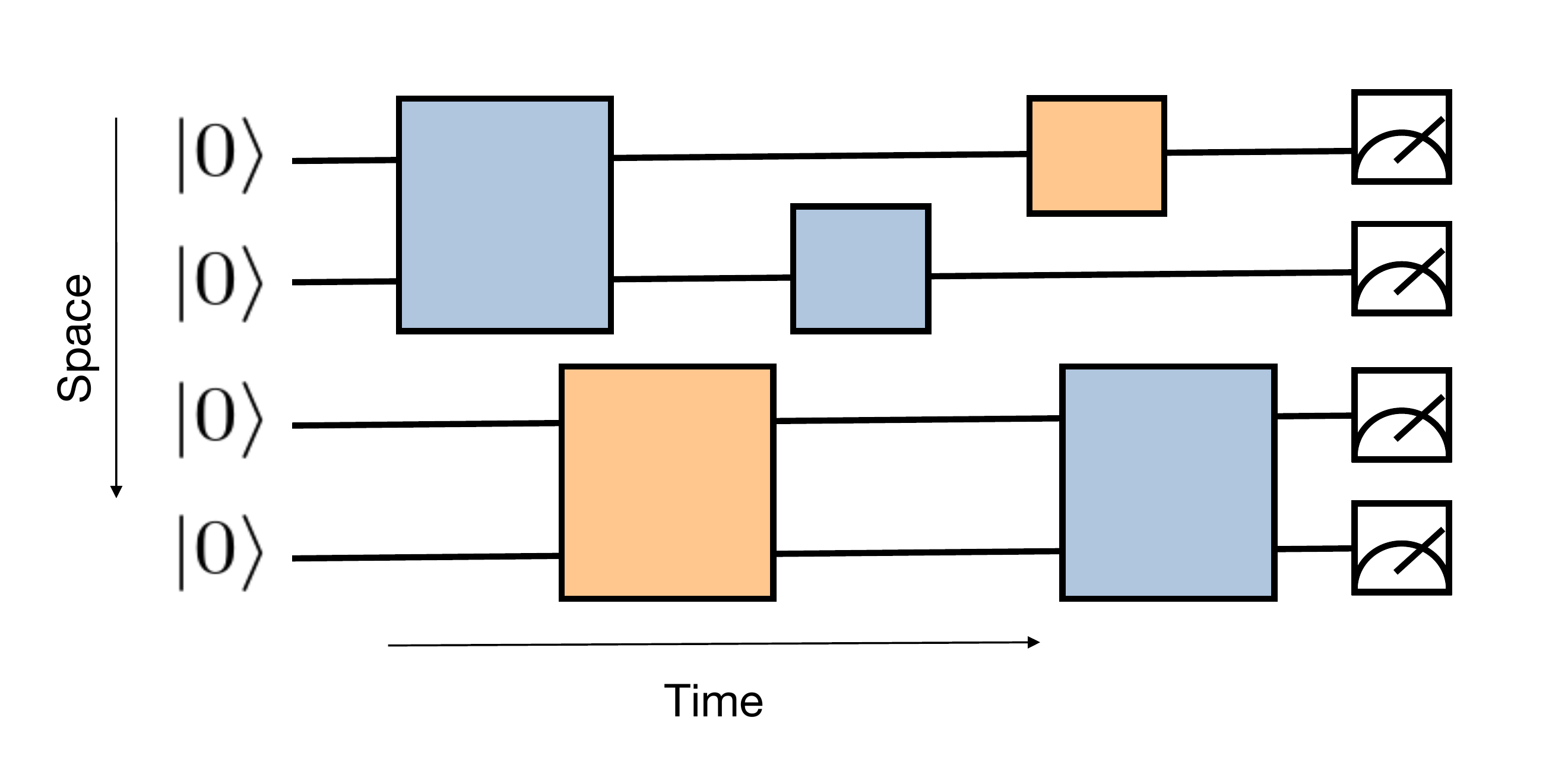}
        \caption{A sample quantum circuit in four qubits initialised in the ground state $\ket0$. The gates are applied chronologically from left to right, representing the arrow of time, followed by the measurement. Two qubit gates create entanglement, one qubit gates create superposition.}
    \label{fig:circuit}
\end{figure}

We will not need to know here which particular gates can or are often used, so we will imagine them as abstract and arbitrary. Quantum gates compose operations in a structured pattern forming quantum circuits. In Fig.~\ref{fig:circuit} the coloured boxes represent one qubit gates and two qubit gates. The boxes which go through two lines are two qubit gates, while the ones which go through only one line are one qubit gates.

In general, two-qubit gates create entanglement, which requires either a physical connection between pairs of qubits or some other communication which mediates the interaction.

\subsubsection{Measurement}
 In Fig.~\ref{fig:circuit}, the final symbol on the quantum circuit is the \textit{measurement}. This is how quantum data is \textit{read}.
  Measurement transforms qubits to bits. It is both probabilistic and irreversible, destroying any entanglement or superpositions in the process. For a general state as in Eq.~\eqref{eq:general state}, the outcome of the measurements is a single binary string $b$ or its corresponding basis state $|b\rangle $. The probability of observing that outcome is $|\alpha_b|^2$. A consequence of this is that quantum superpositions cannot be \textit{read} in the conventional sense. However, repeatedly measuring many equally prepared copies of quantum data can give sufficient statistical information to reconstruct it --- a process referred to as \textit{tomography}. 
     
One of the most fundamental facts about qubits is that no procedure exists which can create copies of them. This fact is often referred to as the \textit{no cloning theorem}. Since many communication protocols are predicated on creating and distributing copies of classical data, no-cloning presents an immediate challenge to naive generalizations.

\subsection{Distributed Quantum Computing}
The core concept of distributed computation naturally extends from classical to quantum computing. The underlying idea is that of using multiple quantum processors to process quantum information (input), towards producing one single output. By connecting multiple quantum devices over a network, one can achieve architectural scalability by scaling-out. The same principles of scaling as in Fig.~\ref{fig:diagram} can apply to quantum devices as well. Here we overview the main techniques that facilitate and promote distributed quantum computation. The idea of a scalable quantum architecture peaks with the ambitious project of the \textit{quantum internet} \citep{kimble_quantum_2008, wehner_quantum_2018, cacciapuoti_quantum_2020, cuomo_towards_2020, rohde_quantum_2021} as one of the main goals for distributed quantum computing. 

The quantum internet implies quantum devices connected in a quantum network style with classical and quantum communication links. This network will thus allow the communication between qubits on different devices apart from each other. A crucial element in the functionality of the quantum internet are quantum repeaters. Like classical repeaters, their role is to propagate the signal into the further nodes. In the same reasoning, the are placed between the nodes. However, unlike classical repeaters, quantum repeaters operate very differently in how they perforate the signal. Quantum repeaters perform the so-called \textit{entanglement swapping} protocol which allows for entanglement distribution. There exist quantum protocols that facilitate the exchange of classical information such as quantum key distribution and superdense coding \citep{cacciapuoti_quantum_2020}. Whereas quantum communication can occur over classical channels via quantum teleportation \citep{nielsen_quantum_2011}. Informally, teleportation requires two classical bits and an entangled pair of qubits to be transmitted between the sender and the receiver. Other than the hardware challenges which currently hinder most of the quantum research, the state of the development of the quantum internet remains with many interesting open challenges \citep{cacciapuoti_quantum_2020}.

At present day, there exist small quantum devices that can be accessed via the cloud \citep{ibm_cloud}. These cloud-based devices offer access to quantum computation via the internet. Classically, cloud-based based approaches are certainly convenient due to their complete computation infrastructure accessible via the internet. However, a lot of the discussion around cloud computation revolves around the security of the network \citep{almorsy_analysis_2016}. On the quantum front, there exists the idea of blind quantum computation \citep{arrighi_quantum_2003, broadbent_universal_2009} which provides a barrier of encryption to either of the nodes accessing the information transmitted. In this protocol which is applicable in a cloud-based environment, the server receives an encrypted algorithm from the client. In this way, the protocol provides security under the assumption of the hidden calculations. However there are certain aspects the server will know about the calculation such as the bandwidth of the calculation size and allocated resources for execution. Much of the current research in this area is focused on the verifiable aspects of the blind computation \citep{fitzsimons_private_2016}.

The long-term vision of a quantum network, where superposition and entanglement are preserved, results in what can simply be interpreted as a single (albeit very large) quantum processor. Ensuring that processor works well will surely require concepts properly termed \textit{distributed quantum computation} in analogy with the classical techniques they will borrow from. But, here we are interested in the bottom-up problem, wherein we assume at some point in the nearer future we will have access to multiple small QPUs, not necessarily connected to a quantum internet, and ask: \textit{can we use these in parallel to train a QNN?}

\section{Quantum Machine Learning}\label{sec:qml}
\subsection{A brief history of QML}
Quantum machine learning encompasses a variety of algorithms that are, broadly speaking, of variational nature, as opposed to the more popular quantum algorithms, such as Shor's algorithm \citep{shor_polynomial-time_1995} that are deterministic in nature. Other kinds of algorithms which can be called deterministic include Refs.~\citep{deutsch_quantum_1985, deutsch_rapid_1992, grover_fast_1996, cleve_quantum_1997, brassard_quantum_1998, montanaro_quantum_2016}. Here we are concerned with the variational ones. Quantum machine learning is the emerging relationship between quantum computing and machine learning. Collectively, the term QML, is used interchangeably in several distinct scenarios regarding the direction of the field and the components used. The directions can take the form of quantum phenomena improving machine learning algorithms, or machine learning algorithms further improving quantum algorithms and designs. The two components needed for this scenario to work --- data and algorithms --- in either case, can be quantum or classical. Below we take a look at the four main paradigms.

\begin{figure}
    \centering
    \includegraphics[width=0.65\textwidth]{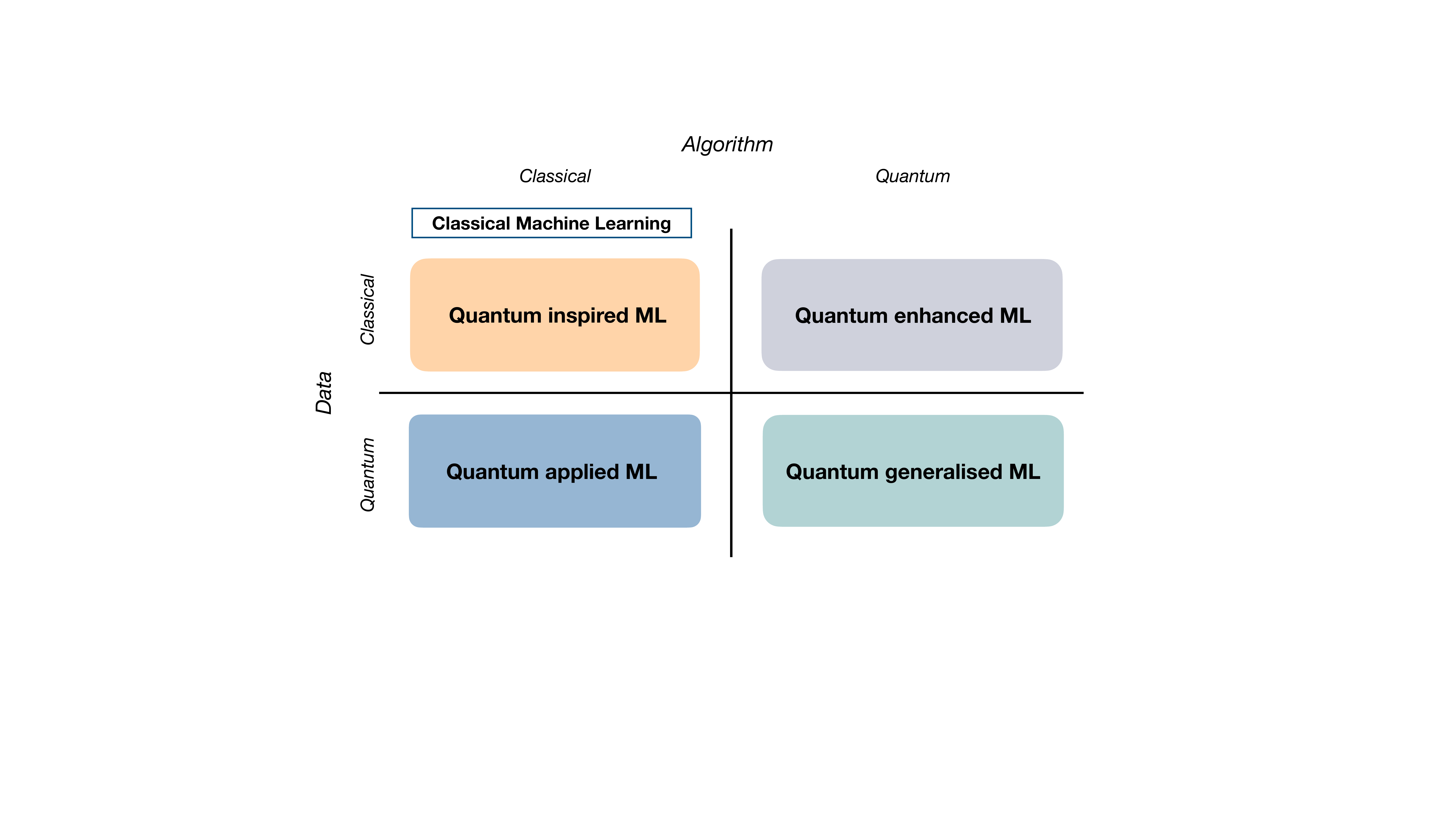}
    \caption{The four quantum machine learning development paradigms compared against data and algorithms type, either of which is considered to be classical or quantum. Classical ML on the upper-left corner for context.}
    \label{fig:qml}
\end{figure}

\subsubsection{Four paradigms}
The first big chunk and usually the entry point in QML is called \textit{quantum-enhanced machine learning}. In this scenario, machine learning analysis of classical data is processed on a quantum computer. In Ref. \citep{dunjko_quantum-enhanced_2016} propose an agent-environment paradigm in four scenarios in which either is \textit{Classical} or \textit{Quantum} \textit{(CC, CQ, QC, QQ)} (Fig.~\ref{fig:qml}) as an attempt to give this new field more organization and perhaps a direction. The context of quantum-enhanced machine learning is desirable due to the power of quantum computers to work with complex linear and matrix computations, as well as the idea of quantum parallelism. The inspiration stems from the fact that the large amount of data needed for machine learning algorithms to yield better results will harness this power, consequentially leading to improvements in runtime and convergence time \citep{lloyd_quantum_2013}. That is also the main goal of this type of setup --- speedups. However, in this case, data needs to be encoded into a quantum state, then queried and retrieved from a quantum RAM --- that introduces issues of its own such as whether the time cost of this action is too high to pay for, in turn, quantum speedups. 

The second direction, \textit{quantum-applied machine learning} is concerned with finding optimal ways to apply machine learning in quantum experiments with the goal of enhancing their performance or finding solutions. These various applications encompass accomplishments beyond quantum computing application and in particle physics, quantum many-body physics, chemical and material physics, and more \citep{carleo_machine_2019, dawid_modern_2022}. To zoom in, some important implementations in quantum computing that have shown promising results take place in quantum control \citep{bukov_reinforcement_2018, niu_universal_2019}, quantum error-correction \citep{nautrup_optimizing_2019, torlai_neural_2017}, quantum state tomography \citep{torlai_neural-network_2018, xu_neural_2018}.

The third paradigm, \textit{quantum-inspired machine learning}, comes up with new ways to design and evaluate classical machine learning algorithms, that are primarily inspired by quantum theory. As reviewed in Ref. \citep{arrazola_quantum-inspired_2020}, the complexity gap between classical and quantum algorithms keeps changing with the new algorithms coming into the picture and the complexity bounds are still somewhere between polynomial and exponential. Due to their relevance in machine learning algorithms, the study in question reviews the ``flagship'' algorithms of quantum computing – quantum algorithms for linear algebra \citep{harrow_quantum_2009, kerenidis_quantum_2016}. Ref. \citep{arrazola_quantum-inspired_2020} questions whether these asymptotic bounds achieved via quantum processing in several quantum-inspired algorithms may be useful in practical real-life applications. The study in Ref. \citep{tang_quantum-inspired_2018} explores the realm of linear-algebraic operations applied in recommendation systems which builds on the work of in Ref. \citep{kerenidis_quantum_2016} that proves exponential improvements over classical algorithms for recommendation systems. However, Ref. \citep{tang_quantum-inspired_2018} narrows that gap by proving that another class of classical algorithms reaches the same exponential improvements. 

The fourth category, \textit{quantum-generalized machine learning} or \textit{fully-quantum machine learning} is the case where the data, as well as the infrastructure, are bona fide quantum. Given the still lagging state-of-the-art of the two components, this approach remains rather futuristic, to be answered at its full scale at this point in time. Nevertheless, among the first attempts to generalize classical machine learning models have been proposed in line with unsupervised classification protocols for quantum data \citep{sentis_unsupervised_2019} and quantum anomaly detection \citep{liu_quantum_2018}, among others.

\subsubsection{Translational QML}
Several of the classical machine learning algorithms have been appropriated in the quantum realm: quantum support vector machines \citep{anguita_quantum_2003}, quantum principal component analysis \citep{lloyd_quantum-principal_2013}, quantum reinforcement learning \citep{dong_quantum_2008}, quantum algorithms for clustering \citep{aimeur_machine_2006}, quantum recommendation systems \citep{kerenidis_quantum_2016} and many others. A notable subroutine on which many such QML works are based on is the so-called HHL algorithm \citep{harrow_quantum_2009}, which proposes a solution to the linear systems of equations using quantum operations. In turn, HHL achieves exponential improvement in time complexity over the best known classical algorithm for the same task. However, there are certain strict conditions that must be met that could otherwise hinder the time advantage. For an analysis of its caveats see Ref. \citep{aaronson_read_2015}, while for an overview of the HHL in some QML methods, see Ref. \citep{duan_survey_2020}.

More recently, QML research has slightly shifted focus beyond beyond computational complexity comparisons with the classical counterparts, to the flavour of building \textit{better} quantum models \citep{schuld_quantum_2022}. In this context, several works \citep{holmes_connecting_2022, sim_expressibility_2019, abbas_power_2021, wright_capacity_2019, du_expressive_2020, banchi_generalization_2021, hubregtsen_evaluation_2021} explore expressibility, generalization power and trainability of a model --- all crucial elements when building robust learning algorithms. Attention is also given to the complexity bounds that shift between classical and quantum data for quantum models \citep{liu_quantum_2018, huang_power_2021, huang_information_2021}.

\subsection{Quantum Neural Networks}
Quantum neural networks represent a class of hybrid quantum-classical models that are executed in both quantum processors as well as classical processors to perform one single task. QNNs are currently one of the most trending topics in quantum machine learning \citep{beer_training_2020}. They are often interchangeably referred to as variational or parameterized quantum circuits (VQCs or PQCs) \citep{mcclean_theory_2016, bharti_noisy_2022, cerezo_variational_2021}. Several studies review more in-depth the increasing body of proposed methods for implementing a QNN or similar model classes \citep{schuld_quest_2014, benedetti_parameterized_2019, mangini_quantum_2021}.

Some of the first works that address the question of quantum neural networks do so from a biological perspective extending on the idea of cognitive perspectives \citep{kak_quantum_1995, chrisley_quantum_1995, lewenstein_quantum_1994}. Others similarly early ones do so from a hardware perspective \citep{behrman_simulations_2000, ventura_quantum_2000}. However, with more contemporary approaches concerning QNNs, its definition has evolved with now to refer to tangents in classical artificial neural network research due to their parameters which require optimization via a training procedure.

\begin{figure}
    \centering
    \includegraphics[width=0.65\textwidth]{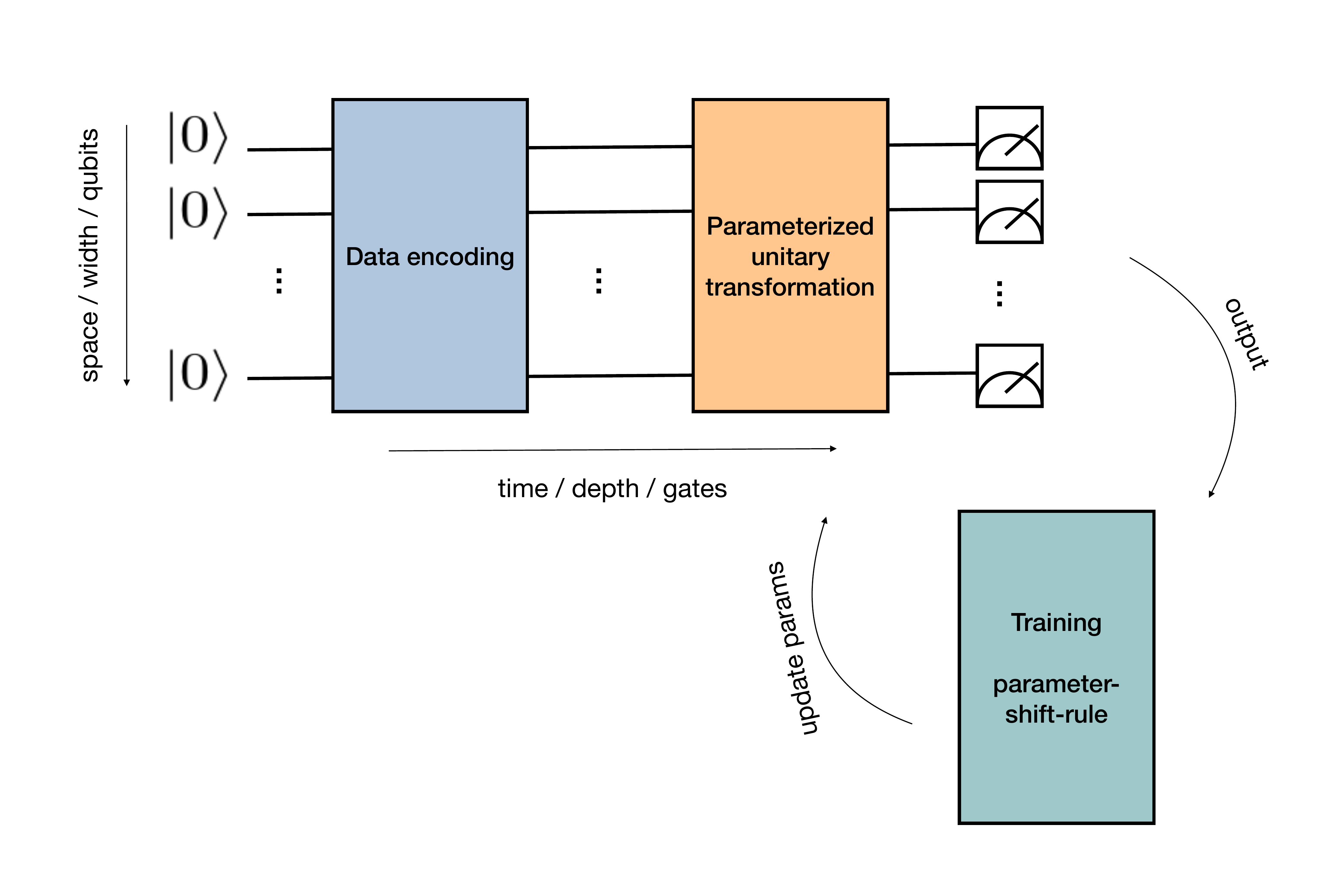}
    \caption{A basic structure of the parameterized quantum circuits with qubits and gates analogy, involving the data encoding stage, the ansatz to be optimized, measurement and an optimization scheme.}
    \label{fig:pqc}
\end{figure}

The QNN architecture has a structure which loosely resembles that of classical neural networks, depicted in Fig.~\ref{fig:pqc}, hence the analogous name. Evidently, when working with quantum data, a preliminary step is to encode the classical data into quantum states. Otherwise, the first step of the QNN training procedure is to define a cost function $C$, which as in the classical case, maps the actual parameter values to the predicted ones. This step is then followed by the circuit with parameters $U(\theta)$ which need to be optimized using an optimization strategy --- often referred to as ansatz or the parameterized quantum circuit (i.e., \citep{kandala_hardware_2017}). This step of the procedure resembles the multi-layered architecture of neural networks, as the ansatz can be composed of multiple layers with the same architecture. The estimation of the gradients $C(\theta)$ occurs in a quantum machine. The optimization task is thus to minimize the value of the cost function. This is followed by the measurement step which is used to introduce non-linearity. The output of the measurement is then compared with the cost function dependent on the task via the training procedure and then the parameters are updated accordingly. Different types of classical optimizers are used for training $\theta$ often based on the gradient descent methods \citep{cerezo_variational_2021, sweke_stochastic_2020}.

Evidently, it is natural that there remain several open issues in quantum neural networks research. One of the main challenges for QNNs remains the linear-nonlinear compatibility between neural network computation and quantum mechanics. Neural network computation is done in a non-linear fashion, that is, the activation function which triggers each neuron is non-linear, otherwise the idea of layers in neural networks would serve no purpose. On the other hand, quantum systems behave in a linear way, which gives rise to the first incompatibility. Among other works and proposals in response to this caveat, Ref. \citep{cao_quantum_2017} designs a quantum neuron as a building block to quantum neural networks based on the so-called repeat-until-success technique to get past the linearities of quantum circuits. Several other fundamental issues are discussed and summarized in Ref. \citep{allcock_quantum_2019} including, the sequential nature of training neural networks, which clashes with the parallel processing power of quantum algorithms. In essence, taking advantage of quantum superposition and managing to parallelize the training of neural networks is a step in the right direction, however in training neural networks data is calculated and stored at many intermediate steps --- an inherent property of the backpropagation algorithm. A recent approach suggest using the rule called parameter-shift, which mimics the way backpropagation works \citep{mitarai_quantum_2018, schuld_evaluating_2019, schuld_circuit_2020}. Finally, the parameters needed for training needs to be encoded in quantum states, a process which is time-consuming, and the topic of further discussion in Sec.~\ref{section:data}. 

To add to the discussion, QNNs are prone to the so-called barren plateaus phenomenon \citep{mcclean_barren_2018, cerezo_cost_2021} which entail flat region in the optimization landscape for even modest numbers of qubits and gates. Although there has been progress towards escaping this phenomenon be it by initialisation methods \citep{grant_initialization_2019} or newer QNN architecture that are not prone to barren plateaus \citep{pesah_absence_2021}. Despite the inherent drawbacks, there are continuous attempts to resolve these issues and unite the two paradigms of AI and QC due to the seemingly promising rewards. To support this, there have been a number of notable works which go along the lines of optimizing versions of parameterized quantum circuits for quantum data \citep{cong_quantum_2019, beer_training_2020}. Furthermore, to align with the deep learning architectures, there are several proposals that extend the main deep architectures into quantum structures such as RNNs \citep{bausch_recurrent_2020}, CNNs\citep{cong_quantum_2019, henderson_quanvolutional_2019, kerenidis_quantum_2020}, GANs \citep{lloyd_quantum_2018, dallaire_quantum_2018, zoufal_quantum_2019} and more \citep{mangini_quantum_2021}. Further enhancing the capabilities of these structures is one potential avenue where the research will go. In the context of QNNs as well, there is an emphasis on the expressibility, trainability and generalization power of these model classes.

However, whatever direction QNN research and applications take, the need to scale-out will soon become apparent, which brings us to distributed QNNs.

\section{Data Parallelism: Splitting the Dataset}\label{section:data}
\begin{wrapfigure}{R}{0.50\textwidth}
    \centering
    \includegraphics[width=0.48\textwidth]{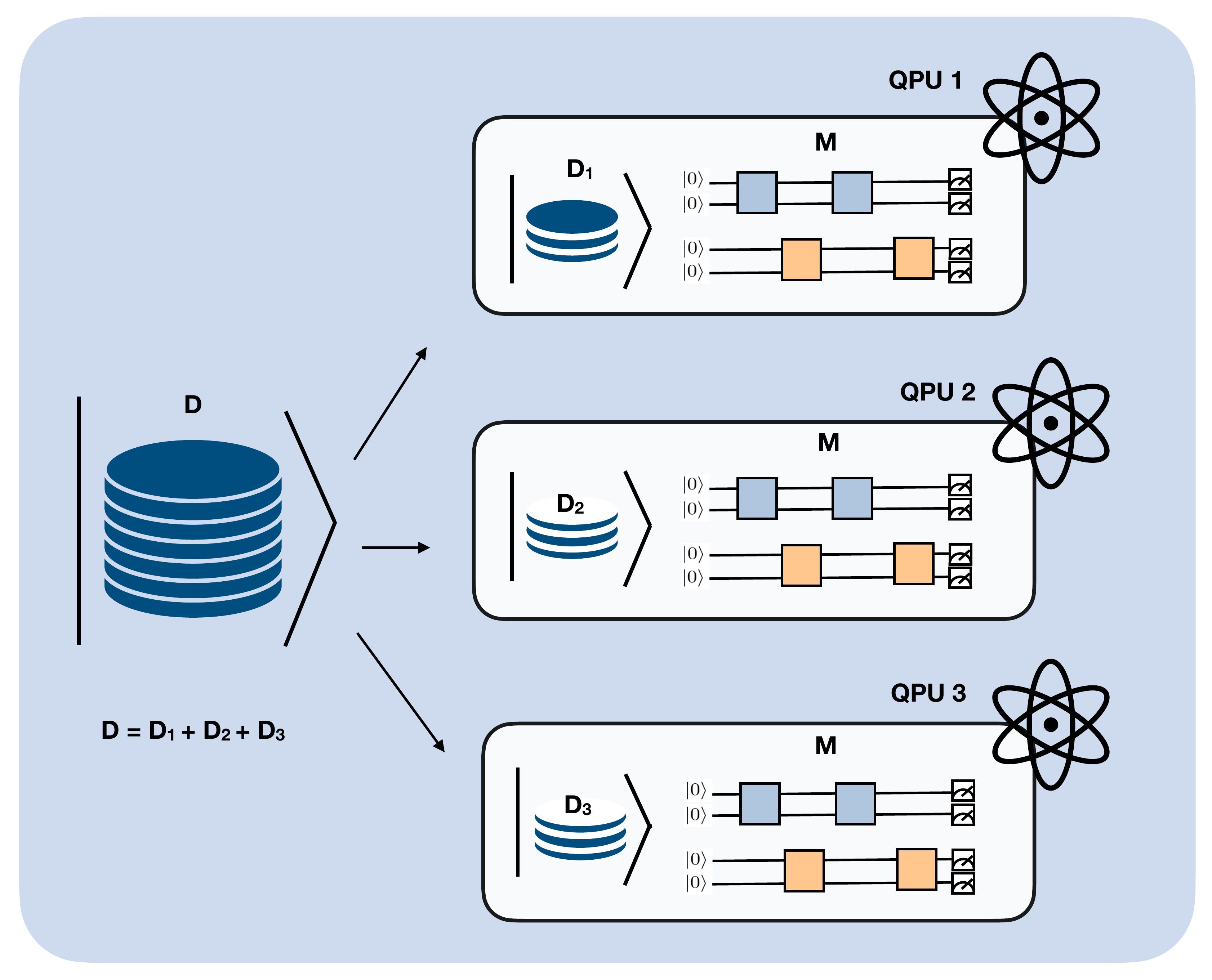}
    \caption{The split of the quantum dataset across three quantum nodes (QPUs). The quantum model is kept unchanged and loaded into all three nodes.}
    \vspace{-10pt}
    \label{fig:qdata}
\end{wrapfigure}
The concept of data in quantum processing is very different from that in the classical world. Straightforwardly, quantum data is the data which is output from any quantum computer or quantum processor. To explain it, one can contrast it with how classical data works \citep{resch_quantum_2019}. Classical data can be saved in permanent storage, moved, and copied as needed. On the other hand, quantum data is rather short-lived. Its lifetime ends with the end of the execution of a program. A very different property is that quantum data cannot be copied as per the no-cloning theorem. The no-cloning theorem does not allow the creation of an identical copy of an arbitrary quantum state \citep{wootters_single_1982}. The discussion on quantum data is tightly linked to its processing mechanism, such as a quantum random access memory (QRAM) \citep{giovannetti_architectures_2008, giovannetti_quantum_2008, arunachalam_robustness_2015}. Being able to retain quantum states longer or query them requires storage capacities to be put in place. This becomes particularly relevant in the discussion in quantum machine learning. In what we call quantum-enhanced machine learning, classical data needs to \textit{a priori} be encoded into quantum states, which inherently is a time-costly process \citep{aaronson_read_2015}. Additionally, for many of the proposed approaches, the presence of a QRAM is a mandatory feature. On the other hand, fully quantum machine learning that operates with quantum data is starting to sprout, and there are reasons to believe that it will be more of an effective direction, as it removes the need for quantum pre-processing.

As is the case with the enhanced QML algorithms workflow, the dataset first needs to be encoded into quantum states \citep{schuld_supervised_2018}. In this context this is the case when working with classical data and quantum algorithms (CQ). As such, data encoding is a crucial part in designing quantum machine learning algorithms. There exist several methods of encoding classical data applied across several works in QML make use of data encodings frameworks \citep{farhi_classification_2018, rebentrost_quantum_2014, havlicek_supervised_2019, harrow_quantum_2009, schuld_circuit_2020, wiebe_quantum_2012, lloyd_quantum_2020, schuld_quantum_2019,  skolik_layerwise_2021} and further push the on-going research in this domain. Some of the most explored encodings in the context of QML include basis encoding, amplitude encoding, angle encoding (tensor product encoding) and hamiltonian encoding. We refer the reader to other encodings as well as more in-depth analysis in Refs. \citep{schuld_supervised_2018, weigold_data_2020, weigold_expanding_2021}. Depending on the purpose of the computation, there are certain techniques better suited than the others. Ref. \citep{weigold_data_2020} concludes that amplitude encoding allows for compact storage and as such can be useful for storing a large amount of data in a small number of qubits. Whereas basis encoding is preferred should arithmetic computations take place. Ref. \citep{larose_robust_2020} explores several encoding types in a noiseless environment as well as under the influence of noise for binary quantum classifiers. In general, encoding data into quantum states is far from being a straightforward process. This part of the workflow is often a bottleneck \citep{aaronson_read_2015} in achieving practical advantage. The research is still on-going and crucial to the success of quantum machine learning algorithms. Moreover, the question of data encoding is relevant beyond QML, such as in quantum simulations, another promising area of research.

Below we discuss two main data encoding techniques and how they would perform in data distributed equivalents.

First and foremost, let $D$ be a classical dataset of size $M$, where each data point $x$ is an $N$-feature vector.

\begin{equation}
D = \{x^{(1)}, ... x^{(m)}, ..., x^{(M)}\}\label{eq:data}
\end{equation}

The process of data distribution begins with splitting $D$ into $L$ available nodes:

\begin{subequations} \label{eq:data-dist}
\begin{align}
D_1 &= \{x^{(1)},\dots, x^{(K)}\} \label{eq:dist1}, \\
D_2 &= \{x^{(K+1)},\dots, x^{(2K)}\} \label{eq:dist2}, \\
&\vdots \nonumber\\
D_j &= \{x^{((j-1)K+1)},\dots, x^{(jK)}\} \label{eq:dist4}, \\
&\vdots \nonumber\\
D_L &= \{x^{(L-1)},\dots, x^{(M)}\}. \label{eq:dist3}
\end{align}
\end{subequations}

where $D = D_1 + D_2 + \dots + D_L$. Each of the $L$ splits of data is processed in a different node. They each hold an equal amount $K$ of different data points from the same dataset. Here we consider classical data to be quantum data after it is encoded into quantum states. When $D$ is encoded into quantum states using either of the available encoding techniques, what is produced is quantum data. We can denote the obtained quantum dataset with $\ket{D}$.

\subsection{Basis encoding with data distribution} \label{subsec:basis}
The first encoding technique we consider is the basis encoding. This procedure has two substantial steps. Prior to encoding, each data point needs to be approximated to some finite precision in bits. Typically, single bit precision is assumed for brevity. Otherwise, a constant number of extra qubits is required. We will follow convention here and assumed each feature is specified by a single bit such that each data point is an $N$-bit string. 

Each of the data points is encoded in a computational basis state uniquely defined by its bit string. The entire dataset is encoded as a uniform superposition of these computational states. The dataset defined in Eq.~\eqref{eq:data} in the basis encoding will result in the following quantum data:
\begin{equation}
    \ket D = \frac{1}{\sqrt{M}} \sum_{m=1}^M \ket {x^{(m)}},
\end{equation}
where $x^{(m)}$ represents a single random data point in the dataset. To encode the classical dataset $D$ into a quantum dataset, $N$ qubits are required (and a constant factor more if the features are represented with more bits). Preparing $\ket{D}$ requires $O(NM)$ gates \citep{ventura_quantum_2000}. 

Examples of basis encoding \citep{wiebe_key_2020} of QML techniques employed for different tasks include neural networks for classification \citep{farhi_classification_2018, schuld_implementing_2017}, quantum data compression \citep{romero_quantum_2017}, quantum Boltzmann machines \citep{amin_quantum_2018}, to name a few.

In the distributed context, assuming the split according to Eq.~\eqref{eq:data-dist}, to encode each of the portions of the dataset (i.e, Eq.~\eqref{eq:dist4}) it also requires $N$ qubits for each of the dataset chunks. We consider $\ket{D}$ to be one quantum state on $LN$ qubits. This will result in:
\begin{equation}
    \ket{D} = \ket{D_1} \otimes \ket{D_2} \otimes \dots \otimes \ket{D_j} \otimes \dots \ket{D_L} \label{eq:data-split}
\end{equation}
where each $\ket{D_j}$ is,
\begin{equation}
    \ket{D_j} = \frac{1}{\sqrt{K}} \sum_{m=(j-1)K+1}^{jK} \ket {x^{(m)}}.
\end{equation}

The preparation of each $\{D_j\}$ requires $O(NK)$ gates since each partition contains $K$ data points. In total, across all partitions, $O(NKL)$ gates are needed. Replacing the parameters from $K L = M$ yields $O(NM)$ gates, the same as in the undistributed scenario. This is not surprising, of course, but one still wonders what has been gained.

There are a few observations we can make. Firstly, using this approach of data encoding to perform data distribution, in the end, requires more qubits than the single node approach. While single node dataset requires $N$ qubits, $L$ splits of the dataset require $LN$ qubits. This way, the number of qubits required grows with the number of splits. However, the \textit{total} number of the gates $NM$ remains the same. In the end, what has been achieved with this splitting technique is the reduction of gates \textit{per node}, precisely by a factor of $L$.

Therefore, the positive aspect yielded in this procedure is the lower depth of state preparation per each individual split in comparison to the preparation of the larger circuit. Lower depth circuits obviously require less time to implement, but also incur fewer errors, which again translates to time in the error-corrected regime, but is far more relevant in the NISQ era. Errors grow at least linearly in the depth of the circuit, hence so-called \textit{shallow circuits} are of great interest, a fact we will discuss later.

More subtle is the notion of \textit{quantumness} in the distributed approach. While it is clear in splitting we may have lost the naive parallelism afforded by quantum data, it is also likely that a significant amount of entanglement will also be lacking. This can be naively inferred to as \textit{less} quantum as a solution, but not necessarily less powerful. As these considerations will be relevant to all splitting procedures we consider, further discussion of parallelism and entanglement will be deferred to Sec.~\ref{sec:parallelism}.

\subsection{Amplitude encoding with data distribution} \label{subsec:amplitude}
Amplitude encoding is another widespread encoding technique that is a widely used in the context of quantum machine learning. This technique uses amplitudes of the quantum state to encode the dataset \citep{schuld_supervised_2018}.

All the data samples with their attributes are concatenated and can be constructed as,
\begin{equation}
    \alpha = \left(x_1^{(1)}, \ldots, x_N^{(1)}, x_1^{(2)}, \ldots, x_N^{(2)}, \ldots x_1^{(j)}, \ldots, x_N^{(j)},\ldots, x_1^{(M)}, \ldots, x_N^{(M)} \right),
\end{equation}
which is a single vector of length $MN$. The dataset $D$ encoded in amplitudes can be characterized as:
\begin{equation}
    \ket{D} = \frac{1}{|\alpha|}\sum_{i=1}^{MN} \alpha_i \ket{i},
\end{equation}
where $|\alpha|$ is the normalization, or length of the vector $\alpha$:
\begin{equation}
    |\alpha|^2 = \sum_{i=1}^{MN}\alpha_i^2,
\end{equation}
which is necessary recalling that all quantum states require normalization.

Amplitude encoding is certainly a more compact way of encoding data in comparison to the basis encoding given that it requires $\log (MN)$ qubits to encode the dataset defined in Eq.~\eqref{eq:data}. Regarding the splitting of the dataset, here, as well, we assume Eq.~\eqref{eq:data-split}. The full dataset will be a tensor product of quantum splits which encodes each of the subsets of data.

Assuming $L$ splits, and applying Eq.~\eqref{eq:data-dist} and Eq.~\eqref{eq:data-split} invokes the following in amplitude encoding, a $\ket{D_j}$ split will be characterized as:
\begin{equation}
    D_j = \{x^{(j-1)(K+1)}, \dots, x^{(jK)}\}.
\end{equation}
then $\alpha_j$:
\begin{equation}
    \alpha_j = (x_1^{(j-1)K+1}, \dots, x_N^{(j-1)K+1}, \dots,   x_1^{(jK}, \dots, x_N^{(jK})
\end{equation}
where $\alpha \in \mathbb{R}^{KN}$. A split $D_j$ can then be written:
\begin{align}
    \ket{D_j} = \frac{1}{\abs{\alpha_j}} \sum_{i}^{KN} \alpha_{j,i} \ket{i}.
\end{align}

Consequently, encoding each of the splits $L$ requires $\log(KN)$ qubits. Each split has $K$ data points with $N$ features. Assuming an equal split of the dataset where $K = \frac{M}{L}$, all the splits $L$ together subsequently yield $L \log(\frac{M}{L} N)$. Assuming $M\gg L$, the total number of qubits is $L \log (MN)$, and again we see that data splitting has increase the number of qubits need by a factor of $L$.

One thing to be noted about the splitting of the data vector in amplitude encoding is on the role of the normalization constant. Distribution of the dataset will result in $L$ different normalization constants per data split, which may in turn disproportionately change the structure of data in $L$ different ways. Of course, it could also be the case that the variance in magnitude of the of normalization constant is insignificant, which we might expect for very large data sets.

An obvious splitting technique in amplitude encoding is between data samples, such that each node receives a state $|D_j\rangle$ consisting of the feature vector $x^{(j)}$ of single data point from $D$. Interestingly, arriving at this splitting is natural when starting from the hybrid quantum-classical approach to QNNs. There, the dataset is paired with labels $y^{(j)}$, for each $x^{(j)}$, which is compared to output of the QML model in sequential fashion. One notable exception is a recent set of simulated experiments \citep{huang_power_2021} which make natural uses of TensorFlow's built-in distributed computing ecosystem to distribute the dataset over $30$ nodes.  

\subsection{Data parallelism discussion \label{sec:data_C}}

Basis and amplitude encoding are the prototypical techniques for constructing quantum data. Due to the current limitations of quantum hardware, other more ``natural'' encodings have been considered dubbed \textit{hardware-efficient}. Angle encoding \citep{schuld_quantum_2019, skolik_layerwise_2021, haug_large_2021}, which is done at the single qubit level and hence does not entangle states within feature vectors, is more akin to basis encoding. Whereas, encoding at the Hamiltonian level \citep{havlicek_supervised_2019, wecker_progress_2015, cade_strategies_2020, wiersema_exploring_2020} typically involves two-qubit entangling gates and, in the context of parallelization, more akin to amplitude encoding.  

The angle encoding technique was recently used in data parallelization experiments to distinguish letters from the MNIST database \citep{du_accelerating_2021}. There, the authors devised a protocol to execute multiple rounds of local gradient training before communicating with a central node which averaged the current parameter values before redistributing them. The experiments investigated accuracy versus number of local gradient evaluations, finding fewer local gradient evaluations to perform better independent of the number of local nodes. The overall speedup to a given accuracy threshold, however, scaled linearly with the number of local nodes.  

A generic approach to encoding classical data is to consider, 
\begin{equation}
    |D_j\rangle = U_{\textrm{enc}}(x^{(j)})|0\rangle,
\end{equation}
where $U_{\textrm{enc}}$ is some encoding circuit. In this context we are somewhat constrained in types of data splitting we ought to consider. By the very nature of the set-up, we already have an implicit splitting between feature vectors. As noted, this is typically processed in series, but in the hybrid quantum-classical setting can be naively distributed using existing classical protocols. In this style of splitting, no entanglement is generated \textit{across} features, while \textit{intra}data entanglement would presumably persist. However, we do note that even within this paradigm, \textit{quantum} training (with access to QRAM, for example) may recover \textit{inter}data entanglement \citep{liao_quantum_2021, pascanu_difficulty_2013, shang_global_1996}. On the other hand, further splitting \textit{within} each feature vector could be considered. However, detailed knowledge of $U_{\textrm{enc}}$ would be required, and this may consist of removing entanglement between features, which is likely the only advantage the QML model is empowered by --- be it computational or expressive. We mention the possibility, though, as such a split might properly be considered a \textit{model} splitting rather than a data splitting, which is an excellent segue. 

\section{Model Parallelism: Splitting the Model}
\label{section:model}
Model parallelism makes use of the idea of distributing the neural network and its parameters. In quantum machine learning, a model can be understood as a parameterized quantum circuit --- i.e., a quantum circuit with variably specified gates. How these circuits are ``split'' is superficially the same as how models are split classically, but differs greatly in the details.

\begin{figure}
    \centering
    \includegraphics[width=\textwidth]{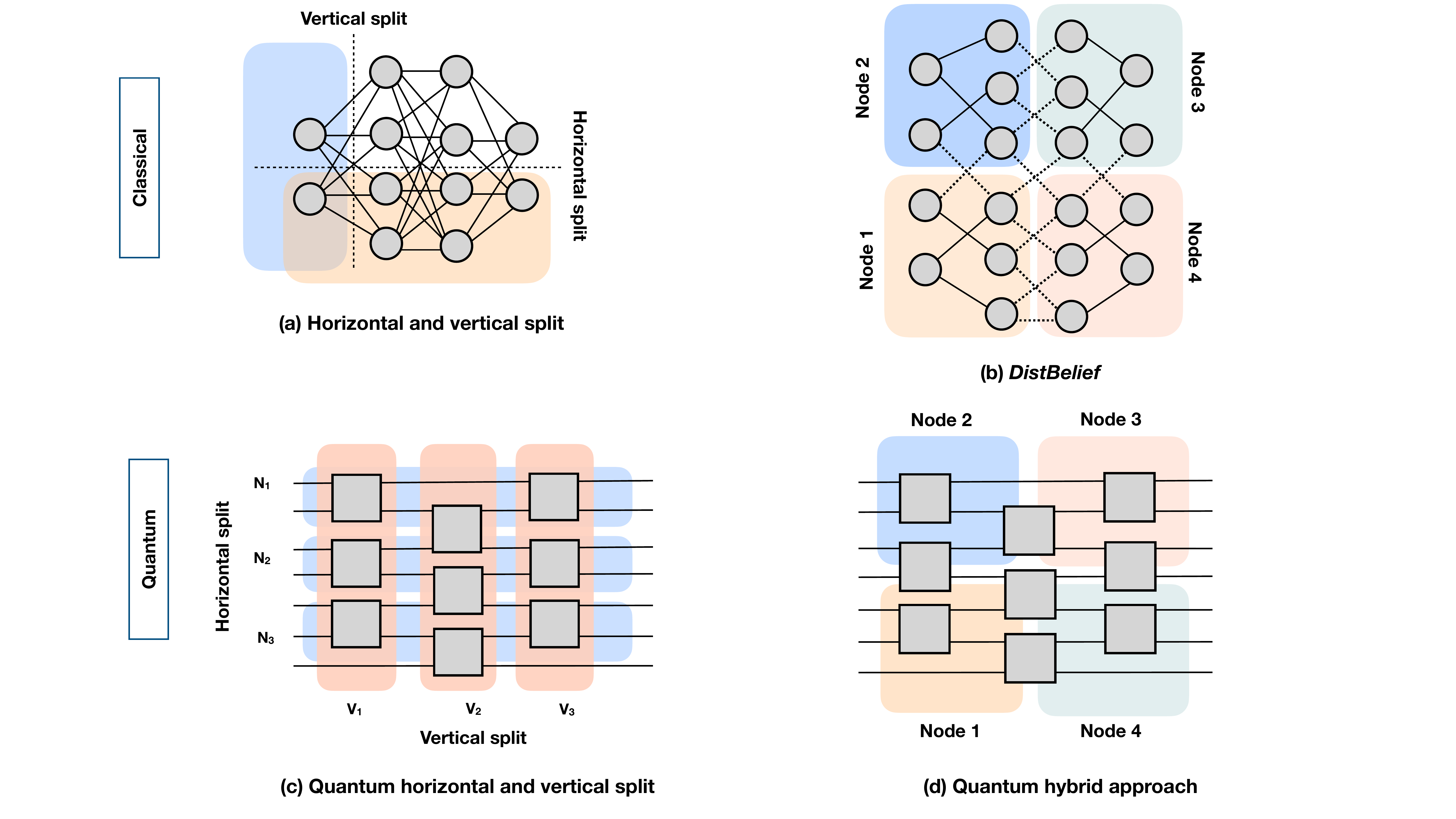}
    \caption{Model parallelism example architectures. $a)$ Horizontal and vertical split of a neural network model. $b)$ A visual representation of the DistBelief \citep{dean_large_2012} architecture that features both data and model parallelism as an example architecture of hybrid parallelism. Here the dataset as well as the model are split across the available nodes. $c)$ Quantum-inspired horizontal and vertical split of a quantum circuit. The horizontally cut sub-circuits $N_n$ would require classical communication among the nodes. The vertical cuts $V_n$ would require quantum tomography. $d)$ A visualisation of the quantum-inspired hybrid approach following the architecture of DistBelief in $b$. The quantum circuit is split across $4$ devices using both data and model parallelism.}
    \label{fig:model}
\end{figure}

Classically, we can point out two types of model splits: horizontal and vertical splitting as in Fig.~\ref{fig:model}\textcolor{red}{a}. In horizontal splitting, it is the layers of the network that are split. While vertical splitting is applied between the layers, leaving individual layers unaffected. The latter feature makes vertical splitting a more versatile technique. This is why classical vertical splitting is generally preferred over horizontal splitting \citep{langer_distributed_2020}. This, however, cannot be used as a naive heuristic for quantum scenarios, as we will see below.

That existing quantum literature in distributed QNNs explores horizontal splits rather than vertical splits. Interestingly, there are certain limitations in the classical analogue which make the horizontal splitting approach the last resort to turn to for distribution \citep{langer_distributed_2020}. In addition, it is often left implicit that model parallelism does not always yield concurrent working nodes due to the inherent property of data dependency in neural networks.

The straightforward model architectures we have considered in Fig.~\ref{fig:model}\textcolor{red}{c} involves splitting the quantum circuit horizontally or vertically, although several other different split methods can be approximated. Intuitively, splitting the quantum circuit vertically may not necessarily yield an advantage. As will be discussed below, each gate or wire cut incurs a cost that grows exponentially.

In the case where a splitting strategy is restricted to communicate classical information between nodes, merging the different parts of the circuit requires the exponentially difficult task of quantum tomography. A strictly vertical split then is maximally inefficient. As in the literature, then, we will mostly focus on horizontal splitting.

At present, there exist several architectures which can be considered horizontal splits on the model. Generally speaking, these works correspond to a few major classes of horizontal splitting, as we will describe below. The taxonomy can be of different flavours, however below we choose to differentiate among incoherent and coherent splitting, presence of communication throughout the calculation, and sampling.

\begin{figure}
    \centering
    \includegraphics[width=0.65\textwidth]{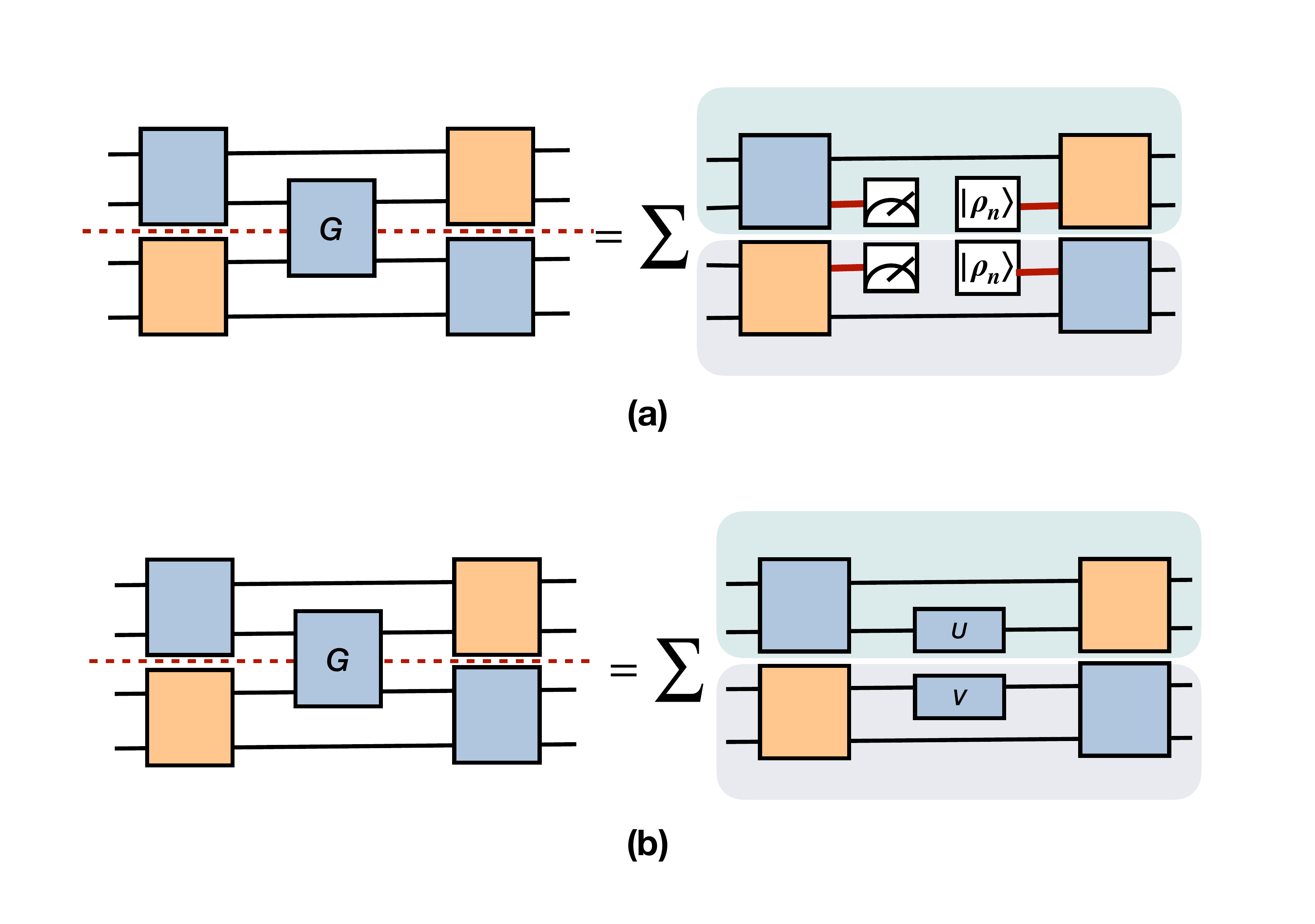}
    \caption{The two of the main paradigms of horizontal splitting introduced in the quantum circuit cutting literature. In $a)$ is depicted the paradigm of splitting gates (or wires) incoherently (classically) as in Ref. \citep{peng_simulating_2020}. Whereas the architecture in In $b)$ splits the gates coherently (quantumly) as in Ref. \citep{marshall_high_2022}.}
    \label{fig:horizontal}
\end{figure}

We summarize the general techniques in Fig~\ref{fig:horizontal}.
Consider the gate $G$ as a two-qubit gate whose action is to be split across two separate quantum computing nodes. For any $G$ there are a number of recipes that allow exact or approximate emulation using only pairs of gates $L_k$ that act separately on each node. The labels carry some "physical" meaning here in that gates which act across subsystems are referred to as "global" while gates that act individually on subsystems are termed "local". Properly, the action of the global gate $G$ can be computed as a \textit{sum} of locally acting gates $L_k$:  
\begin{equation}\label{eq:gate_split}
    G = \sum_{k} c_k L_k,
\end{equation}
where $c_k$ are known real-valued coefficients. Each term in the sum requires a unique computation, the results of which need to be combined in post-processing. The key differentiating factor for how this is accomplished is whether it achieved using quantum measurements or not.

\subsection{Incoherent versus coherent splitting}
In horizontal model splitting, as depicted in Fig~\ref{fig:horizontal}, either there is measurement or there is no measurement. Not depicted, however, is idea of simply cutting a wire \citep{peng_simulating_2020}, which is analogous to the measurement-based gate splitting scenario, which we discuss first. Note that, in the jargon of quantum physics, things which \textit{preserve} quantum information are termed \textit{coherent} while things that do not are \textit{in}coherent.

\subsubsection{Incoherent splitting}
Incoherent splitting refers to methods as depicted in Fig~\ref{fig:horizontal}\textcolor{red}{a}. In these schemes, the overall global circuit is simulated by a sequence of local circuits that include a quantum measurement on the qubits affected by the cut. Since a measurement is an operation which destroys quantum information --- transforming it into classical information --- this approach is called incoherent splitting. We note that it has elsewhere been referred to as ``time-like''  splitting \citep{mitarai_constructing_2021}, but we will avoid this terminology as it references a concept in relativistic physics.     

Ideas similar to classical horizontal parallelism have been proposed in the quantum circuits literature, oftentimes unrelated to the quantum machine learning literature. An equivalent to the idea of a horizontal circuit split has been proposed in Ref. \citep{peng_simulating_2020}. This solution is offered precisely to get across the limited number of qubits available in individual devices. Overall, the scheme involves some classical computation to cut and distribute circuit descriptions among the nodes and also to post-process the results of measurements. The bulk of the overhead occurs in the number of quantum circuits that need to be run, which grows exponentially with the number of cuts. However, as they point out, actual overhead could be much less depending on the structure of the circuit. In the extreme case where two clusters of qubits have no entanglement across them, the splitting can be achieved with no overhead at all. Some applications (e.g. Hamiltonian simulation) which fall between these extremes are discussed. 

These splitting methods were used in Ref. \citep{chen_64qubit_2018} to simulate random 56 qubit quantum circuits of depth 22 on a single personal computer. In Ref. \citep{eddins_doubling_2022}, the technique was referred to as \textit{entanglement forging} and used to enact 10 qubit quantum circuits with a single 5 qubit nuclear magnetic resonance (NMR) quantum processor. 

Since measurements are made in an incoherent splitting procedure, the problem of data analysis can be considered a statistical one. In response to this, Ref. \citep{perlin_quantum_2020} introduced a \textit{maximum likelihood} tomography to approximate the result of the measurements which need to be performed across all circuit splits. They found a slightly enhanced performance in some numerical experiments over the naive recombination of the measurement data.

As noted in the canonical work of Ref. \citep{peng_simulating_2020}, the success of splitting techniques will depend highly the the existent structure of the circuit to be split, assuming an optimal (or at least sensibly obvious) choice of split location. The examples considered possessed a clustering structure where the cut location would obviously correspond to cluster links. If such structure needs to be first found, the classical pre-processing may become difficult. In Ref. \citep{tang_cutqc_2020}, the authors introduce an automated tool, called CutQC, which uses the framework of integer programming to optimize the location of circuit cuts (effectively minimizing the number required). They demonstrate the tool with various simulations of quantum algorithms by obtaining not only orders of magnitude speed-ups in simulation time, but also the ability to go well beyond was is simulatable classically. In particular, they demonstrate the simulation of 100-qubit algorithms, whereas full circuit simulations on typical classical hardware are limited to roughly 30 qubits. Similarly, Ref. \citep{saleem_divide_2021} use a graph-based approach to optimize the cut locations in the wire-cutting scenario.

The most recent example of incoherent wire-splitting is Ref. \citep{lowe_fast_2022}, which utilizes randomized measurements and one-way classical communication to create a conceptually simple splitting procedure which again has an exponential overhead in the number of wire cut. The authors were able to classically simulate a 129-qubit QAOA circuit using this technique.

\subsubsection{Coherent splitting}
In contrast to incoherent splitting, Fig~\ref{fig:horizontal}\textcolor{red}{b} depicts the same cut location, but a simulation strategy that preserves quantum information by using quantum gates rather than measurements. Since quantum information is preserved in each circuit, this is dubbed coherent. It was first introduced in Ref. \citep{mitarai_constructing_2021} where it was called ``space-like'' cutting.  

While conceptually similar to Ref. \citep{peng_simulating_2020}, the coherent splitting technique \citep{mitarai_constructing_2021} can achieve some quantifiable advantages depending on which gates are cut. In essence, efficiency comes down to how many terms are in the sum of Eq.~\eqref{eq:gate_split}, and that depends on which $G$ is being split and how many times. Moreover, in a real device, it may be more applicable to enact single qubit gates than perform measurements. The same authors improved upon the gate decomposition technique, substantially reducing the number local terms $L_k$ in Eq.~\eqref{eq:gate_split}. They also introduced a novel sampling technique we discuss further below.

More recently, Ref. \citep{piveteau_circuit_2022} proposed a method called ``circuit knitting'' which again is motivated by the promise of using present day quantum processors through the partitioning of large quantum circuits into smaller sub-circuits. The resulting output of each sub-circuit is then ``knitted'' using classical communication. This work is conceptually different from all those previously discussed in that those works considered only classical communication in the final post-processing of the data. Ref.  \citep{piveteau_circuit_2022} found that classical communication is advantageous when multiple instances of the same gate will be split. However, prior entanglement between the nodes is necessary to realize this advantage.

These ideas have further motivated explorations in specifically splitting QNNs in the context of quantum machine learning \citep{marshall_high_2022}. Rather than minimizing the number of cuts as in previous work, Ref. \citep{marshall_high_2022} focuses on minimizing the size of the sub-circuits needed to approximate the result. They further test this hybrid circuit cutting architecture in the MNIST dataset, but training a quantum classifier with 64 qubits using eight nodes (each, of course having 8 qubits). Such a simulation directly on 64 qubits would be infeasible both in classical simulation and in today's quantum hardware. This work point out an important additional features of the idea of distribution in QNNs. Notably, the problem of barren plateaus is eased because sub-circuits have a smaller number of qubits leading to larger gradients. This has been corroborated in Ref. \citep{tuysuz_classical_2022}, which explores parallel execution and combination of small sub-circuits in QNNs, finding an avoidance of barren plateaus.

\subsection{Model parallelism discussion}

Here we have made a distinction between incoherent and coherent splitting techniques. There are two points to make in this regard. First, it is not likely that one technique is strictly better than that other and the use of either technique will depend heavily on the context of the circuits being executed. Indeed, Ref. \citep{mitarai_overhead_2021} considers the case of a hybrid technique in which multiple splits in a single circuit may use a combination of both incoherent and coherent splitting. The second point is that this dichotomy is not the only current differentiator in the horizontal splitting techniques existing in the literature.

We have alluded to two other distinctions already. The first is whether or not communication is used in the protocol. The existence of communication is not strictly optional, but borne out of necessity of cutting procedure. Communication takes place in Refs. \citep{piveteau_circuit_2022, lowe_fast_2022} in order to \textit{merge} the different sub-circuits on the fly rather than in post-processing. Again, since the cutting protocol will depend on the context, the existence of communication will as well. Indeed the context may even preclude communication due to technical limitations. 

The second distinction requires more thought. Since the coefficients $c_k$ are generically negative numbers, the total sum may require high precision to accurately calculate exactly or estimate. To get around this, let us rewrite Eq.~\eqref{eq:gate_split} as follows,
\begin{equation}\label{eq:gate_split_2}
    G = N \sum_{k} \frac{|c_k|}{N}\frac{c_k}{|c_k|} L_k,
\end{equation}
where $N = \sum_k |c_k|$. From here we can see that the quantities $p_k = |c_k|/N$ form a discrete probability distribution. By treating the application of $L_k$ as a random variable, the expectation of $G$ can be Monte Carlo sampled, which may result in fewer circuits to run. This was first considered by Ref. \citep{mitarai_overhead_2021}, where it was pointed out that the quantity $N^2$ corresponds to the overhead incurred due to splitting. Since there is no loss in generality in making this move, and standard probabilistic bounds can be straightforwardly applied, it is likely that this \textit{sampling-based} splitting approach will be more favored.

Recall that in data parallelism, the same classical protocols applied to quantum data parallelism. However, we see that for model splitting, the resultant quantum protocol is fundamentally different as it requires potentially many different models to be run serially and then post-processed. Whereas, in classical horizontal splitting, the models do not actually change --- communication protocols are introduced to retain the capacity of the full model. It thus not likely that powerful classical hybrid protocols utilizing both vertical and horizontal model parallelism, such as DistBelief \citep{dean_large_2012} depicted in Fig.~\ref{fig:model}\textcolor{red}{b}, will be generalized to the quantum setting. However, we can naively infer a quantum hybrid architecture in which both quantum data and quantum model are split as in Fig.~\ref{fig:model}\textcolor{red}{d}. Such a model might be realized when distribution can be achieved with efficient entanglement distribution or a fully coherent quantum communication network is available.

Finally, we mentioned the implied classical-quantum hybrid horizontal splitting technique of Ref. \citep{bravyi_trading_2016}, which makes use of both quantum and classical resources to simulate large quantum systems with ``virtual qubits'' running in parallel to a quantum computer.

\section{Discussion}\label{section:discussion}
This overview paper was an introduction of the distributed techniques present in classical deep learning as applied to the novel field of quantum neural networks. In this final discussion we mention some related ideas and comment on the nascent topics of NISQ and quantum software before concluding.

\subsection{Related techniques}

Beyond data parallelism, there exist other strategies that go along the lines of optimizing the amount of data for learning algorithms. Data reduction techniques, often known as coreset techniques in classical machine learning \citep{bachem_practical_2017}, are a prime example. The main idea behind coresets is that for a dataset $D$, there exists a subset $S$ which approximates $D$ for a particular task. Importance sampling is typically the first step in the process of finding such $S$. Coreset techniques are typically algorithm-type specific, however can also be generalized. Making the same parallels to quantum computation, the size of the datasets remains an obvious problem when qubits are at a premium. The question of whether one can use coreset techniques in hybrid quantum-classical architectures has been explored in \citep{harrow_small_2020, tomesh_coreset_2021}. Ref. \citep{harrow_small_2020} work presents several examples of hybrid algorithms making use of data reduction techniques, notably in clustering, regression and boosting. Ref. \citep{tomesh_coreset_2021} builds on this work by extending it to the realm of variational algorithms. The works in question can certainly be part of the greater solution to handling large quantum datasets. Data reduction techniques are not typically discussed in the context of distributed learning. The underlying principles that guide data reduction techniques are not necessarily similar. 

QNNs require classical optimization, which has been studied in a parallelized context in Ref. \cite{self_variational_2021}, where the procedure is called \textit{information sharing}. The novelty of their proposal is in its application to a particular style of optimizing, but the general idea could be consider a form of a distributed QNN where the \textit{parameters} are distributed across nodes. Although most optimizers are adaptive --- meaning the QNNs parameters at one point in time depending on measurement results at a previous point it time --- many optimizers require evaluation of costs at several simultaneous parameter values. This suggests an obvious form of distribution, as in Ref. \cite{self_variational_2021}.

\subsection{NISQ and beyond}\label{sec:parallelism}

The primary motivation for many of the existing QML techniques is to serve the needs, or limitations, of the NISQ era. Recall, NISQ devices are both small (in qubit number) and noisy (limiting circuit depth) \citep{preskill_quantum_2018, chen_complexity_2022}, which is why shallow circuits are of great interest \cite{bravyi_quantum_2018, dunjko_machine_2018, arute_quantum_2019}. Many of the techniques we have discussed above are suitable in the NISQ regime, not requiring a fixed (large) number of densely interacting qubits and not restricted to noiseless computation.

Combining the training of QNNs across many of the nodes may yield advantages be it in terms of time complexity, or in terms of generalisation power, scalability, and explainability. Of the two main architectures, in classical deep learning, data parallelism is the more explored one. That is for several reasons. Firstly, it is practically easier to split the dataset rather than the model. When it comes to splitting the model there exist different strategies which can be more suited to the task at hand. Secondly, it is widely accepted that data parallelism allows better cluster utilization \citep{langer_distributed_2020}. 

Naive data splitting is straightforwardly applied when using mature AI software packages \cite{huang_power_2021}. We expect such techniques to find use in the first NISQ implementations of QNNs. However, the restricted size of NISQ devices will also see the use of horizontal splitting techniques. Incoherent splitting is likely more suited to early NISQ devices since no new capabilities are required. In addition to reducing the demand on the number of qubits, splitting techniques can reduce the depth of the circuit as well --- depending of course on the structure.  

One may wonder why such techniques have not been widely adopted and applied to existing devices? Typically, the larger the dataset is, the more relevant it becomes to distribute the dataset. Distributing smaller datasets may not yield obvious advantages. In the initial works that here we consider equivalent to model distribution, there are assumptions that can be made on the complexity of the datasets. For instance, \citep{marshall_high_2022} observes that synthetic quantum data performs better in their technique than classical high-dimensional data. As it currently stands, the number of qubits available may be sufficient, but the level of noise needs to be reduced to allow for sufficiently deep circuits. Determining what types of data and what structural features of circuits are most suited to splitting is open area of research. What is clear, however, is that a principled approach to the development of software for this purpose is needed.

\subsection{Software tools}\label{subsec:software}
There exist a number of software packages and libraries that implement distributed deep learning strategies. The Tensorflow software \citep{abadi_tensorflow_2015}, for instance, implements distributed training techniques as an out-of-the-box feature. In TensorFlow there are several strategies for distributing over resources \citep{dist_tensorflow}. For instance, the \textit{MirroredStrategy} as a distributed technique makes use of all the available central processing units (CPU) or GPU resources in a single device. \textit{MultiWorkerMirroredStrategy} is a synchronous distributed strategy that makes use of multiple devices, each potentially containing multiple GPUs or CPUs. All of the aforementioned strategies support synchronous training. A few all-reduce implementations of choice are also available. Other than the methods above, there exists the \textit{ParameterServerStrategy} which implements asynchronous communication. In addition to Tensorflow, there exist a number of other libraries one of which is the Horovod library \citep{sergeev_horovod_2018}, that implements the ring-allreduce algorithm across a distributed cluster using NCCL \citep{nccl} as the communication library. Other software that facilitate the same principles are MXNet \citep{chen_mxnet_2015}, PyTorch \citep{paszke_automatic_2017}, CNTK \citep{seide_cntk_2016} etc. A more complete list of software available that support distributed deep learning can be found in Table 3 in Ref. \citep{mayer_scalable_2019}.

\begin{figure}
    \centering
    \includegraphics[width=\textwidth]{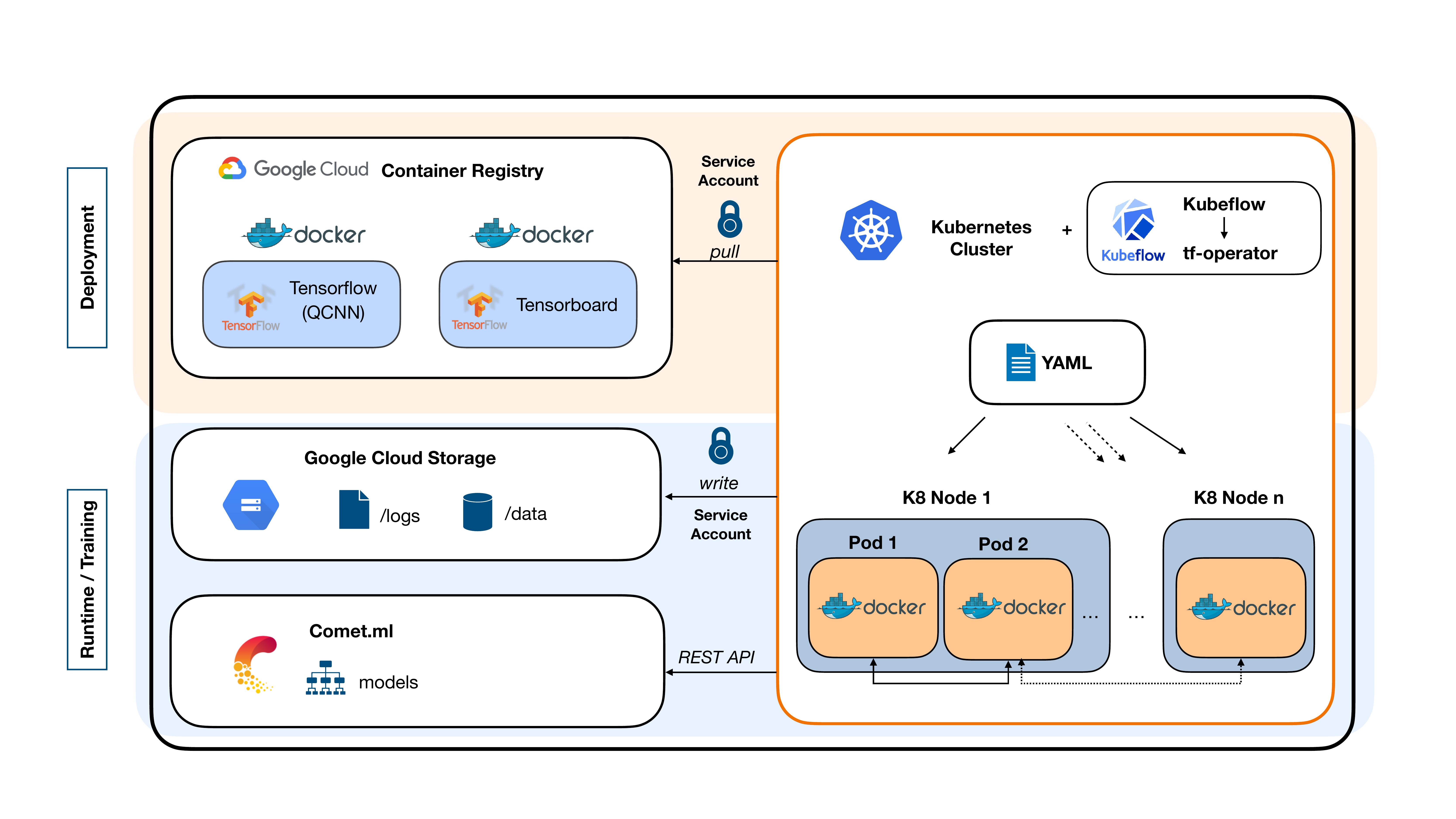}
    \caption{Architecture stack of an example software infrastructure running numerical experiments as in \citep{xing_training_2021}. The process is initiated by Kubernetes \textit{pulling} the two docker images from the container registry: Tensorflow QCNN image and the Tensorboard image (Tensorboard is Tensorflows' graphical user interface of results overview). The images were built locally and uploaded to the container registry. After pulling the images Kubernetes creates pods (which can be thought of as actual processors or in software as \textit{jobs}), the number of which is determined by the number of the \textit{replica} parameter. Parallelism is managed by the number of pods. Pods are then deployed in Kubernetes nodes. Pod distribution over nodes is transparently managed by Kubernetes. At the end of the job, the master pod which writes to Google Cloud Storage bucket. At the same time, results get uploaded to Comet.ml via REST APIs. Note that Tensorboard can be used as a visual tool for reading the results without the need for external software. TFJob operator, part of the Kubeflow project, is Kubernetes' custom resource that facilitates the deployment of tensorflow instances in a Kubernetes cluster. The entire process can be categorized into two parts: the \textit{deployment} phase responsible for the setup, and the \textit{runtime} or in our scenarios of ML optimization, the \textit{training} phase.}
    \label{fig:arch}
\end{figure}

Tensorflow Quantum \citep{broughton_tensorflow_2020} is the quantum machine learning extension that allows simulating hybrid quantum circuit models. As an example in terms of how the distributed implementation would look like in quantum neural networks, TensorFlow Quantum provides a blog-post setting up the architecture \citep{xing_training_2021} for distributed training of QNNs. This example implements the \textit{MultiWorkerMirroredStrategy}. The backbone architecture of these experiments is the quantum convolutional neural network (QCNN) architecture developed in Ref. \citep{cong_quantum_2019}. Fig.~\ref{fig:arch} gives an overview of the architecture stack enabling such experiments. Further on the the quantum front, quantum software such as Qiskit \citep{aleksandrowicz_qiskit_2019} or Pennylane \citep{bergholm_pennylane_2018} can be used either for simulations or experiments on actual quantum devices \citep{fingerhuth_open_2018}. Table 2 in Ref. \citep{benedetti_parameterized_2019} summarizes some of the works which implement QNNs across different quantum hardware platforms for different tasks \citep{otterbach_unsupervised_2017, riste_demonstration_2017, grant_hierarchical_2018, tacchino_artificial_2019, benedetti_generative_2019, coyle_born_2020, rocchetto_experimental_2019, ding_experimental_2019}.

In relation to the techniques useful for the distributed approaches, Ref. \citep{tang_cutqc_2020} mentioned above, develops CutQC which is a software package that automates the location of wire cuts when splitting a quantum circuit. In similar lines, Ref. \citep{parekh_quantum_2021} builds Interlin-q on top of the real time quantum networks simulator QuNetSim \citep{diadamo_qunetsim_2021}. Interlin-q is a software package which helps in designing distributed algorithms. It follows a general centralised architecture where a client node is responsible for propagating information to the computing nodes, via a middle point controller node.

\subsection{Closing remarks} \label{subsec:closing}

A lot of work in circuit cutting schemes give special attention to the role of entanglement when distributing qubits. The question of entanglement is heavily addressed in distributed quantum computing and quantum internet research \citep{cirac_distributed_1999, gyongyosi_entanglement_2019}. The first step in distributed quantum networks is to establish entanglement via entanglement distribution techniques. There exist several strategies for entanglement distribution which can be cost effective \citep{streltsov_quantum_2012}. However, entanglement can be fragile and lost over time and as such it needs robust techniques for its preservation over long-distances. This is a crucial element to consider in distributed quantum machine learning schemes as well. In the avenue of quantum neural networks, Ref. \citep{sharma_reformulation_2022} demonstrates the crucial role of entanglement in training QNNs. The work in question expands on the data as a resource in classical machine learning, by demonstrating entanglement as an asset in the quantum setting. More concretely, the presence of entanglement reduces the demand for quantum data, something that was once thought to be of necessity in exponential order. Following this, it is pivotal to account for entanglement as a resource in distributed training. Such accounting directly relates to the overhead in distributed quantum learning.

One possible avenue of development in the distributed scenario is to see how splitting of the circuit plays out in different proposed quantum deep learning architectures, beyond the feed-forward QNNs. Some of the possible questions we are interested in exploring include a comparison of classical versus quantum training time and training accuracy. Tangentially, we note that all of the proposals for QNNs assume the circuit model of quantum computation. Following classical DDL literature, there remain several scenarios to consider in the quantum front regarding parameter scheduling, architectural centralisation and further into communication protocols. It may also be more natural to consider distributed quantum computing in alternative models such as measurement-based quantum computing \citep{raussendorf_measurement_2003}. 

A natural descendent of the distributed architectures is the concept of federated learning \citep{mcmahan_communication_2017} which makes use of end-user data locally rather than assuming one single central storage. This paradigm is proposed to mitigate privacy and security issues that concern centralized training architectures. Federated learning from a quantum perspective has been initiated in Refs. \citep{chehimi_quantum_2021, chen_federated_2021}. These techniques may eventually overlap with the distributed QNN approach we considered here.

In a recent issue of Quantum Science and Technology, leaders in the field of quantum computing were asked, "What would you do with 1000 qubits?" \citep{morello_what_2018}. As in one response \citep{perdomo_opportunities_2018}, many have suggested QML as one of the first applications that may provide an advantage over classical techniques. However, this question --- and much of the reaction --- was posed before distributed techniques became popular in the quantum information community. Anything you can do with 1000 qubits, you can do with ten 100-qubit devices and a bit of time. So perhaps this threshold is much closer than many suspect.

\paragraph*{Acknowledgements}
We thank Mária Kieferová for the discussions. LP would like to extend the gratitude to Dan Browne and his research group for the cordiality at the University College London, UK, during the writing of this manuscript. LP was supported by the Sydney Quantum Academy, Sydney, NSW, Australia.

\bibliography{dqcnn}

\begin{thebibliography}{212}
\expandafter\ifx\csname natexlab\endcsname\relax\def\natexlab#1{#1}\fi
\expandafter\ifx\csname bibnamefont\endcsname\relax
  \def\bibnamefont#1{#1}\fi
\expandafter\ifx\csname bibfnamefont\endcsname\relax
  \def\bibfnamefont#1{#1}\fi
\expandafter\ifx\csname citenamefont\endcsname\relax
  \def\citenamefont#1{#1}\fi
\expandafter\ifx\csname url\endcsname\relax
  \def\url#1{\texttt{#1}}\fi
\expandafter\ifx\csname urlprefix\endcsname\relax\def\urlprefix{URL }\fi
\providecommand{\bibinfo}[2]{#2}
\providecommand{\eprint}[2][]{\url{#2}}

\bibitem[{\citenamefont{Russell and Norvig}(2010)}]{russell_artificial_2010}
\bibinfo{author}{\bibfnamefont{S.}~\bibnamefont{Russell}} \bibnamefont{and}
  \bibinfo{author}{\bibfnamefont{P.}~\bibnamefont{Norvig}},
  \emph{\bibinfo{title}{Artificial Intelligence: A Modern Approach}}
  (\bibinfo{publisher}{Prentice Hall}, \bibinfo{year}{2010}),
  \bibinfo{edition}{3rd} ed.

\bibitem[{\citenamefont{Mitchell}(1997)}]{mitchell_machine_1997}
\bibinfo{author}{\bibfnamefont{T.~M.} \bibnamefont{Mitchell}},
  \emph{\bibinfo{title}{Machine Learning}} (\bibinfo{publisher}{McGraw-Hill,
  Inc.}, \bibinfo{address}{USA}, \bibinfo{year}{1997}), \bibinfo{edition}{1st}
  ed., ISBN \bibinfo{isbn}{0070428077}.

\bibitem[{\citenamefont{Bishop}(2006)}]{bishop_pattern_2006}
\bibinfo{author}{\bibfnamefont{C.~M.} \bibnamefont{Bishop}},
  \emph{\bibinfo{title}{Pattern {Recognition} and {Machine} {Learning}
  ({Information} {Science} and {Statistics})}}
  (\bibinfo{publisher}{Springer-Verlag}, \bibinfo{address}{Berlin, Heidelberg},
  \bibinfo{year}{2006}), ISBN \bibinfo{isbn}{978-0-387-31073-2}.

\bibitem[{\citenamefont{LeCun et~al.}(2015)\citenamefont{LeCun, Bengio, and
  Hinton}}]{lecun_deep_2015}
\bibinfo{author}{\bibfnamefont{Y.}~\bibnamefont{LeCun}},
  \bibinfo{author}{\bibfnamefont{Y.}~\bibnamefont{Bengio}}, \bibnamefont{and}
  \bibinfo{author}{\bibfnamefont{G.}~\bibnamefont{Hinton}},
  \bibinfo{journal}{Nature} \textbf{\bibinfo{volume}{521}},
  \bibinfo{pages}{436} (\bibinfo{year}{2015}).

\bibitem[{\citenamefont{Goodfellow et~al.}(2016)\citenamefont{Goodfellow,
  Bengio, and Courville}}]{goodfellow_deep_2016}
\bibinfo{author}{\bibfnamefont{I.}~\bibnamefont{Goodfellow}},
  \bibinfo{author}{\bibfnamefont{Y.}~\bibnamefont{Bengio}}, \bibnamefont{and}
  \bibinfo{author}{\bibfnamefont{A.}~\bibnamefont{Courville}},
  \emph{\bibinfo{title}{Deep Learning}} (\bibinfo{publisher}{MIT Press},
  \bibinfo{year}{2016}), \bibinfo{note}{\url{http://www.deeplearningbook.org}}.

\bibitem[{\citenamefont{Schmidhuber}(2015)}]{schmidhuber_deep_2015}
\bibinfo{author}{\bibfnamefont{J.}~\bibnamefont{Schmidhuber}},
  \bibinfo{journal}{Neural Networks} \textbf{\bibinfo{volume}{61}},
  \bibinfo{pages}{85} (\bibinfo{year}{2015}).

\bibitem[{\citenamefont{Nielsen and Chuang}(2011)}]{nielsen_quantum_2011}
\bibinfo{author}{\bibfnamefont{M.~A.} \bibnamefont{Nielsen}} \bibnamefont{and}
  \bibinfo{author}{\bibfnamefont{I.~L.} \bibnamefont{Chuang}},
  \emph{\bibinfo{title}{Quantum Computation and Quantum Information: 10th
  Anniversary Edition}} (\bibinfo{publisher}{Cambridge University Press},
  \bibinfo{year}{2011}), ISBN \bibinfo{isbn}{1107002176}.

\bibitem[{\citenamefont{Biamonte et~al.}(2016)\citenamefont{Biamonte, Wittek,
  Pancotti, Rebentrost, Wiebe, and Lloyd}}]{biamonte_quantum_2017}
\bibinfo{author}{\bibfnamefont{J.}~\bibnamefont{Biamonte}},
  \bibinfo{author}{\bibfnamefont{P.}~\bibnamefont{Wittek}},
  \bibinfo{author}{\bibfnamefont{N.}~\bibnamefont{Pancotti}},
  \bibinfo{author}{\bibfnamefont{P.}~\bibnamefont{Rebentrost}},
  \bibinfo{author}{\bibfnamefont{N.}~\bibnamefont{Wiebe}}, \bibnamefont{and}
  \bibinfo{author}{\bibfnamefont{S.}~\bibnamefont{Lloyd}},
  \bibinfo{journal}{Nature} \textbf{\bibinfo{volume}{549}},
  \bibinfo{pages}{195} (\bibinfo{year}{2016}).

\bibitem[{\citenamefont{Schuld and Petruccione}(2018)}]{schuld_supervised_2018}
\bibinfo{author}{\bibfnamefont{M.}~\bibnamefont{Schuld}} \bibnamefont{and}
  \bibinfo{author}{\bibfnamefont{F.}~\bibnamefont{Petruccione}},
  \emph{\bibinfo{title}{Supervised {Learning} with {Quantum} {Computers}}},
  Quantum {Science} and {Technology} (\bibinfo{publisher}{Springer
  International Publishing}, \bibinfo{year}{2018}), ISBN
  \bibinfo{isbn}{978-3-319-96423-2}.

\bibitem[{\citenamefont{Wittek}(2014)}]{wittek_quantum_2014}
\bibinfo{author}{\bibfnamefont{P.}~\bibnamefont{Wittek}},
  \emph{\bibinfo{title}{{Quantum Machine Learning: What Quantum Computing Means
  to Data Mining}}} (\bibinfo{publisher}{Elsevier Science},
  \bibinfo{year}{2014}), ISBN \bibinfo{isbn}{9780128009536},
  \urlprefix\url{https://books.google.com.au/books?id=PwUongEACAAJ}.

\bibitem[{\citenamefont{Ciliberto et~al.}(2018)\citenamefont{Ciliberto,
  Herbster, Ialongo, Pontil, Rocchetto, Severini, and
  Wossnig}}]{ciliberto_quantum_2018}
\bibinfo{author}{\bibfnamefont{C.}~\bibnamefont{Ciliberto}},
  \bibinfo{author}{\bibfnamefont{M.}~\bibnamefont{Herbster}},
  \bibinfo{author}{\bibfnamefont{A.~D.} \bibnamefont{Ialongo}},
  \bibinfo{author}{\bibfnamefont{M.}~\bibnamefont{Pontil}},
  \bibinfo{author}{\bibfnamefont{A.}~\bibnamefont{Rocchetto}},
  \bibinfo{author}{\bibfnamefont{S.}~\bibnamefont{Severini}}, \bibnamefont{and}
  \bibinfo{author}{\bibfnamefont{L.}~\bibnamefont{Wossnig}},
  \bibinfo{journal}{Proceedings of the Royal Society A: Mathematical, Physical
  and Engineering Sciences} \textbf{\bibinfo{volume}{474}},
  \bibinfo{pages}{20170551} (\bibinfo{year}{2018}).

\bibitem[{\citenamefont{Cerezo et~al.}(2022)\citenamefont{Cerezo, Verdon,
  Huang, Cincio, and Coles}}]{cerezo_challenges_2022}
\bibinfo{author}{\bibfnamefont{M.}~\bibnamefont{Cerezo}},
  \bibinfo{author}{\bibfnamefont{G.}~\bibnamefont{Verdon}},
  \bibinfo{author}{\bibfnamefont{H.-Y.} \bibnamefont{Huang}},
  \bibinfo{author}{\bibfnamefont{L.}~\bibnamefont{Cincio}}, \bibnamefont{and}
  \bibinfo{author}{\bibfnamefont{P.}~\bibnamefont{Coles}},
  \bibinfo{journal}{Nature Computational Science}  (\bibinfo{year}{2022}).

\bibitem[{\citenamefont{A{\"i}meur et~al.}(2006)\citenamefont{A{\"i}meur,
  Brassard, and Gambs}}]{aimeur_machine_2006}
\bibinfo{author}{\bibfnamefont{E.}~\bibnamefont{A{\"i}meur}},
  \bibinfo{author}{\bibfnamefont{G.}~\bibnamefont{Brassard}}, \bibnamefont{and}
  \bibinfo{author}{\bibfnamefont{S.}~\bibnamefont{Gambs}}, in
  \emph{\bibinfo{booktitle}{Advances in Artificial Intelligence}}
  (\bibinfo{publisher}{Springer Berlin Heidelberg}, \bibinfo{year}{2006}), pp.
  \bibinfo{pages}{431--442}.

\bibitem[{\citenamefont{Dunjko and Briegel}(2018)}]{dunjko_machine_2018}
\bibinfo{author}{\bibfnamefont{V.}~\bibnamefont{Dunjko}} \bibnamefont{and}
  \bibinfo{author}{\bibfnamefont{H.~J.} \bibnamefont{Briegel}},
  \bibinfo{journal}{Reports on Progress in Physics}
  \textbf{\bibinfo{volume}{81}}, \bibinfo{pages}{074001}
  (\bibinfo{year}{2018}).

\bibitem[{\citenamefont{Benedetti
  et~al.}(2019{\natexlab{a}})\citenamefont{Benedetti, Lloyd, Sack, and
  Fiorentini}}]{benedetti_parameterized_2019}
\bibinfo{author}{\bibfnamefont{M.}~\bibnamefont{Benedetti}},
  \bibinfo{author}{\bibfnamefont{E.}~\bibnamefont{Lloyd}},
  \bibinfo{author}{\bibfnamefont{S.}~\bibnamefont{Sack}}, \bibnamefont{and}
  \bibinfo{author}{\bibfnamefont{M.}~\bibnamefont{Fiorentini}},
  \bibinfo{journal}{Quantum Science and Technology}
  \textbf{\bibinfo{volume}{4}} (\bibinfo{year}{2019}{\natexlab{a}}).

\bibitem[{\citenamefont{Preskill}(2018)}]{preskill_quantum_2018}
\bibinfo{author}{\bibfnamefont{J.}~\bibnamefont{Preskill}},
  \bibinfo{journal}{Quantum} \textbf{\bibinfo{volume}{2}}, \bibinfo{pages}{79}
  (\bibinfo{year}{2018}), ISSN \bibinfo{issn}{2521-327X}.

\bibitem[{\citenamefont{Verbraeken et~al.}(2020)\citenamefont{Verbraeken,
  Wolting, Katzy, Kloppenburg, Verbelen, and
  Rellermeyer}}]{verbraeken_survey_2020}
\bibinfo{author}{\bibfnamefont{J.}~\bibnamefont{Verbraeken}},
  \bibinfo{author}{\bibfnamefont{M.}~\bibnamefont{Wolting}},
  \bibinfo{author}{\bibfnamefont{J.}~\bibnamefont{Katzy}},
  \bibinfo{author}{\bibfnamefont{J.}~\bibnamefont{Kloppenburg}},
  \bibinfo{author}{\bibfnamefont{T.}~\bibnamefont{Verbelen}}, \bibnamefont{and}
  \bibinfo{author}{\bibfnamefont{J.~S.} \bibnamefont{Rellermeyer}},
  \bibinfo{journal}{ACM Computing Surveys} \textbf{\bibinfo{volume}{53}}
  (\bibinfo{year}{2020}), ISSN \bibinfo{issn}{0360-0300}.

\bibitem[{\citenamefont{Ben-Nun and Hoefler}(2019)}]{bennun_demystifying_2019}
\bibinfo{author}{\bibfnamefont{T.}~\bibnamefont{Ben-Nun}} \bibnamefont{and}
  \bibinfo{author}{\bibfnamefont{T.}~\bibnamefont{Hoefler}},
  \bibinfo{journal}{ACM Computing Surveys} \textbf{\bibinfo{volume}{52}}
  (\bibinfo{year}{2019}).

\bibitem[{\citenamefont{Chahal et~al.}(2020)\citenamefont{Chahal, Grover, Dey,
  and Shah}}]{chahal_hitchiker_2020}
\bibinfo{author}{\bibfnamefont{K.~S.} \bibnamefont{Chahal}},
  \bibinfo{author}{\bibfnamefont{M.~S.} \bibnamefont{Grover}},
  \bibinfo{author}{\bibfnamefont{K.}~\bibnamefont{Dey}}, \bibnamefont{and}
  \bibinfo{author}{\bibfnamefont{R.~R.} \bibnamefont{Shah}},
  \bibinfo{journal}{Journal of Parallel and Distributed Computing}
  \textbf{\bibinfo{volume}{137}}, \bibinfo{pages}{65} (\bibinfo{year}{2020}),
  ISSN \bibinfo{issn}{0743-7315}.

\bibitem[{\citenamefont{Mayer and Jacobsen}(2020)}]{mayer_scalable_2019}
\bibinfo{author}{\bibfnamefont{R.}~\bibnamefont{Mayer}} \bibnamefont{and}
  \bibinfo{author}{\bibfnamefont{H.-A.} \bibnamefont{Jacobsen}},
  \bibinfo{journal}{ACM Computing Surveys} \textbf{\bibinfo{volume}{53}}
  (\bibinfo{year}{2020}).

\bibitem[{\citenamefont{Langer et~al.}(2020)\citenamefont{Langer, He, Rahayu,
  and Xue}}]{langer_distributed_2020}
\bibinfo{author}{\bibfnamefont{M.}~\bibnamefont{Langer}},
  \bibinfo{author}{\bibfnamefont{Z.}~\bibnamefont{He}},
  \bibinfo{author}{\bibfnamefont{W.}~\bibnamefont{Rahayu}}, \bibnamefont{and}
  \bibinfo{author}{\bibfnamefont{Y.}~\bibnamefont{Xue}}, \bibinfo{journal}{IEEE
  Transactions on Parallel and Distributed Systems}
  \textbf{\bibinfo{volume}{31}}, \bibinfo{pages}{2802} (\bibinfo{year}{2020}).

\bibitem[{\citenamefont{Goyal et~al.}(2017)\citenamefont{Goyal, Dollár,
  Girshick, Noordhuis, Wesolowski, Kyrola, Tulloch, Jia, and
  He}}]{goyal_accurate_2018}
\bibinfo{author}{\bibfnamefont{P.}~\bibnamefont{Goyal}},
  \bibinfo{author}{\bibfnamefont{P.}~\bibnamefont{Dollár}},
  \bibinfo{author}{\bibfnamefont{R.}~\bibnamefont{Girshick}},
  \bibinfo{author}{\bibfnamefont{P.}~\bibnamefont{Noordhuis}},
  \bibinfo{author}{\bibfnamefont{L.}~\bibnamefont{Wesolowski}},
  \bibinfo{author}{\bibfnamefont{A.}~\bibnamefont{Kyrola}},
  \bibinfo{author}{\bibfnamefont{A.}~\bibnamefont{Tulloch}},
  \bibinfo{author}{\bibfnamefont{Y.}~\bibnamefont{Jia}}, \bibnamefont{and}
  \bibinfo{author}{\bibfnamefont{K.}~\bibnamefont{He}}, \bibinfo{journal}{arXiv
  preprint arXiv:1706.02677}  (\bibinfo{year}{2017}).

\bibitem[{\citenamefont{Deng et~al.}(2009)\citenamefont{Deng, Dong, Socher, Li,
  Li, and Fei-Fei}}]{deng_imagenet_2009}
\bibinfo{author}{\bibfnamefont{J.}~\bibnamefont{Deng}},
  \bibinfo{author}{\bibfnamefont{W.}~\bibnamefont{Dong}},
  \bibinfo{author}{\bibfnamefont{R.}~\bibnamefont{Socher}},
  \bibinfo{author}{\bibfnamefont{L.-J.} \bibnamefont{Li}},
  \bibinfo{author}{\bibfnamefont{K.}~\bibnamefont{Li}}, \bibnamefont{and}
  \bibinfo{author}{\bibfnamefont{L.}~\bibnamefont{Fei-Fei}}, in
  \emph{\bibinfo{booktitle}{2009 IEEE Conference on Computer Vision and Pattern
  Recognition}} (\bibinfo{year}{2009}), pp. \bibinfo{pages}{248--255}.

\bibitem[{\citenamefont{McCulloch and Pitts}(1943)}]{mcculloch_logical_1943}
\bibinfo{author}{\bibfnamefont{W.~S.} \bibnamefont{McCulloch}}
  \bibnamefont{and} \bibinfo{author}{\bibfnamefont{W.}~\bibnamefont{Pitts}},
  \bibinfo{journal}{The bulletin of mathematical biophysics}
  \textbf{\bibinfo{volume}{5}}, \bibinfo{pages}{115} (\bibinfo{year}{1943}),
  ISSN \bibinfo{issn}{1522-9602}.

\bibitem[{\citenamefont{Rumelhart et~al.}(1986)\citenamefont{Rumelhart, Hinton,
  and Williams}}]{rumelhart_learning_1986}
\bibinfo{author}{\bibfnamefont{D.~E.} \bibnamefont{Rumelhart}},
  \bibinfo{author}{\bibfnamefont{G.~E.} \bibnamefont{Hinton}},
  \bibnamefont{and} \bibinfo{author}{\bibfnamefont{R.~J.}
  \bibnamefont{Williams}}, \bibinfo{journal}{Nature}
  \textbf{\bibinfo{volume}{323}}, \bibinfo{pages}{533} (\bibinfo{year}{1986}).

\bibitem[{\citenamefont{Hinton et~al.}(2006)\citenamefont{Hinton, Osindero, and
  Teh}}]{hinton_fast_2006}
\bibinfo{author}{\bibfnamefont{G.~E.} \bibnamefont{Hinton}},
  \bibinfo{author}{\bibfnamefont{S.}~\bibnamefont{Osindero}}, \bibnamefont{and}
  \bibinfo{author}{\bibfnamefont{Y.-W.} \bibnamefont{Teh}},
  \bibinfo{journal}{Neural Computation} \textbf{\bibinfo{volume}{18}},
  \bibinfo{pages}{1527–1554} (\bibinfo{year}{2006}), ISSN
  \bibinfo{issn}{0899-7667},
  \urlprefix\url{https://doi.org/10.1162/neco.2006.18.7.1527}.

\bibitem[{\citenamefont{Kingma and Ba}(2015)}]{kingma_adam_2015}
\bibinfo{author}{\bibfnamefont{D.~P.} \bibnamefont{Kingma}} \bibnamefont{and}
  \bibinfo{author}{\bibfnamefont{J.}~\bibnamefont{Ba}}, in
  \emph{\bibinfo{booktitle}{3rd International Conference on Learning
  Representations, {ICLR} 2015, San Diego, CA, USA, May 7-9, 2015, Conference
  Track Proceedings}}, edited by
  \bibinfo{editor}{\bibfnamefont{Y.}~\bibnamefont{Bengio}} \bibnamefont{and}
  \bibinfo{editor}{\bibfnamefont{Y.}~\bibnamefont{LeCun}}
  (\bibinfo{year}{2015}), \urlprefix\url{http://arxiv.org/abs/1412.6980}.

\bibitem[{\citenamefont{Ruder}(2016)}]{ruder_overview_2016}
\bibinfo{author}{\bibfnamefont{S.}~\bibnamefont{Ruder}},
  \bibinfo{journal}{arXiv preprint arXiv:1609.04747}  (\bibinfo{year}{2016}).

\bibitem[{\citenamefont{Rosenblatt}(1957)}]{rosenblatt_perceptron_1957}
\bibinfo{author}{\bibfnamefont{F.}~\bibnamefont{Rosenblatt}},
  \emph{\bibinfo{title}{The Perceptron, a {Perceiving} and {Recognizing}
  {Automaton}, {Project} {Para}}}, vol. \bibinfo{volume}{85, Issues 460-461} of
  \emph{\bibinfo{series}{Report: Cornell Aeronautical Laboratory}}
  (\bibinfo{publisher}{Cornell Aeronautical Laboratory}, \bibinfo{year}{1957}).

\bibitem[{\citenamefont{Hopfield}(1982)}]{hopfield_neural_1982}
\bibinfo{author}{\bibfnamefont{J.~J.} \bibnamefont{Hopfield}},
  \bibinfo{journal}{Proceedings of the National Academy of Sciences}
  \textbf{\bibinfo{volume}{79}}, \bibinfo{pages}{2554} (\bibinfo{year}{1982}).

\bibitem[{\citenamefont{Szeliski}(2010)}]{szeliski_computer_2010}
\bibinfo{author}{\bibfnamefont{R.}~\bibnamefont{Szeliski}},
  \emph{\bibinfo{title}{{Computer Vision: Algorithms and Applications}}}
  (\bibinfo{publisher}{Springer Science \& Business Media},
  \bibinfo{year}{2010}), ISBN \bibinfo{isbn}{1848829345}.

\bibitem[{\citenamefont{Krizhevsky et~al.}(2012)\citenamefont{Krizhevsky,
  Sutskever, and Hinton}}]{krizhevsky_imagenet_2012}
\bibinfo{author}{\bibfnamefont{A.}~\bibnamefont{Krizhevsky}},
  \bibinfo{author}{\bibfnamefont{I.}~\bibnamefont{Sutskever}},
  \bibnamefont{and} \bibinfo{author}{\bibfnamefont{G.~E.}
  \bibnamefont{Hinton}}, in \emph{\bibinfo{booktitle}{Advances in Neural
  Information Processing Systems 25}} (\bibinfo{publisher}{Curran Associates,
  Inc.}, \bibinfo{year}{2012}), pp. \bibinfo{pages}{1097--1105}.

\bibitem[{\citenamefont{Yin et~al.}(2017)\citenamefont{Yin, Kann, Yu, and
  Sch{\"u}tze}}]{yin_comparative_2017}
\bibinfo{author}{\bibfnamefont{W.}~\bibnamefont{Yin}},
  \bibinfo{author}{\bibfnamefont{K.}~\bibnamefont{Kann}},
  \bibinfo{author}{\bibfnamefont{M.}~\bibnamefont{Yu}}, \bibnamefont{and}
  \bibinfo{author}{\bibfnamefont{H.}~\bibnamefont{Sch{\"u}tze}},
  \bibinfo{journal}{arXiv preprint arXiv:1702.01923}  (\bibinfo{year}{2017}).

\bibitem[{\citenamefont{Zinkevich et~al.}(2010)\citenamefont{Zinkevich, Weimer,
  Li, and Smola}}]{zinkevich_parallelized_2010}
\bibinfo{author}{\bibfnamefont{M.}~\bibnamefont{Zinkevich}},
  \bibinfo{author}{\bibfnamefont{M.}~\bibnamefont{Weimer}},
  \bibinfo{author}{\bibfnamefont{L.}~\bibnamefont{Li}}, \bibnamefont{and}
  \bibinfo{author}{\bibfnamefont{A.}~\bibnamefont{Smola}}, in
  \emph{\bibinfo{booktitle}{Advances in Neural Information Processing Systems}}
  (\bibinfo{publisher}{Curran Associates, Inc.}, \bibinfo{year}{2010}),
  vol.~\bibinfo{volume}{23}.

\bibitem[{\citenamefont{Recht et~al.}(2011)\citenamefont{Recht, Re, Wright, and
  Niu}}]{niu_hogwild_2011}
\bibinfo{author}{\bibfnamefont{B.}~\bibnamefont{Recht}},
  \bibinfo{author}{\bibfnamefont{C.}~\bibnamefont{Re}},
  \bibinfo{author}{\bibfnamefont{S.}~\bibnamefont{Wright}}, \bibnamefont{and}
  \bibinfo{author}{\bibfnamefont{F.}~\bibnamefont{Niu}}, in
  \emph{\bibinfo{booktitle}{Advances in Neural Information Processing
  Systems}}, edited by
  \bibinfo{editor}{\bibfnamefont{J.}~\bibnamefont{Shawe-Taylor}},
  \bibinfo{editor}{\bibfnamefont{R.}~\bibnamefont{Zemel}},
  \bibinfo{editor}{\bibfnamefont{P.}~\bibnamefont{Bartlett}},
  \bibinfo{editor}{\bibfnamefont{F.}~\bibnamefont{Pereira}}, \bibnamefont{and}
  \bibinfo{editor}{\bibfnamefont{K.~Q.} \bibnamefont{Weinberger}}
  (\bibinfo{publisher}{Curran Associates, Inc.}, \bibinfo{year}{2011}),
  vol.~\bibinfo{volume}{24}.

\bibitem[{\citenamefont{Dean et~al.}(2012)\citenamefont{Dean, Corrado, Monga,
  Chen, Devin, Le, Mao, Ranzato, Senior, Tucker et~al.}}]{dean_large_2012}
\bibinfo{author}{\bibfnamefont{J.}~\bibnamefont{Dean}},
  \bibinfo{author}{\bibfnamefont{G.~S.} \bibnamefont{Corrado}},
  \bibinfo{author}{\bibfnamefont{R.}~\bibnamefont{Monga}},
  \bibinfo{author}{\bibfnamefont{K.}~\bibnamefont{Chen}},
  \bibinfo{author}{\bibfnamefont{M.}~\bibnamefont{Devin}},
  \bibinfo{author}{\bibfnamefont{Q.~V.} \bibnamefont{Le}},
  \bibinfo{author}{\bibfnamefont{M.~Z.} \bibnamefont{Mao}},
  \bibinfo{author}{\bibfnamefont{M.}~\bibnamefont{Ranzato}},
  \bibinfo{author}{\bibfnamefont{A.}~\bibnamefont{Senior}},
  \bibinfo{author}{\bibfnamefont{P.}~\bibnamefont{Tucker}},
  \bibnamefont{et~al.}, in \emph{\bibinfo{booktitle}{Advances in neural
  information processing systems}} (\bibinfo{year}{2012}), pp.
  \bibinfo{pages}{1223--1231},
  \urlprefix\url{http://papers.nips.cc/paper/4687-large-scale-distributed-deep-networks.pdf}.

\bibitem[{\citenamefont{Coates et~al.}(2013)\citenamefont{Coates, Huval, Wang,
  Wu, Catanzaro, and Andrew}}]{coates_deep_2013}
\bibinfo{author}{\bibfnamefont{A.}~\bibnamefont{Coates}},
  \bibinfo{author}{\bibfnamefont{B.}~\bibnamefont{Huval}},
  \bibinfo{author}{\bibfnamefont{T.}~\bibnamefont{Wang}},
  \bibinfo{author}{\bibfnamefont{D.}~\bibnamefont{Wu}},
  \bibinfo{author}{\bibfnamefont{B.}~\bibnamefont{Catanzaro}},
  \bibnamefont{and} \bibinfo{author}{\bibfnamefont{N.}~\bibnamefont{Andrew}},
  in \emph{\bibinfo{booktitle}{Proceedings of the 30th International Conference
  on Machine Learning}} (\bibinfo{publisher}{PMLR}, \bibinfo{year}{2013}),
  vol.~\bibinfo{volume}{28} of \emph{\bibinfo{series}{Proceedings of Machine
  Learning Research}}, pp. \bibinfo{pages}{1337--1345}.

\bibitem[{\citenamefont{Gholami et~al.}(2018)\citenamefont{Gholami, Azad, Jin,
  Keutzer, and Buluc}}]{gholami_integrated_2017}
\bibinfo{author}{\bibfnamefont{A.}~\bibnamefont{Gholami}},
  \bibinfo{author}{\bibfnamefont{A.}~\bibnamefont{Azad}},
  \bibinfo{author}{\bibfnamefont{P.}~\bibnamefont{Jin}},
  \bibinfo{author}{\bibfnamefont{K.}~\bibnamefont{Keutzer}}, \bibnamefont{and}
  \bibinfo{author}{\bibfnamefont{A.}~\bibnamefont{Buluc}}, in
  \emph{\bibinfo{booktitle}{Proceedings of the 30th on Symposium on Parallelism
  in Algorithms and Architectures}} (\bibinfo{publisher}{Association for
  Computing Machinery}, \bibinfo{year}{2018}), p. \bibinfo{pages}{77–86},
  ISBN \bibinfo{isbn}{9781450357999}.

\bibitem[{\citenamefont{Huang et~al.}(2019)\citenamefont{Huang, Cheng, Bapna,
  Firat, Chen, Chen, Lee, Ngiam, Le, Wu et~al.}}]{huang_gpipe_2019}
\bibinfo{author}{\bibfnamefont{Y.}~\bibnamefont{Huang}},
  \bibinfo{author}{\bibfnamefont{Y.}~\bibnamefont{Cheng}},
  \bibinfo{author}{\bibfnamefont{A.}~\bibnamefont{Bapna}},
  \bibinfo{author}{\bibfnamefont{O.}~\bibnamefont{Firat}},
  \bibinfo{author}{\bibfnamefont{D.}~\bibnamefont{Chen}},
  \bibinfo{author}{\bibfnamefont{M.}~\bibnamefont{Chen}},
  \bibinfo{author}{\bibfnamefont{H.}~\bibnamefont{Lee}},
  \bibinfo{author}{\bibfnamefont{J.}~\bibnamefont{Ngiam}},
  \bibinfo{author}{\bibfnamefont{Q.~V.} \bibnamefont{Le}},
  \bibinfo{author}{\bibfnamefont{Y.}~\bibnamefont{Wu}}, \bibnamefont{et~al.},
  in \emph{\bibinfo{booktitle}{Advances in Neural Information Processing
  Systems}} (\bibinfo{publisher}{Curran Associates, Inc.},
  \bibinfo{year}{2019}), vol.~\bibinfo{volume}{32}.

\bibitem[{\citenamefont{Narayanan et~al.}(2019)\citenamefont{Narayanan, Harlap,
  Phanishayee, Seshadri, Devanur, Ganger, Gibbons, and
  Zaharia}}]{narayanan_pipedream_2019}
\bibinfo{author}{\bibfnamefont{D.}~\bibnamefont{Narayanan}},
  \bibinfo{author}{\bibfnamefont{A.}~\bibnamefont{Harlap}},
  \bibinfo{author}{\bibfnamefont{A.}~\bibnamefont{Phanishayee}},
  \bibinfo{author}{\bibfnamefont{V.}~\bibnamefont{Seshadri}},
  \bibinfo{author}{\bibfnamefont{N.~R.} \bibnamefont{Devanur}},
  \bibinfo{author}{\bibfnamefont{G.~R.} \bibnamefont{Ganger}},
  \bibinfo{author}{\bibfnamefont{P.~B.} \bibnamefont{Gibbons}},
  \bibnamefont{and} \bibinfo{author}{\bibfnamefont{M.}~\bibnamefont{Zaharia}},
  in \emph{\bibinfo{booktitle}{Proceedings of the 27th ACM Symposium on
  Operating Systems Principles}} (\bibinfo{publisher}{Association for Computing
  Machinery}, \bibinfo{year}{2019}), SOSP '19, p. \bibinfo{pages}{1–15}, ISBN
  \bibinfo{isbn}{9781450368735}.

\bibitem[{\citenamefont{Xing et~al.}(2015)\citenamefont{Xing, Ho, Xie, and
  Dai}}]{xing_strategies_2015}
\bibinfo{author}{\bibfnamefont{E.~P.} \bibnamefont{Xing}},
  \bibinfo{author}{\bibfnamefont{Q.}~\bibnamefont{Ho}},
  \bibinfo{author}{\bibfnamefont{P.}~\bibnamefont{Xie}}, \bibnamefont{and}
  \bibinfo{author}{\bibfnamefont{W.}~\bibnamefont{Dai}},
  \bibinfo{journal}{arXiv preprint arXiv:1512.09295}  (\bibinfo{year}{2015}).

\bibitem[{\citenamefont{Jia et~al.}(2019)\citenamefont{Jia, Zaharia, and
  Aiken}}]{jia_beyond_2019}
\bibinfo{author}{\bibfnamefont{Z.}~\bibnamefont{Jia}},
  \bibinfo{author}{\bibfnamefont{M.}~\bibnamefont{Zaharia}}, \bibnamefont{and}
  \bibinfo{author}{\bibfnamefont{A.}~\bibnamefont{Aiken}}, in
  \emph{\bibinfo{booktitle}{Proceedings of Machine Learning and Systems}},
  edited by \bibinfo{editor}{\bibfnamefont{A.}~\bibnamefont{Talwalkar}},
  \bibinfo{editor}{\bibfnamefont{V.}~\bibnamefont{Smith}}, \bibnamefont{and}
  \bibinfo{editor}{\bibfnamefont{M.}~\bibnamefont{Zaharia}}
  (\bibinfo{year}{2019}), vol.~\bibinfo{volume}{1}, pp. \bibinfo{pages}{1--13},
  \urlprefix\url{https://proceedings.mlsys.org/paper/2019/file/c74d97b01eae257e44aa9d5bade97baf-Paper.pdf}.

\bibitem[{\citenamefont{Li et~al.}(2014)\citenamefont{Li, Andersen, Park,
  Smola, Ahmed, Josifovski, Long, Shekita, and Su}}]{li_scaling_2014}
\bibinfo{author}{\bibfnamefont{M.}~\bibnamefont{Li}},
  \bibinfo{author}{\bibfnamefont{D.~G.} \bibnamefont{Andersen}},
  \bibinfo{author}{\bibfnamefont{J.~W.} \bibnamefont{Park}},
  \bibinfo{author}{\bibfnamefont{A.}~\bibnamefont{Smola}},
  \bibinfo{author}{\bibfnamefont{A.}~\bibnamefont{Ahmed}},
  \bibinfo{author}{\bibfnamefont{V.}~\bibnamefont{Josifovski}},
  \bibinfo{author}{\bibfnamefont{J.}~\bibnamefont{Long}},
  \bibinfo{author}{\bibfnamefont{E.~J.} \bibnamefont{Shekita}},
  \bibnamefont{and} \bibinfo{author}{\bibfnamefont{B.-Y.} \bibnamefont{Su}}, in
  \emph{\bibinfo{booktitle}{Proceedings of the 2014 International Conference on
  Big Data Science and Computing}} (\bibinfo{publisher}{Association for
  Computing Machinery}, \bibinfo{address}{New York, NY, USA},
  \bibinfo{year}{2014}), BigDataScience '14, ISBN
  \bibinfo{isbn}{9781450328913},
  \urlprefix\url{https://doi.org/10.1145/2640087.2644155}.

\bibitem[{\citenamefont{Gupta et~al.}(2017)\citenamefont{Gupta, Zhang, and
  Wang}}]{gupta_model_2017}
\bibinfo{author}{\bibfnamefont{S.}~\bibnamefont{Gupta}},
  \bibinfo{author}{\bibfnamefont{W.}~\bibnamefont{Zhang}}, \bibnamefont{and}
  \bibinfo{author}{\bibfnamefont{F.}~\bibnamefont{Wang}}, in
  \emph{\bibinfo{booktitle}{Proceedings of the Twenty-Sixth International Joint
  Conference on Artificial Intelligence, {IJCAI-17}}} (\bibinfo{year}{2017}),
  pp. \bibinfo{pages}{4854--4858},
  \urlprefix\url{https://doi.org/10.24963/ijcai.2017/681}.

\bibitem[{\citenamefont{Sergeev and Balso}(2018)}]{sergeev_horovod_2018}
\bibinfo{author}{\bibfnamefont{A.}~\bibnamefont{Sergeev}} \bibnamefont{and}
  \bibinfo{author}{\bibfnamefont{M.~D.} \bibnamefont{Balso}},
  \bibinfo{journal}{arXiv preprint arXiv:1802.05799}  (\bibinfo{year}{2018}).

\bibitem[{\citenamefont{Daily et~al.}(2018)\citenamefont{Daily, Vishnu, Siegel,
  Warfel, and Amatya}}]{daily_gossipgrad_2018}
\bibinfo{author}{\bibfnamefont{J.}~\bibnamefont{Daily}},
  \bibinfo{author}{\bibfnamefont{A.}~\bibnamefont{Vishnu}},
  \bibinfo{author}{\bibfnamefont{C.}~\bibnamefont{Siegel}},
  \bibinfo{author}{\bibfnamefont{T.}~\bibnamefont{Warfel}}, \bibnamefont{and}
  \bibinfo{author}{\bibfnamefont{V.}~\bibnamefont{Amatya}},
  \bibinfo{journal}{arXiv preprint arXiv:1803.05880}  (\bibinfo{year}{2018}),
  \urlprefix\url{https://arxiv.org/abs/1803.05880}.

\bibitem[{\citenamefont{Lian et~al.}(2017)\citenamefont{Lian, Zhang, Zhang,
  Hsieh, Zhang, and Liu}}]{lian_can_2017}
\bibinfo{author}{\bibfnamefont{X.}~\bibnamefont{Lian}},
  \bibinfo{author}{\bibfnamefont{C.}~\bibnamefont{Zhang}},
  \bibinfo{author}{\bibfnamefont{H.}~\bibnamefont{Zhang}},
  \bibinfo{author}{\bibfnamefont{C.-J.} \bibnamefont{Hsieh}},
  \bibinfo{author}{\bibfnamefont{W.}~\bibnamefont{Zhang}}, \bibnamefont{and}
  \bibinfo{author}{\bibfnamefont{J.}~\bibnamefont{Liu}}, in
  \emph{\bibinfo{booktitle}{Proceedings of the 31st International Conference on
  Neural Information Processing Systems}} (\bibinfo{publisher}{Curran
  Associates Inc.}, \bibinfo{address}{Red Hook, NY, USA},
  \bibinfo{year}{2017}), NIPS'17, p. \bibinfo{pages}{5336–5346}, ISBN
  \bibinfo{isbn}{9781510860964}.

\bibitem[{\citenamefont{Iandola et~al.}(2016)\citenamefont{Iandola, Moskewicz,
  and Keutzer}}]{iandola_firecaffe_2016}
\bibinfo{author}{\bibfnamefont{F.}~\bibnamefont{Iandola}},
  \bibinfo{author}{\bibfnamefont{M.}~\bibnamefont{Moskewicz}},
  \bibnamefont{and} \bibinfo{author}{\bibfnamefont{K.}~\bibnamefont{Keutzer}},
  in \emph{\bibinfo{booktitle}{Proceedings of the IEEE Conference on Computer
  Vision and Pattern Recognition (CVPR)}} (\bibinfo{year}{2016}), pp.
  \bibinfo{pages}{2592--2600}.

\bibitem[{\citenamefont{Keuper and Pfreundt}(2015)}]{keuper_asynchronous_2015}
\bibinfo{author}{\bibfnamefont{J.}~\bibnamefont{Keuper}} \bibnamefont{and}
  \bibinfo{author}{\bibfnamefont{F.-J.} \bibnamefont{Pfreundt}}, in
  \emph{\bibinfo{booktitle}{Proceedings of the Workshop on Machine Learning in
  High-Performance Computing Environments}} (\bibinfo{publisher}{Association
  for Computing Machinery}, \bibinfo{address}{New York, NY, USA},
  \bibinfo{year}{2015}), MLHPC '15, ISBN \bibinfo{isbn}{9781450340069},
  \urlprefix\url{https://doi.org/10.1145/2834892.2834893}.

\bibitem[{\citenamefont{Ho et~al.}(2013)\citenamefont{Ho, Cipar, Cui, Lee, Kim,
  Gibbons, Gibson, Ganger, and Xing}}]{ho_more_2013}
\bibinfo{author}{\bibfnamefont{Q.}~\bibnamefont{Ho}},
  \bibinfo{author}{\bibfnamefont{J.}~\bibnamefont{Cipar}},
  \bibinfo{author}{\bibfnamefont{H.}~\bibnamefont{Cui}},
  \bibinfo{author}{\bibfnamefont{S.}~\bibnamefont{Lee}},
  \bibinfo{author}{\bibfnamefont{J.~K.} \bibnamefont{Kim}},
  \bibinfo{author}{\bibfnamefont{P.~B.} \bibnamefont{Gibbons}},
  \bibinfo{author}{\bibfnamefont{G.~A.} \bibnamefont{Gibson}},
  \bibinfo{author}{\bibfnamefont{G.}~\bibnamefont{Ganger}}, \bibnamefont{and}
  \bibinfo{author}{\bibfnamefont{E.~P.} \bibnamefont{Xing}},
  \bibinfo{journal}{Advances in neural information processing systems}
  \textbf{\bibinfo{volume}{26}} (\bibinfo{year}{2013}).

\bibitem[{\citenamefont{Thakur et~al.}(2005)\citenamefont{Thakur, Rabenseifner,
  and Gropp}}]{thakur_optimization_2005}
\bibinfo{author}{\bibfnamefont{R.}~\bibnamefont{Thakur}},
  \bibinfo{author}{\bibfnamefont{R.}~\bibnamefont{Rabenseifner}},
  \bibnamefont{and} \bibinfo{author}{\bibfnamefont{W.}~\bibnamefont{Gropp}},
  \bibinfo{journal}{The International Journal of High Performance Computing
  Applications} \textbf{\bibinfo{volume}{19}}, \bibinfo{pages}{49}
  (\bibinfo{year}{2005}).

\bibitem[{\citenamefont{Patarasuk and Yuan}(2009)}]{patarasuk_bandwidth_2009}
\bibinfo{author}{\bibfnamefont{P.}~\bibnamefont{Patarasuk}} \bibnamefont{and}
  \bibinfo{author}{\bibfnamefont{X.}~\bibnamefont{Yuan}},
  \bibinfo{journal}{Journal of Parallel and Distributed Computing}
  \textbf{\bibinfo{volume}{69}}, \bibinfo{pages}{117} (\bibinfo{year}{2009}).

\bibitem[{\citenamefont{Walker et~al.}(1996)\citenamefont{Walker, Walker,
  Dongarra, and Dongarra}}]{walker_mpi_1996}
\bibinfo{author}{\bibfnamefont{D.~W.} \bibnamefont{Walker}},
  \bibinfo{author}{\bibfnamefont{D.~W.} \bibnamefont{Walker}},
  \bibinfo{author}{\bibfnamefont{J.~J.} \bibnamefont{Dongarra}},
  \bibnamefont{and} \bibinfo{author}{\bibfnamefont{J.~J.}
  \bibnamefont{Dongarra}}, \bibinfo{journal}{Supercomputer}
  \textbf{\bibinfo{volume}{12}}, \bibinfo{pages}{56} (\bibinfo{year}{1996}).

\bibitem[{\citenamefont{Benioff}(1980)}]{benioff_computer_1980}
\bibinfo{author}{\bibfnamefont{P.}~\bibnamefont{Benioff}},
  \bibinfo{journal}{Journal of Statistical Physics}
  \textbf{\bibinfo{volume}{22}}, \bibinfo{pages}{563} (\bibinfo{year}{1980}).

\bibitem[{\citenamefont{Feynman}(1982)}]{feynman_simulating_1982}
\bibinfo{author}{\bibfnamefont{R.~P.} \bibnamefont{Feynman}},
  \bibinfo{journal}{International Journal of Theoretical Physics}
  \textbf{\bibinfo{volume}{21}}, \bibinfo{pages}{467} (\bibinfo{year}{1982}).

\bibitem[{\citenamefont{Prati et~al.}(2017)\citenamefont{Prati, Rotta,
  Sebastiano, and Charbon}}]{prati_quantum_2017}
\bibinfo{author}{\bibfnamefont{E.}~\bibnamefont{Prati}},
  \bibinfo{author}{\bibfnamefont{D.}~\bibnamefont{Rotta}},
  \bibinfo{author}{\bibfnamefont{F.}~\bibnamefont{Sebastiano}},
  \bibnamefont{and} \bibinfo{author}{\bibfnamefont{E.}~\bibnamefont{Charbon}},
  in \emph{\bibinfo{booktitle}{2017 IEEE International Conference on Rebooting
  Computing (ICRC)}} (\bibinfo{year}{2017}), pp. \bibinfo{pages}{1--4}.

\bibitem[{\citenamefont{Markov}(2014)}]{markov_limits_2014}
\bibinfo{author}{\bibfnamefont{I.}~\bibnamefont{Markov}},
  \bibinfo{journal}{Nature} \textbf{\bibinfo{volume}{512}},
  \bibinfo{pages}{147} (\bibinfo{year}{2014}).

\bibitem[{\citenamefont{Montanaro}(2016)}]{montanaro_quantum_2016}
\bibinfo{author}{\bibfnamefont{A.}~\bibnamefont{Montanaro}},
  \bibinfo{journal}{npj Quantum Information} \textbf{\bibinfo{volume}{2}},
  \bibinfo{pages}{15023} (\bibinfo{year}{2016}), ISSN
  \bibinfo{issn}{2056-6387}.

\bibitem[{\citenamefont{Kimble}(2008)}]{kimble_quantum_2008}
\bibinfo{author}{\bibfnamefont{H.~J.} \bibnamefont{Kimble}},
  \bibinfo{journal}{Nature} \textbf{\bibinfo{volume}{453}},
  \bibinfo{pages}{1023} (\bibinfo{year}{2008}).

\bibitem[{\citenamefont{Wehner et~al.}(2018)\citenamefont{Wehner, Elkouss, and
  Hanson}}]{wehner_quantum_2018}
\bibinfo{author}{\bibfnamefont{S.}~\bibnamefont{Wehner}},
  \bibinfo{author}{\bibfnamefont{D.}~\bibnamefont{Elkouss}}, \bibnamefont{and}
  \bibinfo{author}{\bibfnamefont{R.}~\bibnamefont{Hanson}},
  \bibinfo{journal}{Science} \textbf{\bibinfo{volume}{362}},
  \bibinfo{pages}{eaam9288} (\bibinfo{year}{2018}).

\bibitem[{\citenamefont{Cacciapuoti et~al.}(2020)\citenamefont{Cacciapuoti,
  Caleffi, Tafuri, Cataliotti, Gherardini, and
  Bianchi}}]{cacciapuoti_quantum_2020}
\bibinfo{author}{\bibfnamefont{A.~S.} \bibnamefont{Cacciapuoti}},
  \bibinfo{author}{\bibfnamefont{M.}~\bibnamefont{Caleffi}},
  \bibinfo{author}{\bibfnamefont{F.}~\bibnamefont{Tafuri}},
  \bibinfo{author}{\bibfnamefont{F.~S.} \bibnamefont{Cataliotti}},
  \bibinfo{author}{\bibfnamefont{S.}~\bibnamefont{Gherardini}},
  \bibnamefont{and} \bibinfo{author}{\bibfnamefont{G.}~\bibnamefont{Bianchi}},
  \bibinfo{journal}{{IEEE} Network} \textbf{\bibinfo{volume}{34}},
  \bibinfo{pages}{137} (\bibinfo{year}{2020}).

\bibitem[{\citenamefont{Cuomo et~al.}(2020)\citenamefont{Cuomo, Caleffi, and
  Cacciapuoti}}]{cuomo_towards_2020}
\bibinfo{author}{\bibfnamefont{D.}~\bibnamefont{Cuomo}},
  \bibinfo{author}{\bibfnamefont{M.}~\bibnamefont{Caleffi}}, \bibnamefont{and}
  \bibinfo{author}{\bibfnamefont{A.~S.} \bibnamefont{Cacciapuoti}},
  \bibinfo{journal}{{IET} Quantum Communication} \textbf{\bibinfo{volume}{1}},
  \bibinfo{pages}{3} (\bibinfo{year}{2020}).

\bibitem[{\citenamefont{Rohde}(2021)}]{rohde_quantum_2021}
\bibinfo{author}{\bibfnamefont{P.~P.} \bibnamefont{Rohde}},
  \emph{\bibinfo{title}{The Quantum Internet: The Second Quantum Revolution}}
  (\bibinfo{publisher}{Cambridge University Press}, \bibinfo{year}{2021}).

\bibitem[{ibm()}]{ibm_cloud}
\emph{\bibinfo{title}{{IBM} {Quantum} {Experience}}},
  \bibinfo{note}{https://quantum-computing.ibm.com. Last Accessed 06.2022},
  \urlprefix\url{https://quantum-computing.ibm.com}.

\bibitem[{\citenamefont{Almorsy et~al.}(2016)\citenamefont{Almorsy, Grundy, and
  Müller}}]{almorsy_analysis_2016}
\bibinfo{author}{\bibfnamefont{M.}~\bibnamefont{Almorsy}},
  \bibinfo{author}{\bibfnamefont{J.}~\bibnamefont{Grundy}}, \bibnamefont{and}
  \bibinfo{author}{\bibfnamefont{I.}~\bibnamefont{Müller}},
  \bibinfo{journal}{arXiv preprint arXiv:1609.01107}  (\bibinfo{year}{2016}).

\bibitem[{\citenamefont{Arrighi and Salvail}(2006)}]{arrighi_quantum_2003}
\bibinfo{author}{\bibfnamefont{P.}~\bibnamefont{Arrighi}} \bibnamefont{and}
  \bibinfo{author}{\bibfnamefont{L.}~\bibnamefont{Salvail}},
  \bibinfo{journal}{International Journal of Quantum Information}
  \textbf{\bibinfo{volume}{4}}, \bibinfo{pages}{883} (\bibinfo{year}{2006}).

\bibitem[{\citenamefont{Broadbent et~al.}(2009)\citenamefont{Broadbent,
  Fitzsimons, and Kashefi}}]{broadbent_universal_2009}
\bibinfo{author}{\bibfnamefont{A.}~\bibnamefont{Broadbent}},
  \bibinfo{author}{\bibfnamefont{J.}~\bibnamefont{Fitzsimons}},
  \bibnamefont{and} \bibinfo{author}{\bibfnamefont{E.}~\bibnamefont{Kashefi}},
  in \emph{\bibinfo{booktitle}{2009 50th Annual IEEE Symposium on Foundations
  of Computer Science}} (\bibinfo{year}{2009}), pp. \bibinfo{pages}{517--526}.

\bibitem[{\citenamefont{Fitzsimons}(2016)}]{fitzsimons_private_2016}
\bibinfo{author}{\bibfnamefont{J.~F.} \bibnamefont{Fitzsimons}}
  (\bibinfo{year}{2016}).

\bibitem[{\citenamefont{Shor}(1997)}]{shor_polynomial-time_1995}
\bibinfo{author}{\bibfnamefont{P.~W.} \bibnamefont{Shor}},
  \bibinfo{journal}{SIAM J. Comput.} \textbf{\bibinfo{volume}{26}}
  (\bibinfo{year}{1997}), ISSN \bibinfo{issn}{0097-5397}.

\bibitem[{\citenamefont{Deutsch}(1985)}]{deutsch_quantum_1985}
\bibinfo{author}{\bibfnamefont{D.}~\bibnamefont{Deutsch}},
  \bibinfo{journal}{Proceedings of the Royal Society of London. A. Mathematical
  and Physical Sciences} \textbf{\bibinfo{volume}{400}} (\bibinfo{year}{1985}),
  ISSN \bibinfo{issn}{2053-9169}.

\bibitem[{\citenamefont{Deutsch and Jozsa}(1992)}]{deutsch_rapid_1992}
\bibinfo{author}{\bibfnamefont{D.}~\bibnamefont{Deutsch}} \bibnamefont{and}
  \bibinfo{author}{\bibfnamefont{R.}~\bibnamefont{Jozsa}},
  \bibinfo{journal}{Proceedings of the Royal Society of London. Series A:
  Mathematical and Physical Sciences} \textbf{\bibinfo{volume}{439}}
  (\bibinfo{year}{1992}).

\bibitem[{\citenamefont{Grover}(1997)}]{grover_fast_1996}
\bibinfo{author}{\bibfnamefont{L.~K.} \bibnamefont{Grover}},
  \bibinfo{journal}{Physical Review Letters} \textbf{\bibinfo{volume}{79}},
  \bibinfo{pages}{325–328} (\bibinfo{year}{1997}), ISSN
  \bibinfo{issn}{1079-7114}.

\bibitem[{\citenamefont{Cleve et~al.}(1997)\citenamefont{Cleve, Ekert,
  Macchiavello, and Mosca}}]{cleve_quantum_1997}
\bibinfo{author}{\bibfnamefont{R.}~\bibnamefont{Cleve}},
  \bibinfo{author}{\bibfnamefont{A.}~\bibnamefont{Ekert}},
  \bibinfo{author}{\bibfnamefont{C.}~\bibnamefont{Macchiavello}},
  \bibnamefont{and} \bibinfo{author}{\bibfnamefont{M.}~\bibnamefont{Mosca}},
  \bibinfo{journal}{Proceedings of the Royal Society A: Mathematical, Physical
  and Engineering Sciences} \textbf{\bibinfo{volume}{454}}
  (\bibinfo{year}{1997}).

\bibitem[{\citenamefont{Brassard et~al.}(1998)\citenamefont{Brassard, Hoyer,
  and Tapp}}]{brassard_quantum_1998}
\bibinfo{author}{\bibfnamefont{G.}~\bibnamefont{Brassard}},
  \bibinfo{author}{\bibfnamefont{P.}~\bibnamefont{Hoyer}}, \bibnamefont{and}
  \bibinfo{author}{\bibfnamefont{A.}~\bibnamefont{Tapp}},
  \bibinfo{journal}{Automata Languages and Programming}
  \textbf{\bibinfo{volume}{1443}} (\bibinfo{year}{1998}).

\bibitem[{\citenamefont{Dunjko et~al.}(2016)\citenamefont{Dunjko, Taylor, and
  Briegel}}]{dunjko_quantum-enhanced_2016}
\bibinfo{author}{\bibfnamefont{V.}~\bibnamefont{Dunjko}},
  \bibinfo{author}{\bibfnamefont{J.~M.} \bibnamefont{Taylor}},
  \bibnamefont{and} \bibinfo{author}{\bibfnamefont{H.~J.}
  \bibnamefont{Briegel}}, \bibinfo{journal}{Physical Review Letters}
  \textbf{\bibinfo{volume}{117}} (\bibinfo{year}{2016}), ISSN
  \bibinfo{issn}{1079-7114}.

\bibitem[{\citenamefont{Lloyd et~al.}(2013)\citenamefont{Lloyd, Mohseni, and
  Rebentrost}}]{lloyd_quantum_2013}
\bibinfo{author}{\bibfnamefont{S.}~\bibnamefont{Lloyd}},
  \bibinfo{author}{\bibfnamefont{M.}~\bibnamefont{Mohseni}}, \bibnamefont{and}
  \bibinfo{author}{\bibfnamefont{P.}~\bibnamefont{Rebentrost}},
  \bibinfo{journal}{arXiv preprint arXiv:1307.0411}  (\bibinfo{year}{2013}).

\bibitem[{\citenamefont{Carleo et~al.}(2019)\citenamefont{Carleo, Cirac,
  Cranmer, Daudet, Schuld, Tishby, Vogt-Maranto, and
  Zdeborov\'a}}]{carleo_machine_2019}
\bibinfo{author}{\bibfnamefont{G.}~\bibnamefont{Carleo}},
  \bibinfo{author}{\bibfnamefont{I.}~\bibnamefont{Cirac}},
  \bibinfo{author}{\bibfnamefont{K.}~\bibnamefont{Cranmer}},
  \bibinfo{author}{\bibfnamefont{L.}~\bibnamefont{Daudet}},
  \bibinfo{author}{\bibfnamefont{M.}~\bibnamefont{Schuld}},
  \bibinfo{author}{\bibfnamefont{N.}~\bibnamefont{Tishby}},
  \bibinfo{author}{\bibfnamefont{L.}~\bibnamefont{Vogt-Maranto}},
  \bibnamefont{and}
  \bibinfo{author}{\bibfnamefont{L.}~\bibnamefont{Zdeborov\'a}},
  \bibinfo{journal}{Reviews of Modern Physics} \textbf{\bibinfo{volume}{91}}
  (\bibinfo{year}{2019}).

\bibitem[{\citenamefont{Dawid et~al.}(2022)\citenamefont{Dawid, Arnold,
  Requena, Gresch, Płodzień, Donatella, Nicoli, Stornati, Koch, Büttner
  et~al.}}]{dawid_modern_2022}
\bibinfo{author}{\bibfnamefont{A.}~\bibnamefont{Dawid}},
  \bibinfo{author}{\bibfnamefont{J.}~\bibnamefont{Arnold}},
  \bibinfo{author}{\bibfnamefont{B.}~\bibnamefont{Requena}},
  \bibinfo{author}{\bibfnamefont{A.}~\bibnamefont{Gresch}},
  \bibinfo{author}{\bibfnamefont{M.}~\bibnamefont{Płodzień}},
  \bibinfo{author}{\bibfnamefont{K.}~\bibnamefont{Donatella}},
  \bibinfo{author}{\bibfnamefont{K.~A.} \bibnamefont{Nicoli}},
  \bibinfo{author}{\bibfnamefont{P.}~\bibnamefont{Stornati}},
  \bibinfo{author}{\bibfnamefont{R.}~\bibnamefont{Koch}},
  \bibinfo{author}{\bibfnamefont{M.}~\bibnamefont{Büttner}},
  \bibnamefont{et~al.}, \bibinfo{journal}{arXiv preprint arXiv:2204.04198}
  (\bibinfo{year}{2022}), \urlprefix\url{https://arxiv.org/abs/2204.04198}.

\bibitem[{\citenamefont{Bukov et~al.}(2018)\citenamefont{Bukov, Day, Sels,
  Weinberg, Polkovnikov, and Mehta}}]{bukov_reinforcement_2018}
\bibinfo{author}{\bibfnamefont{M.}~\bibnamefont{Bukov}},
  \bibinfo{author}{\bibfnamefont{A.~G.~R.} \bibnamefont{Day}},
  \bibinfo{author}{\bibfnamefont{D.}~\bibnamefont{Sels}},
  \bibinfo{author}{\bibfnamefont{P.}~\bibnamefont{Weinberg}},
  \bibinfo{author}{\bibfnamefont{A.}~\bibnamefont{Polkovnikov}},
  \bibnamefont{and} \bibinfo{author}{\bibfnamefont{P.}~\bibnamefont{Mehta}},
  \bibinfo{journal}{Physical Review X} \textbf{\bibinfo{volume}{8}}
  (\bibinfo{year}{2018}).

\bibitem[{\citenamefont{Niu et~al.}(2019)\citenamefont{Niu, Boixo, Smelyanskiy,
  and Neven}}]{niu_universal_2019}
\bibinfo{author}{\bibfnamefont{M.~Y.} \bibnamefont{Niu}},
  \bibinfo{author}{\bibfnamefont{S.}~\bibnamefont{Boixo}},
  \bibinfo{author}{\bibfnamefont{V.~N.} \bibnamefont{Smelyanskiy}},
  \bibnamefont{and} \bibinfo{author}{\bibfnamefont{H.}~\bibnamefont{Neven}},
  \bibinfo{journal}{npj Quantum Information} \textbf{\bibinfo{volume}{5}},
  \bibinfo{pages}{1} (\bibinfo{year}{2019}).

\bibitem[{\citenamefont{Nautrup et~al.}(2019)\citenamefont{Nautrup, Delfosse,
  Dunjko, Briegel, and Friis}}]{nautrup_optimizing_2019}
\bibinfo{author}{\bibfnamefont{H.~P.} \bibnamefont{Nautrup}},
  \bibinfo{author}{\bibfnamefont{N.}~\bibnamefont{Delfosse}},
  \bibinfo{author}{\bibfnamefont{V.}~\bibnamefont{Dunjko}},
  \bibinfo{author}{\bibfnamefont{H.~J.} \bibnamefont{Briegel}},
  \bibnamefont{and} \bibinfo{author}{\bibfnamefont{N.}~\bibnamefont{Friis}},
  \bibinfo{journal}{Quantum} \textbf{\bibinfo{volume}{3}}, \bibinfo{pages}{215}
  (\bibinfo{year}{2019}), ISSN \bibinfo{issn}{2521-327X}.

\bibitem[{\citenamefont{Torlai and Melko}(2017)}]{torlai_neural_2017}
\bibinfo{author}{\bibfnamefont{G.}~\bibnamefont{Torlai}} \bibnamefont{and}
  \bibinfo{author}{\bibfnamefont{R.~G.} \bibnamefont{Melko}},
  \bibinfo{journal}{Physical Review Letters} \textbf{\bibinfo{volume}{119}},
  \bibinfo{pages}{030501} (\bibinfo{year}{2017}).

\bibitem[{\citenamefont{Torlai et~al.}(2018)\citenamefont{Torlai, Mazzola,
  Carrasquilla, Troyer, Melko, and Carleo}}]{torlai_neural-network_2018}
\bibinfo{author}{\bibfnamefont{G.}~\bibnamefont{Torlai}},
  \bibinfo{author}{\bibfnamefont{G.}~\bibnamefont{Mazzola}},
  \bibinfo{author}{\bibfnamefont{J.}~\bibnamefont{Carrasquilla}},
  \bibinfo{author}{\bibfnamefont{M.}~\bibnamefont{Troyer}},
  \bibinfo{author}{\bibfnamefont{R.}~\bibnamefont{Melko}}, \bibnamefont{and}
  \bibinfo{author}{\bibfnamefont{G.}~\bibnamefont{Carleo}},
  \bibinfo{journal}{Nature Physics} \textbf{\bibinfo{volume}{14}}
  (\bibinfo{year}{2018}).

\bibitem[{\citenamefont{Xu and Xu}(2018)}]{xu_neural_2018}
\bibinfo{author}{\bibfnamefont{Q.}~\bibnamefont{Xu}} \bibnamefont{and}
  \bibinfo{author}{\bibfnamefont{S.}~\bibnamefont{Xu}}, \bibinfo{journal}{arXiv
  preprint arXiv:1811.06654}  (\bibinfo{year}{2018}).

\bibitem[{\citenamefont{Arrazola et~al.}(2020)\citenamefont{Arrazola, Delgado,
  Bardhan, and Lloyd}}]{arrazola_quantum-inspired_2020}
\bibinfo{author}{\bibfnamefont{J.~M.} \bibnamefont{Arrazola}},
  \bibinfo{author}{\bibfnamefont{A.}~\bibnamefont{Delgado}},
  \bibinfo{author}{\bibfnamefont{B.~R.} \bibnamefont{Bardhan}},
  \bibnamefont{and} \bibinfo{author}{\bibfnamefont{S.}~\bibnamefont{Lloyd}},
  \bibinfo{journal}{{Quantum}} \textbf{\bibinfo{volume}{4}},
  \bibinfo{pages}{307} (\bibinfo{year}{2020}), ISSN \bibinfo{issn}{2521-327X},
  \urlprefix\url{https://doi.org/10.22331/q-2020-08-13-307}.

\bibitem[{\citenamefont{Harrow et~al.}(2009)\citenamefont{Harrow, Hassidim, and
  Lloyd}}]{harrow_quantum_2009}
\bibinfo{author}{\bibfnamefont{A.~W.} \bibnamefont{Harrow}},
  \bibinfo{author}{\bibfnamefont{A.}~\bibnamefont{Hassidim}}, \bibnamefont{and}
  \bibinfo{author}{\bibfnamefont{S.}~\bibnamefont{Lloyd}},
  \bibinfo{journal}{Physical Review Letters} \textbf{\bibinfo{volume}{103}}
  (\bibinfo{year}{2009}), ISSN \bibinfo{issn}{0031-9007, 1079-7114}.

\bibitem[{\citenamefont{Kerenidis and Prakash}(2016)}]{kerenidis_quantum_2016}
\bibinfo{author}{\bibfnamefont{I.}~\bibnamefont{Kerenidis}} \bibnamefont{and}
  \bibinfo{author}{\bibfnamefont{A.}~\bibnamefont{Prakash}},
  \bibinfo{journal}{arXiv preprint arXiv:1603.08675}  (\bibinfo{year}{2016}).

\bibitem[{\citenamefont{Tang}(2019)}]{tang_quantum-inspired_2018}
\bibinfo{author}{\bibfnamefont{E.}~\bibnamefont{Tang}}, in
  \emph{\bibinfo{booktitle}{Proceedings of the 51st Annual ACM SIGACT Symposium
  on Theory of Computing}} (\bibinfo{year}{2019}), STOC 2019, p.
  \bibinfo{pages}{217–228}, ISBN \bibinfo{isbn}{9781450367059}.

\bibitem[{\citenamefont{Sent\'{\i}s et~al.}(2019)\citenamefont{Sent\'{\i}s,
  Monr\`as, Mu\~noz Tapia, Calsamiglia, and Bagan}}]{sentis_unsupervised_2019}
\bibinfo{author}{\bibfnamefont{G.}~\bibnamefont{Sent\'{\i}s}},
  \bibinfo{author}{\bibfnamefont{A.}~\bibnamefont{Monr\`as}},
  \bibinfo{author}{\bibfnamefont{R.}~\bibnamefont{Mu\~noz Tapia}},
  \bibinfo{author}{\bibfnamefont{J.}~\bibnamefont{Calsamiglia}},
  \bibnamefont{and} \bibinfo{author}{\bibfnamefont{E.}~\bibnamefont{Bagan}},
  \bibinfo{journal}{Physical Review X} \textbf{\bibinfo{volume}{9}},
  \bibinfo{pages}{041029} (\bibinfo{year}{2019}).

\bibitem[{\citenamefont{Liu and Rebentrost}(2018)}]{liu_quantum_2018}
\bibinfo{author}{\bibfnamefont{N.}~\bibnamefont{Liu}} \bibnamefont{and}
  \bibinfo{author}{\bibfnamefont{P.}~\bibnamefont{Rebentrost}},
  \bibinfo{journal}{Physical Review A} \textbf{\bibinfo{volume}{97}},
  \bibinfo{pages}{042315} (\bibinfo{year}{2018}).

\bibitem[{\citenamefont{Anguita et~al.}(2003)\citenamefont{Anguita, Ridella,
  Rivieccio, and Zunino}}]{anguita_quantum_2003}
\bibinfo{author}{\bibfnamefont{D.}~\bibnamefont{Anguita}},
  \bibinfo{author}{\bibfnamefont{S.}~\bibnamefont{Ridella}},
  \bibinfo{author}{\bibfnamefont{F.}~\bibnamefont{Rivieccio}},
  \bibnamefont{and} \bibinfo{author}{\bibfnamefont{R.}~\bibnamefont{Zunino}},
  \bibinfo{journal}{Neural Networks} \textbf{\bibinfo{volume}{16}},
  \bibinfo{pages}{763–770} (\bibinfo{year}{2003}), ISSN
  \bibinfo{issn}{0893-6080}.

\bibitem[{\citenamefont{Lloyd et~al.}(2014)\citenamefont{Lloyd, Mohseni, and
  Rebentrost}}]{lloyd_quantum-principal_2013}
\bibinfo{author}{\bibfnamefont{S.}~\bibnamefont{Lloyd}},
  \bibinfo{author}{\bibfnamefont{M.}~\bibnamefont{Mohseni}}, \bibnamefont{and}
  \bibinfo{author}{\bibfnamefont{P.}~\bibnamefont{Rebentrost}},
  \bibinfo{journal}{Nature Physics} \textbf{\bibinfo{volume}{10}},
  \bibinfo{pages}{631–633} (\bibinfo{year}{2014}), ISSN
  \bibinfo{issn}{1745-2481}.

\bibitem[{\citenamefont{Dong et~al.}(2008)\citenamefont{Dong, Chen, Li, and
  Tarn}}]{dong_quantum_2008}
\bibinfo{author}{\bibfnamefont{D.}~\bibnamefont{Dong}},
  \bibinfo{author}{\bibfnamefont{C.}~\bibnamefont{Chen}},
  \bibinfo{author}{\bibfnamefont{H.}~\bibnamefont{Li}}, \bibnamefont{and}
  \bibinfo{author}{\bibfnamefont{T.-J.} \bibnamefont{Tarn}},
  \bibinfo{journal}{IEEE Transactions on Systems, Man, and Cybernetics, Part B
  (Cybernetics)} \textbf{\bibinfo{volume}{38}}, \bibinfo{pages}{1207–1220}
  (\bibinfo{year}{2008}), ISSN \bibinfo{issn}{1083-4419}.

\bibitem[{\citenamefont{Aaronson}(2015)}]{aaronson_read_2015}
\bibinfo{author}{\bibfnamefont{S.}~\bibnamefont{Aaronson}},
  \bibinfo{journal}{Nature Physics} \textbf{\bibinfo{volume}{11}},
  \bibinfo{pages}{291} (\bibinfo{year}{2015}), ISSN \bibinfo{issn}{1745-2473,
  1745-2481}.

\bibitem[{\citenamefont{Duan et~al.}(2020)\citenamefont{Duan, Yuan, Yu, Huang,
  and Hsieh}}]{duan_survey_2020}
\bibinfo{author}{\bibfnamefont{B.}~\bibnamefont{Duan}},
  \bibinfo{author}{\bibfnamefont{J.}~\bibnamefont{Yuan}},
  \bibinfo{author}{\bibfnamefont{C.-H.} \bibnamefont{Yu}},
  \bibinfo{author}{\bibfnamefont{J.}~\bibnamefont{Huang}}, \bibnamefont{and}
  \bibinfo{author}{\bibfnamefont{C.-Y.} \bibnamefont{Hsieh}},
  \bibinfo{journal}{Physics Letters A} \textbf{\bibinfo{volume}{384}},
  \bibinfo{pages}{126595} (\bibinfo{year}{2020}), ISSN
  \bibinfo{issn}{0375-9601}.

\bibitem[{\citenamefont{Schuld and Killoran}(2022)}]{schuld_quantum_2022}
\bibinfo{author}{\bibfnamefont{M.}~\bibnamefont{Schuld}} \bibnamefont{and}
  \bibinfo{author}{\bibfnamefont{N.}~\bibnamefont{Killoran}},
  \bibinfo{journal}{arXiv preprint arXiv:2203.01340}  (\bibinfo{year}{2022}),
  \urlprefix\url{https://arxiv.org/abs/2203.01340}.

\bibitem[{\citenamefont{Holmes et~al.}(2022)\citenamefont{Holmes, Sharma,
  Cerezo, and Coles}}]{holmes_connecting_2022}
\bibinfo{author}{\bibfnamefont{Z.}~\bibnamefont{Holmes}},
  \bibinfo{author}{\bibfnamefont{K.}~\bibnamefont{Sharma}},
  \bibinfo{author}{\bibfnamefont{M.}~\bibnamefont{Cerezo}}, \bibnamefont{and}
  \bibinfo{author}{\bibfnamefont{P.~J.} \bibnamefont{Coles}},
  \bibinfo{journal}{{PRX} Quantum} \textbf{\bibinfo{volume}{3}}
  (\bibinfo{year}{2022}),
  \urlprefix\url{https://doi.org/10.1103%2Fprxquantum.3.010313}.

\bibitem[{\citenamefont{Sim et~al.}(2019)\citenamefont{Sim, Johnson, and
  Aspuru-Guzik}}]{sim_expressibility_2019}
\bibinfo{author}{\bibfnamefont{S.}~\bibnamefont{Sim}},
  \bibinfo{author}{\bibfnamefont{P.~D.} \bibnamefont{Johnson}},
  \bibnamefont{and}
  \bibinfo{author}{\bibfnamefont{A.}~\bibnamefont{Aspuru-Guzik}},
  \bibinfo{journal}{Advanced Quantum Technologies}
  \textbf{\bibinfo{volume}{2}}, \bibinfo{pages}{1900070}
  (\bibinfo{year}{2019}).

\bibitem[{\citenamefont{Abbas et~al.}(2021)\citenamefont{Abbas, Sutter, Zoufal,
  Lucchi, Figalli, and Woerner}}]{abbas_power_2021}
\bibinfo{author}{\bibfnamefont{A.}~\bibnamefont{Abbas}},
  \bibinfo{author}{\bibfnamefont{D.}~\bibnamefont{Sutter}},
  \bibinfo{author}{\bibfnamefont{C.}~\bibnamefont{Zoufal}},
  \bibinfo{author}{\bibfnamefont{A.}~\bibnamefont{Lucchi}},
  \bibinfo{author}{\bibfnamefont{A.}~\bibnamefont{Figalli}}, \bibnamefont{and}
  \bibinfo{author}{\bibfnamefont{S.}~\bibnamefont{Woerner}},
  \bibinfo{journal}{Nature Computational Science} \textbf{\bibinfo{volume}{1}},
  \bibinfo{pages}{403} (\bibinfo{year}{2021}).

\bibitem[{\citenamefont{Wright and McMahon}(2019)}]{wright_capacity_2019}
\bibinfo{author}{\bibfnamefont{L.~G.} \bibnamefont{Wright}} \bibnamefont{and}
  \bibinfo{author}{\bibfnamefont{P.~L.} \bibnamefont{McMahon}},
  \bibinfo{journal}{arXiv preprint arXiv:1908.01364}  (\bibinfo{year}{2019}),
  \urlprefix\url{https://arxiv.org/abs/1908.01364}.

\bibitem[{\citenamefont{Du et~al.}(2020)\citenamefont{Du, Hsieh, Liu, and
  Tao}}]{du_expressive_2020}
\bibinfo{author}{\bibfnamefont{Y.}~\bibnamefont{Du}},
  \bibinfo{author}{\bibfnamefont{M.-H.} \bibnamefont{Hsieh}},
  \bibinfo{author}{\bibfnamefont{T.}~\bibnamefont{Liu}}, \bibnamefont{and}
  \bibinfo{author}{\bibfnamefont{D.}~\bibnamefont{Tao}},
  \bibinfo{journal}{Phys. Rev. Research} \textbf{\bibinfo{volume}{2}},
  \bibinfo{pages}{033125} (\bibinfo{year}{2020}),
  \urlprefix\url{https://link.aps.org/doi/10.1103/PhysRevResearch.2.033125}.

\bibitem[{\citenamefont{Banchi et~al.}(2021)\citenamefont{Banchi, Pereira, and
  Pirandola}}]{banchi_generalization_2021}
\bibinfo{author}{\bibfnamefont{L.}~\bibnamefont{Banchi}},
  \bibinfo{author}{\bibfnamefont{J.}~\bibnamefont{Pereira}}, \bibnamefont{and}
  \bibinfo{author}{\bibfnamefont{S.}~\bibnamefont{Pirandola}},
  \bibinfo{journal}{{PRX} Quantum} \textbf{\bibinfo{volume}{2}}
  (\bibinfo{year}{2021}),
  \urlprefix\url{https://doi.org/10.1103%2Fprxquantum.2.040321}.

\bibitem[{\citenamefont{Hubregtsen et~al.}(2021)\citenamefont{Hubregtsen,
  Pichlmeier, Stecher, and Bertels}}]{hubregtsen_evaluation_2021}
\bibinfo{author}{\bibfnamefont{T.}~\bibnamefont{Hubregtsen}},
  \bibinfo{author}{\bibfnamefont{J.}~\bibnamefont{Pichlmeier}},
  \bibinfo{author}{\bibfnamefont{P.}~\bibnamefont{Stecher}}, \bibnamefont{and}
  \bibinfo{author}{\bibfnamefont{K.}~\bibnamefont{Bertels}},
  \bibinfo{journal}{Quantum Machine Intelligence} \textbf{\bibinfo{volume}{3}}
  (\bibinfo{year}{2021}),
  \urlprefix\url{https://doi.org/10.1007/s42484-021-00038-w}.

\bibitem[{\citenamefont{Huang et~al.}(2021{\natexlab{a}})\citenamefont{Huang,
  Broughton, Mohseni, Babbush, Boixo, Neven, and McClean}}]{huang_power_2021}
\bibinfo{author}{\bibfnamefont{H.-Y.} \bibnamefont{Huang}},
  \bibinfo{author}{\bibfnamefont{M.}~\bibnamefont{Broughton}},
  \bibinfo{author}{\bibfnamefont{M.}~\bibnamefont{Mohseni}},
  \bibinfo{author}{\bibfnamefont{R.}~\bibnamefont{Babbush}},
  \bibinfo{author}{\bibfnamefont{S.}~\bibnamefont{Boixo}},
  \bibinfo{author}{\bibfnamefont{H.}~\bibnamefont{Neven}}, \bibnamefont{and}
  \bibinfo{author}{\bibfnamefont{J.~R.} \bibnamefont{McClean}},
  \bibinfo{journal}{Nature Communications} \textbf{\bibinfo{volume}{12}}
  (\bibinfo{year}{2021}{\natexlab{a}}), ISSN \bibinfo{issn}{2041-1723}.

\bibitem[{\citenamefont{Huang et~al.}(2021{\natexlab{b}})\citenamefont{Huang,
  Kueng, and Preskill}}]{huang_information_2021}
\bibinfo{author}{\bibfnamefont{H.-Y.} \bibnamefont{Huang}},
  \bibinfo{author}{\bibfnamefont{R.}~\bibnamefont{Kueng}}, \bibnamefont{and}
  \bibinfo{author}{\bibfnamefont{J.}~\bibnamefont{Preskill}},
  \bibinfo{journal}{Phys. Rev. Lett.} \textbf{\bibinfo{volume}{126}},
  \bibinfo{pages}{190505} (\bibinfo{year}{2021}{\natexlab{b}}),
  \urlprefix\url{https://link.aps.org/doi/10.1103/PhysRevLett.126.190505}.

\bibitem[{\citenamefont{Beer et~al.}(2020)\citenamefont{Beer, Bondarenko,
  Farrelly, Osborne, Salzmann, Scheiermann, and Wolf}}]{beer_training_2020}
\bibinfo{author}{\bibfnamefont{K.}~\bibnamefont{Beer}},
  \bibinfo{author}{\bibfnamefont{D.}~\bibnamefont{Bondarenko}},
  \bibinfo{author}{\bibfnamefont{T.}~\bibnamefont{Farrelly}},
  \bibinfo{author}{\bibfnamefont{T.}~\bibnamefont{Osborne}},
  \bibinfo{author}{\bibfnamefont{R.}~\bibnamefont{Salzmann}},
  \bibinfo{author}{\bibfnamefont{D.}~\bibnamefont{Scheiermann}},
  \bibnamefont{and} \bibinfo{author}{\bibfnamefont{R.}~\bibnamefont{Wolf}},
  \bibinfo{journal}{Nature Communications} \textbf{\bibinfo{volume}{11}},
  \bibinfo{pages}{808} (\bibinfo{year}{2020}).

\bibitem[{\citenamefont{McClean et~al.}(2016)\citenamefont{McClean, Romero,
  Babbush, and Aspuru-Guzik}}]{mcclean_theory_2016}
\bibinfo{author}{\bibfnamefont{J.~R.} \bibnamefont{McClean}},
  \bibinfo{author}{\bibfnamefont{J.}~\bibnamefont{Romero}},
  \bibinfo{author}{\bibfnamefont{R.}~\bibnamefont{Babbush}}, \bibnamefont{and}
  \bibinfo{author}{\bibfnamefont{A.}~\bibnamefont{Aspuru-Guzik}},
  \bibinfo{journal}{New Journal of Physics} \textbf{\bibinfo{volume}{18}}
  (\bibinfo{year}{2016}).

\bibitem[{\citenamefont{Bharti et~al.}(2022)\citenamefont{Bharti,
  Cervera-Lierta, Kyaw, Haug, Alperin-Lea, Anand, Degroote, Heimonen, Kottmann,
  Menke et~al.}}]{bharti_noisy_2022}
\bibinfo{author}{\bibfnamefont{K.}~\bibnamefont{Bharti}},
  \bibinfo{author}{\bibfnamefont{A.}~\bibnamefont{Cervera-Lierta}},
  \bibinfo{author}{\bibfnamefont{T.~H.} \bibnamefont{Kyaw}},
  \bibinfo{author}{\bibfnamefont{T.}~\bibnamefont{Haug}},
  \bibinfo{author}{\bibfnamefont{S.}~\bibnamefont{Alperin-Lea}},
  \bibinfo{author}{\bibfnamefont{A.}~\bibnamefont{Anand}},
  \bibinfo{author}{\bibfnamefont{M.}~\bibnamefont{Degroote}},
  \bibinfo{author}{\bibfnamefont{H.}~\bibnamefont{Heimonen}},
  \bibinfo{author}{\bibfnamefont{J.~S.} \bibnamefont{Kottmann}},
  \bibinfo{author}{\bibfnamefont{T.}~\bibnamefont{Menke}},
  \bibnamefont{et~al.}, \bibinfo{journal}{Rev. Mod. Phys.}
  \textbf{\bibinfo{volume}{94}}, \bibinfo{pages}{015004}
  (\bibinfo{year}{2022}),
  \urlprefix\url{https://link.aps.org/doi/10.1103/RevModPhys.94.015004}.

\bibitem[{\citenamefont{Cerezo et~al.}(2021{\natexlab{a}})\citenamefont{Cerezo,
  Arrasmith, Babbush, Benjamin, Endo, Fujii, McClean, Mitarai, Yuan, Cincio
  et~al.}}]{cerezo_variational_2021}
\bibinfo{author}{\bibfnamefont{M.}~\bibnamefont{Cerezo}},
  \bibinfo{author}{\bibfnamefont{A.}~\bibnamefont{Arrasmith}},
  \bibinfo{author}{\bibfnamefont{R.}~\bibnamefont{Babbush}},
  \bibinfo{author}{\bibfnamefont{S.~C.} \bibnamefont{Benjamin}},
  \bibinfo{author}{\bibfnamefont{S.}~\bibnamefont{Endo}},
  \bibinfo{author}{\bibfnamefont{K.}~\bibnamefont{Fujii}},
  \bibinfo{author}{\bibfnamefont{J.~R.} \bibnamefont{McClean}},
  \bibinfo{author}{\bibfnamefont{K.}~\bibnamefont{Mitarai}},
  \bibinfo{author}{\bibfnamefont{X.}~\bibnamefont{Yuan}},
  \bibinfo{author}{\bibfnamefont{L.}~\bibnamefont{Cincio}},
  \bibnamefont{et~al.}, \bibinfo{journal}{Nature Reviews Physics}
  \textbf{\bibinfo{volume}{3}}, \bibinfo{pages}{625}
  (\bibinfo{year}{2021}{\natexlab{a}}).

\bibitem[{\citenamefont{Schuld et~al.}(2014)\citenamefont{Schuld, Sinayskiy,
  and Petruccione}}]{schuld_quest_2014}
\bibinfo{author}{\bibfnamefont{M.}~\bibnamefont{Schuld}},
  \bibinfo{author}{\bibfnamefont{I.}~\bibnamefont{Sinayskiy}},
  \bibnamefont{and}
  \bibinfo{author}{\bibfnamefont{F.}~\bibnamefont{Petruccione}},
  \bibinfo{journal}{Quantum Information Processing}
  \textbf{\bibinfo{volume}{13}} (\bibinfo{year}{2014}).

\bibitem[{\citenamefont{Mangini et~al.}(2021)\citenamefont{Mangini, Tacchino,
  Gerace, Bajoni, and Macchiavello}}]{mangini_quantum_2021}
\bibinfo{author}{\bibfnamefont{S.}~\bibnamefont{Mangini}},
  \bibinfo{author}{\bibfnamefont{F.}~\bibnamefont{Tacchino}},
  \bibinfo{author}{\bibfnamefont{D.}~\bibnamefont{Gerace}},
  \bibinfo{author}{\bibfnamefont{D.}~\bibnamefont{Bajoni}}, \bibnamefont{and}
  \bibinfo{author}{\bibfnamefont{C.}~\bibnamefont{Macchiavello}},
  \bibinfo{journal}{Europhysics Letters} \textbf{\bibinfo{volume}{134}},
  \bibinfo{pages}{10002} (\bibinfo{year}{2021}).

\bibitem[{\citenamefont{Kak}(1995)}]{kak_quantum_1995}
\bibinfo{author}{\bibfnamefont{S.}~\bibnamefont{Kak}},
  \bibinfo{journal}{Information Sciences} \textbf{\bibinfo{volume}{83}},
  \bibinfo{pages}{143} (\bibinfo{year}{1995}), ISSN \bibinfo{issn}{0020-0255}.

\bibitem[{\citenamefont{Chrisley}(1995)}]{chrisley_quantum_1995}
\bibinfo{author}{\bibfnamefont{R.}~\bibnamefont{Chrisley}}, in
  \emph{\bibinfo{booktitle}{Proceedings of the international symposium,
  {Saariselka}}} (\bibinfo{year}{1995}), pp. \bibinfo{pages}{4--9}.

\bibitem[{\citenamefont{Lewenstein}(1994)}]{lewenstein_quantum_1994}
\bibinfo{author}{\bibfnamefont{M.}~\bibnamefont{Lewenstein}},
  \bibinfo{journal}{Journal of Modern Optics} \textbf{\bibinfo{volume}{41}},
  \bibinfo{pages}{2491} (\bibinfo{year}{1994}), ISSN \bibinfo{issn}{0950-0340}.

\bibitem[{\citenamefont{Behrman et~al.}(2000)\citenamefont{Behrman, Nash,
  Steck, Chandrashekar, and Skinner}}]{behrman_simulations_2000}
\bibinfo{author}{\bibfnamefont{E.}~\bibnamefont{Behrman}},
  \bibinfo{author}{\bibfnamefont{L.}~\bibnamefont{Nash}},
  \bibinfo{author}{\bibfnamefont{J.}~\bibnamefont{Steck}},
  \bibinfo{author}{\bibfnamefont{V.}~\bibnamefont{Chandrashekar}},
  \bibnamefont{and} \bibinfo{author}{\bibfnamefont{S.}~\bibnamefont{Skinner}},
  \bibinfo{journal}{Information Sciences} \textbf{\bibinfo{volume}{128}},
  \bibinfo{pages}{257} (\bibinfo{year}{2000}), ISSN \bibinfo{issn}{0020-0255},
  \urlprefix\url{https://www.sciencedirect.com/science/article/pii/S0020025500000566}.

\bibitem[{\citenamefont{Ventura and Martinez}(2000)}]{ventura_quantum_2000}
\bibinfo{author}{\bibfnamefont{D.}~\bibnamefont{Ventura}} \bibnamefont{and}
  \bibinfo{author}{\bibfnamefont{T.}~\bibnamefont{Martinez}},
  \bibinfo{journal}{Information Sciences} \textbf{\bibinfo{volume}{126}},
  \bibinfo{pages}{273} (\bibinfo{year}{2000}).

\bibitem[{\citenamefont{Kandala et~al.}(2017)\citenamefont{Kandala, Mezzacapo,
  Temme, Takita, Brink, Chow, and Gambetta}}]{kandala_hardware_2017}
\bibinfo{author}{\bibfnamefont{A.}~\bibnamefont{Kandala}},
  \bibinfo{author}{\bibfnamefont{A.}~\bibnamefont{Mezzacapo}},
  \bibinfo{author}{\bibfnamefont{K.}~\bibnamefont{Temme}},
  \bibinfo{author}{\bibfnamefont{M.}~\bibnamefont{Takita}},
  \bibinfo{author}{\bibfnamefont{M.}~\bibnamefont{Brink}},
  \bibinfo{author}{\bibfnamefont{J.~M.} \bibnamefont{Chow}}, \bibnamefont{and}
  \bibinfo{author}{\bibfnamefont{J.~M.} \bibnamefont{Gambetta}},
  \bibinfo{journal}{Nature} \textbf{\bibinfo{volume}{549}},
  \bibinfo{pages}{242} (\bibinfo{year}{2017}),
  \urlprefix\url{https://doi.org/10.1038%2Fnature23879}.

\bibitem[{\citenamefont{Sweke et~al.}(2020)\citenamefont{Sweke, Wilde, Meyer,
  Schuld, Faehrmann, Meynard-Piganeau, and Eisert}}]{sweke_stochastic_2020}
\bibinfo{author}{\bibfnamefont{R.}~\bibnamefont{Sweke}},
  \bibinfo{author}{\bibfnamefont{F.}~\bibnamefont{Wilde}},
  \bibinfo{author}{\bibfnamefont{J.}~\bibnamefont{Meyer}},
  \bibinfo{author}{\bibfnamefont{M.}~\bibnamefont{Schuld}},
  \bibinfo{author}{\bibfnamefont{P.~K.} \bibnamefont{Faehrmann}},
  \bibinfo{author}{\bibfnamefont{B.}~\bibnamefont{Meynard-Piganeau}},
  \bibnamefont{and} \bibinfo{author}{\bibfnamefont{J.}~\bibnamefont{Eisert}},
  \bibinfo{journal}{Quantum} \textbf{\bibinfo{volume}{4}}, \bibinfo{pages}{314}
  (\bibinfo{year}{2020}),
  \urlprefix\url{https://doi.org/10.22331%2Fq-2020-08-31-314}.

\bibitem[{\citenamefont{Cao et~al.}(2017)\citenamefont{Cao, Guerreschi, and
  Aspuru-Guzik}}]{cao_quantum_2017}
\bibinfo{author}{\bibfnamefont{Y.}~\bibnamefont{Cao}},
  \bibinfo{author}{\bibfnamefont{G.~G.} \bibnamefont{Guerreschi}},
  \bibnamefont{and}
  \bibinfo{author}{\bibfnamefont{A.}~\bibnamefont{Aspuru-Guzik}},
  \bibinfo{journal}{arXiv preprint arXiv:1711.11240}  (\bibinfo{year}{2017}).

\bibitem[{\citenamefont{Allcock et~al.}(2020)\citenamefont{Allcock, Hsieh,
  Kerenidis, and Zhang}}]{allcock_quantum_2019}
\bibinfo{author}{\bibfnamefont{J.}~\bibnamefont{Allcock}},
  \bibinfo{author}{\bibfnamefont{C.-Y.} \bibnamefont{Hsieh}},
  \bibinfo{author}{\bibfnamefont{I.}~\bibnamefont{Kerenidis}},
  \bibnamefont{and} \bibinfo{author}{\bibfnamefont{S.}~\bibnamefont{Zhang}},
  \bibinfo{journal}{ACM Transactions on Quantum Computing}
  \textbf{\bibinfo{volume}{1}} (\bibinfo{year}{2020}), ISSN
  \bibinfo{issn}{2643-6809}, \urlprefix\url{https://doi.org/10.1145/3411466}.

\bibitem[{\citenamefont{Mitarai et~al.}(2018)\citenamefont{Mitarai, Negoro,
  Kitagawa, and Fujii}}]{mitarai_quantum_2018}
\bibinfo{author}{\bibfnamefont{K.}~\bibnamefont{Mitarai}},
  \bibinfo{author}{\bibfnamefont{M.}~\bibnamefont{Negoro}},
  \bibinfo{author}{\bibfnamefont{M.}~\bibnamefont{Kitagawa}}, \bibnamefont{and}
  \bibinfo{author}{\bibfnamefont{K.}~\bibnamefont{Fujii}},
  \bibinfo{journal}{Physical Review A} \textbf{\bibinfo{volume}{98}}
  (\bibinfo{year}{2018}).

\bibitem[{\citenamefont{Schuld et~al.}(2019)\citenamefont{Schuld, Bergholm,
  Gogolin, Izaac, and Killoran}}]{schuld_evaluating_2019}
\bibinfo{author}{\bibfnamefont{M.}~\bibnamefont{Schuld}},
  \bibinfo{author}{\bibfnamefont{V.}~\bibnamefont{Bergholm}},
  \bibinfo{author}{\bibfnamefont{C.}~\bibnamefont{Gogolin}},
  \bibinfo{author}{\bibfnamefont{J.}~\bibnamefont{Izaac}}, \bibnamefont{and}
  \bibinfo{author}{\bibfnamefont{N.}~\bibnamefont{Killoran}},
  \bibinfo{journal}{Physical Review A} \textbf{\bibinfo{volume}{99}}
  (\bibinfo{year}{2019}),
  \urlprefix\url{https://doi.org/10.1103%2Fphysreva.99.032331}.

\bibitem[{\citenamefont{Schuld et~al.}(2020)\citenamefont{Schuld, Bocharov,
  Svore, and Wiebe}}]{schuld_circuit_2020}
\bibinfo{author}{\bibfnamefont{M.}~\bibnamefont{Schuld}},
  \bibinfo{author}{\bibfnamefont{A.}~\bibnamefont{Bocharov}},
  \bibinfo{author}{\bibfnamefont{K.~M.} \bibnamefont{Svore}}, \bibnamefont{and}
  \bibinfo{author}{\bibfnamefont{N.}~\bibnamefont{Wiebe}},
  \bibinfo{journal}{Physical Review A} \textbf{\bibinfo{volume}{101}}
  (\bibinfo{year}{2020}), ISSN \bibinfo{issn}{2469-9934}.

\bibitem[{\citenamefont{McClean et~al.}(2018)\citenamefont{McClean, Boixo,
  Smelyanskiy, Babbush, and Neven}}]{mcclean_barren_2018}
\bibinfo{author}{\bibfnamefont{J.~R.} \bibnamefont{McClean}},
  \bibinfo{author}{\bibfnamefont{S.}~\bibnamefont{Boixo}},
  \bibinfo{author}{\bibfnamefont{V.~N.} \bibnamefont{Smelyanskiy}},
  \bibinfo{author}{\bibfnamefont{R.}~\bibnamefont{Babbush}}, \bibnamefont{and}
  \bibinfo{author}{\bibfnamefont{H.}~\bibnamefont{Neven}},
  \bibinfo{journal}{Nature Communications} \textbf{\bibinfo{volume}{9}}
  (\bibinfo{year}{2018}).

\bibitem[{\citenamefont{Cerezo et~al.}(2021{\natexlab{b}})\citenamefont{Cerezo,
  Sone, Volkoff, Cincio, and Coles}}]{cerezo_cost_2021}
\bibinfo{author}{\bibfnamefont{M.}~\bibnamefont{Cerezo}},
  \bibinfo{author}{\bibfnamefont{A.}~\bibnamefont{Sone}},
  \bibinfo{author}{\bibfnamefont{T.}~\bibnamefont{Volkoff}},
  \bibinfo{author}{\bibfnamefont{L.}~\bibnamefont{Cincio}}, \bibnamefont{and}
  \bibinfo{author}{\bibfnamefont{P.~J.} \bibnamefont{Coles}},
  \bibinfo{journal}{Nature communications} \textbf{\bibinfo{volume}{12}},
  \bibinfo{pages}{1} (\bibinfo{year}{2021}{\natexlab{b}}).

\bibitem[{\citenamefont{Grant et~al.}(2019)\citenamefont{Grant, Wossnig,
  Ostaszewski, and Benedetti}}]{grant_initialization_2019}
\bibinfo{author}{\bibfnamefont{E.}~\bibnamefont{Grant}},
  \bibinfo{author}{\bibfnamefont{L.}~\bibnamefont{Wossnig}},
  \bibinfo{author}{\bibfnamefont{M.}~\bibnamefont{Ostaszewski}},
  \bibnamefont{and}
  \bibinfo{author}{\bibfnamefont{M.}~\bibnamefont{Benedetti}},
  \bibinfo{journal}{Quantum} \textbf{\bibinfo{volume}{3}}
  (\bibinfo{year}{2019}), ISSN \bibinfo{issn}{2521-327X}.

\bibitem[{\citenamefont{Pesah et~al.}(2021)\citenamefont{Pesah, Cerezo, Wang,
  Volkoff, Sornborger, and Coles}}]{pesah_absence_2021}
\bibinfo{author}{\bibfnamefont{A.}~\bibnamefont{Pesah}},
  \bibinfo{author}{\bibfnamefont{M.}~\bibnamefont{Cerezo}},
  \bibinfo{author}{\bibfnamefont{S.}~\bibnamefont{Wang}},
  \bibinfo{author}{\bibfnamefont{T.}~\bibnamefont{Volkoff}},
  \bibinfo{author}{\bibfnamefont{A.~T.} \bibnamefont{Sornborger}},
  \bibnamefont{and} \bibinfo{author}{\bibfnamefont{P.~J.} \bibnamefont{Coles}},
  \bibinfo{journal}{Phys. Rev. X} \textbf{\bibinfo{volume}{11}},
  \bibinfo{pages}{041011} (\bibinfo{year}{2021}),
  \urlprefix\url{https://link.aps.org/doi/10.1103/PhysRevX.11.041011}.

\bibitem[{\citenamefont{Cong et~al.}(2019)\citenamefont{Cong, Choi, and
  Lukin}}]{cong_quantum_2019}
\bibinfo{author}{\bibfnamefont{I.}~\bibnamefont{Cong}},
  \bibinfo{author}{\bibfnamefont{S.}~\bibnamefont{Choi}}, \bibnamefont{and}
  \bibinfo{author}{\bibfnamefont{M.~D.} \bibnamefont{Lukin}},
  \bibinfo{journal}{Nature Physics} \textbf{\bibinfo{volume}{15}},
  \bibinfo{pages}{1273} (\bibinfo{year}{2019}).

\bibitem[{\citenamefont{Bausch}(2020)}]{bausch_recurrent_2020}
\bibinfo{author}{\bibfnamefont{J.}~\bibnamefont{Bausch}}, in
  \emph{\bibinfo{booktitle}{Advances in Neural Information Processing
  Systems}}, edited by
  \bibinfo{editor}{\bibfnamefont{H.}~\bibnamefont{Larochelle}},
  \bibinfo{editor}{\bibfnamefont{M.}~\bibnamefont{Ranzato}},
  \bibinfo{editor}{\bibfnamefont{R.}~\bibnamefont{Hadsell}},
  \bibinfo{editor}{\bibfnamefont{M.}~\bibnamefont{Balcan}}, \bibnamefont{and}
  \bibinfo{editor}{\bibfnamefont{H.}~\bibnamefont{Lin}}
  (\bibinfo{publisher}{Curran Associates, Inc.}, \bibinfo{year}{2020}),
  vol.~\bibinfo{volume}{33}, pp. \bibinfo{pages}{1368--1379},
  \urlprefix\url{https://proceedings.neurips.cc/paper/2020/file/0ec96be397dd6d3cf2fecb4a2d627c1c-Paper.pdf}.

\bibitem[{\citenamefont{Henderson et~al.}(2019)\citenamefont{Henderson, Shakya,
  Pradhan, and Cook}}]{henderson_quanvolutional_2019}
\bibinfo{author}{\bibfnamefont{M.}~\bibnamefont{Henderson}},
  \bibinfo{author}{\bibfnamefont{S.}~\bibnamefont{Shakya}},
  \bibinfo{author}{\bibfnamefont{S.}~\bibnamefont{Pradhan}}, \bibnamefont{and}
  \bibinfo{author}{\bibfnamefont{T.}~\bibnamefont{Cook}},
  \bibinfo{journal}{arXiv preprint arXiv:1904.04767}  (\bibinfo{year}{2019}),
  \urlprefix\url{https://arxiv.org/abs/1904.04767}.

\bibitem[{\citenamefont{Kerenidis et~al.}(2020)\citenamefont{Kerenidis,
  Landman, and Prakash}}]{kerenidis_quantum_2020}
\bibinfo{author}{\bibfnamefont{I.}~\bibnamefont{Kerenidis}},
  \bibinfo{author}{\bibfnamefont{J.}~\bibnamefont{Landman}}, \bibnamefont{and}
  \bibinfo{author}{\bibfnamefont{A.}~\bibnamefont{Prakash}}, in
  \emph{\bibinfo{booktitle}{International Conference on Learning
  Representations}} (\bibinfo{year}{2020}),
  \urlprefix\url{https://openreview.net/forum?id=Hygab1rKDS}.

\bibitem[{\citenamefont{Lloyd and Weedbrook}(2018)}]{lloyd_quantum_2018}
\bibinfo{author}{\bibfnamefont{S.}~\bibnamefont{Lloyd}} \bibnamefont{and}
  \bibinfo{author}{\bibfnamefont{C.}~\bibnamefont{Weedbrook}},
  \bibinfo{journal}{Physical Review Letters} \textbf{\bibinfo{volume}{121}}
  (\bibinfo{year}{2018}), ISSN \bibinfo{issn}{0031-9007, 1079-7114}.

\bibitem[{\citenamefont{Dallaire-Demers and
  Killoran}(2018)}]{dallaire_quantum_2018}
\bibinfo{author}{\bibfnamefont{P.-L.} \bibnamefont{Dallaire-Demers}}
  \bibnamefont{and} \bibinfo{author}{\bibfnamefont{N.}~\bibnamefont{Killoran}},
  \bibinfo{journal}{Physical Review A} \textbf{\bibinfo{volume}{98}},
  \bibinfo{pages}{012324} (\bibinfo{year}{2018}).

\bibitem[{\citenamefont{Zoufal et~al.}(2019)\citenamefont{Zoufal, Lucchi, and
  Woerner}}]{zoufal_quantum_2019}
\bibinfo{author}{\bibfnamefont{C.}~\bibnamefont{Zoufal}},
  \bibinfo{author}{\bibfnamefont{A.}~\bibnamefont{Lucchi}}, \bibnamefont{and}
  \bibinfo{author}{\bibfnamefont{S.}~\bibnamefont{Woerner}},
  \bibinfo{journal}{npj Quantum Information} \textbf{\bibinfo{volume}{5}}
  (\bibinfo{year}{2019}),
  \urlprefix\url{https://doi.org/10.1038%2Fs41534-019-0223-2}.

\bibitem[{\citenamefont{Resch and Karpuzcu}(2019)}]{resch_quantum_2019}
\bibinfo{author}{\bibfnamefont{S.}~\bibnamefont{Resch}} \bibnamefont{and}
  \bibinfo{author}{\bibfnamefont{U.~R.} \bibnamefont{Karpuzcu}},
  \bibinfo{journal}{arXiv preprint arXiv:1905.07240}  (\bibinfo{year}{2019}).

\bibitem[{\citenamefont{Wootters and Zurek}(1982)}]{wootters_single_1982}
\bibinfo{author}{\bibfnamefont{W.~K.} \bibnamefont{Wootters}} \bibnamefont{and}
  \bibinfo{author}{\bibfnamefont{W.~H.} \bibnamefont{Zurek}},
  \bibinfo{journal}{Nature} \textbf{\bibinfo{volume}{299}},
  \bibinfo{pages}{802} (\bibinfo{year}{1982}).

\bibitem[{\citenamefont{Giovannetti
  et~al.}(2008{\natexlab{a}})\citenamefont{Giovannetti, Lloyd, and
  Maccone}}]{giovannetti_architectures_2008}
\bibinfo{author}{\bibfnamefont{V.}~\bibnamefont{Giovannetti}},
  \bibinfo{author}{\bibfnamefont{S.}~\bibnamefont{Lloyd}}, \bibnamefont{and}
  \bibinfo{author}{\bibfnamefont{L.}~\bibnamefont{Maccone}},
  \bibinfo{journal}{Physical Review A} \textbf{\bibinfo{volume}{78}}
  (\bibinfo{year}{2008}{\natexlab{a}}), ISSN \bibinfo{issn}{1094-1622}.

\bibitem[{\citenamefont{Giovannetti
  et~al.}(2008{\natexlab{b}})\citenamefont{Giovannetti, Lloyd, and
  Maccone}}]{giovannetti_quantum_2008}
\bibinfo{author}{\bibfnamefont{V.}~\bibnamefont{Giovannetti}},
  \bibinfo{author}{\bibfnamefont{S.}~\bibnamefont{Lloyd}}, \bibnamefont{and}
  \bibinfo{author}{\bibfnamefont{L.}~\bibnamefont{Maccone}},
  \bibinfo{journal}{Physical Review Letters} \textbf{\bibinfo{volume}{100}}
  (\bibinfo{year}{2008}{\natexlab{b}}), ISSN \bibinfo{issn}{1079-7114}.

\bibitem[{\citenamefont{Arunachalam et~al.}(2015)\citenamefont{Arunachalam,
  Gheorghiu, Jochym-O'Connor, Mosca, and
  Srinivasan}}]{arunachalam_robustness_2015}
\bibinfo{author}{\bibfnamefont{S.}~\bibnamefont{Arunachalam}},
  \bibinfo{author}{\bibfnamefont{V.}~\bibnamefont{Gheorghiu}},
  \bibinfo{author}{\bibfnamefont{T.}~\bibnamefont{Jochym-O'Connor}},
  \bibinfo{author}{\bibfnamefont{M.}~\bibnamefont{Mosca}}, \bibnamefont{and}
  \bibinfo{author}{\bibfnamefont{P.~V.} \bibnamefont{Srinivasan}},
  \bibinfo{journal}{New Journal of Physics} \textbf{\bibinfo{volume}{17}},
  \bibinfo{pages}{123010} (\bibinfo{year}{2015}),
  \urlprefix\url{https://doi.org/10.1088%2F1367-2630%2F17%2F12%2F123010}.

\bibitem[{\citenamefont{Farhi and Neven}(2018)}]{farhi_classification_2018}
\bibinfo{author}{\bibfnamefont{E.}~\bibnamefont{Farhi}} \bibnamefont{and}
  \bibinfo{author}{\bibfnamefont{H.}~\bibnamefont{Neven}},
  \bibinfo{journal}{arXiv preprint arXiv:1802.06002}  (\bibinfo{year}{2018}).

\bibitem[{\citenamefont{Rebentrost et~al.}(2014)\citenamefont{Rebentrost,
  Mohseni, and Lloyd}}]{rebentrost_quantum_2014}
\bibinfo{author}{\bibfnamefont{P.}~\bibnamefont{Rebentrost}},
  \bibinfo{author}{\bibfnamefont{M.}~\bibnamefont{Mohseni}}, \bibnamefont{and}
  \bibinfo{author}{\bibfnamefont{S.}~\bibnamefont{Lloyd}},
  \bibinfo{journal}{Phys. Rev. Lett.} \textbf{\bibinfo{volume}{113}},
  \bibinfo{pages}{130503} (\bibinfo{year}{2014}),
  \urlprefix\url{https://link.aps.org/doi/10.1103/PhysRevLett.113.130503}.

\bibitem[{\citenamefont{Havlíček et~al.}(2019)\citenamefont{Havlíček,
  Córcoles, Temme, Harrow, Kandala, Chow, and
  Gambetta}}]{havlicek_supervised_2019}
\bibinfo{author}{\bibfnamefont{V.}~\bibnamefont{Havlíček}},
  \bibinfo{author}{\bibfnamefont{A.~D.} \bibnamefont{Córcoles}},
  \bibinfo{author}{\bibfnamefont{K.}~\bibnamefont{Temme}},
  \bibinfo{author}{\bibfnamefont{A.~W.} \bibnamefont{Harrow}},
  \bibinfo{author}{\bibfnamefont{A.}~\bibnamefont{Kandala}},
  \bibinfo{author}{\bibfnamefont{J.~M.} \bibnamefont{Chow}}, \bibnamefont{and}
  \bibinfo{author}{\bibfnamefont{J.~M.} \bibnamefont{Gambetta}},
  \bibinfo{journal}{Nature} \textbf{\bibinfo{volume}{567}},
  \bibinfo{pages}{209} (\bibinfo{year}{2019}), ISSN \bibinfo{issn}{1476-4687}.

\bibitem[{\citenamefont{Wiebe et~al.}(2012)\citenamefont{Wiebe, Braun, and
  Lloyd}}]{wiebe_quantum_2012}
\bibinfo{author}{\bibfnamefont{N.}~\bibnamefont{Wiebe}},
  \bibinfo{author}{\bibfnamefont{D.}~\bibnamefont{Braun}}, \bibnamefont{and}
  \bibinfo{author}{\bibfnamefont{S.}~\bibnamefont{Lloyd}},
  \bibinfo{journal}{Phys. Rev. Lett.} \textbf{\bibinfo{volume}{109}},
  \bibinfo{pages}{050505} (\bibinfo{year}{2012}),
  \urlprefix\url{https://link.aps.org/doi/10.1103/PhysRevLett.109.050505}.

\bibitem[{\citenamefont{Lloyd et~al.}(2020)\citenamefont{Lloyd, Schuld, Ijaz,
  Izaac, and Killoran}}]{lloyd_quantum_2020}
\bibinfo{author}{\bibfnamefont{S.}~\bibnamefont{Lloyd}},
  \bibinfo{author}{\bibfnamefont{M.}~\bibnamefont{Schuld}},
  \bibinfo{author}{\bibfnamefont{A.}~\bibnamefont{Ijaz}},
  \bibinfo{author}{\bibfnamefont{J.}~\bibnamefont{Izaac}}, \bibnamefont{and}
  \bibinfo{author}{\bibfnamefont{N.}~\bibnamefont{Killoran}},
  \bibinfo{journal}{arXiv preprint arXiv:2001.03622}  (\bibinfo{year}{2020}).

\bibitem[{\citenamefont{Schuld and Killoran}(2019)}]{schuld_quantum_2019}
\bibinfo{author}{\bibfnamefont{M.}~\bibnamefont{Schuld}} \bibnamefont{and}
  \bibinfo{author}{\bibfnamefont{N.}~\bibnamefont{Killoran}},
  \bibinfo{journal}{Phys. Rev. Lett.} \textbf{\bibinfo{volume}{122}},
  \bibinfo{pages}{040504} (\bibinfo{year}{2019}),
  \urlprefix\url{https://link.aps.org/doi/10.1103/PhysRevLett.122.040504}.

\bibitem[{\citenamefont{Skolik et~al.}(2021)\citenamefont{Skolik, McClean,
  Mohseni, van~der Smagt, and Leib}}]{skolik_layerwise_2021}
\bibinfo{author}{\bibfnamefont{A.}~\bibnamefont{Skolik}},
  \bibinfo{author}{\bibfnamefont{J.~R.} \bibnamefont{McClean}},
  \bibinfo{author}{\bibfnamefont{M.}~\bibnamefont{Mohseni}},
  \bibinfo{author}{\bibfnamefont{P.}~\bibnamefont{van~der Smagt}},
  \bibnamefont{and} \bibinfo{author}{\bibfnamefont{M.}~\bibnamefont{Leib}},
  \bibinfo{journal}{Quantum Machine Intelligence} \textbf{\bibinfo{volume}{3}},
  \bibinfo{pages}{1} (\bibinfo{year}{2021}).

\bibitem[{\citenamefont{Weigold et~al.}(2020)\citenamefont{Weigold, Barzen,
  Leymann, and Salm}}]{weigold_data_2020}
\bibinfo{author}{\bibfnamefont{M.}~\bibnamefont{Weigold}},
  \bibinfo{author}{\bibfnamefont{J.}~\bibnamefont{Barzen}},
  \bibinfo{author}{\bibfnamefont{F.}~\bibnamefont{Leymann}}, \bibnamefont{and}
  \bibinfo{author}{\bibfnamefont{M.}~\bibnamefont{Salm}}, in
  \emph{\bibinfo{booktitle}{Proceedings of the 27th Conference on Pattern
  Languages of Programs}} (\bibinfo{publisher}{The Hillside Group},
  \bibinfo{year}{2020}), PLoP '20, ISBN \bibinfo{isbn}{9781941652169}.

\bibitem[{\citenamefont{Weigold et~al.}(2021)\citenamefont{Weigold, Barzen,
  Leymann, and Salm}}]{weigold_expanding_2021}
\bibinfo{author}{\bibfnamefont{M.}~\bibnamefont{Weigold}},
  \bibinfo{author}{\bibfnamefont{J.}~\bibnamefont{Barzen}},
  \bibinfo{author}{\bibfnamefont{F.}~\bibnamefont{Leymann}}, \bibnamefont{and}
  \bibinfo{author}{\bibfnamefont{M.}~\bibnamefont{Salm}}, in
  \emph{\bibinfo{booktitle}{2021 IEEE 18th International Conference on Software
  Architecture Companion (ICSA-C)}} (\bibinfo{year}{2021}), pp.
  \bibinfo{pages}{95--101}.

\bibitem[{\citenamefont{LaRose and Coyle}(2020)}]{larose_robust_2020}
\bibinfo{author}{\bibfnamefont{R.}~\bibnamefont{LaRose}} \bibnamefont{and}
  \bibinfo{author}{\bibfnamefont{B.}~\bibnamefont{Coyle}},
  \bibinfo{journal}{Physical Review A} \textbf{\bibinfo{volume}{102}}
  (\bibinfo{year}{2020}).

\bibitem[{\citenamefont{Wiebe}(2020)}]{wiebe_key_2020}
\bibinfo{author}{\bibfnamefont{N.}~\bibnamefont{Wiebe}}, \bibinfo{journal}{New
  Journal of Physics} \textbf{\bibinfo{volume}{22}}, \bibinfo{pages}{091001}
  (\bibinfo{year}{2020}),
  \urlprefix\url{https://doi.org/10.1088/1367-2630/abac39}.

\bibitem[{\citenamefont{Schuld et~al.}(2017)\citenamefont{Schuld, Fingerhuth,
  and Petruccione}}]{schuld_implementing_2017}
\bibinfo{author}{\bibfnamefont{M.}~\bibnamefont{Schuld}},
  \bibinfo{author}{\bibfnamefont{M.}~\bibnamefont{Fingerhuth}},
  \bibnamefont{and}
  \bibinfo{author}{\bibfnamefont{F.}~\bibnamefont{Petruccione}},
  \bibinfo{journal}{{EPL} (Europhysics Letters)}
  \textbf{\bibinfo{volume}{119}}, \bibinfo{pages}{60002}
  (\bibinfo{year}{2017}),
  \urlprefix\url{https://doi.org/10.1209%2F0295-5075%2F119%2F60002}.

\bibitem[{\citenamefont{Romero et~al.}(2017)\citenamefont{Romero, Olson, and
  Aspuru-Guzik}}]{romero_quantum_2017}
\bibinfo{author}{\bibfnamefont{J.}~\bibnamefont{Romero}},
  \bibinfo{author}{\bibfnamefont{J.~P.} \bibnamefont{Olson}}, \bibnamefont{and}
  \bibinfo{author}{\bibfnamefont{A.}~\bibnamefont{Aspuru-Guzik}},
  \bibinfo{journal}{Quantum Science and Technology}
  \textbf{\bibinfo{volume}{2}}, \bibinfo{pages}{045001} (\bibinfo{year}{2017}).

\bibitem[{\citenamefont{Amin et~al.}(2018)\citenamefont{Amin, Andriyash, Rolfe,
  Kulchytskyy, and Melko}}]{amin_quantum_2018}
\bibinfo{author}{\bibfnamefont{M.~H.} \bibnamefont{Amin}},
  \bibinfo{author}{\bibfnamefont{E.}~\bibnamefont{Andriyash}},
  \bibinfo{author}{\bibfnamefont{J.}~\bibnamefont{Rolfe}},
  \bibinfo{author}{\bibfnamefont{B.}~\bibnamefont{Kulchytskyy}},
  \bibnamefont{and} \bibinfo{author}{\bibfnamefont{R.}~\bibnamefont{Melko}},
  \bibinfo{journal}{Phys. Rev. X} \textbf{\bibinfo{volume}{8}},
  \bibinfo{pages}{021050} (\bibinfo{year}{2018}),
  \urlprefix\url{https://link.aps.org/doi/10.1103/PhysRevX.8.021050}.

\bibitem[{\citenamefont{Haug et~al.}(2021)\citenamefont{Haug, Self, and
  Kim}}]{haug_large_2021}
\bibinfo{author}{\bibfnamefont{T.}~\bibnamefont{Haug}},
  \bibinfo{author}{\bibfnamefont{C.~N.} \bibnamefont{Self}}, \bibnamefont{and}
  \bibinfo{author}{\bibfnamefont{M.~S.} \bibnamefont{Kim}},
  \bibinfo{journal}{arXiv preprint arXiv:2108.01039}  (\bibinfo{year}{2021}),
  \urlprefix\url{https://arxiv.org/abs/2108.01039}.

\bibitem[{\citenamefont{Wecker et~al.}(2015)\citenamefont{Wecker, Hastings, and
  Troyer}}]{wecker_progress_2015}
\bibinfo{author}{\bibfnamefont{D.}~\bibnamefont{Wecker}},
  \bibinfo{author}{\bibfnamefont{M.~B.} \bibnamefont{Hastings}},
  \bibnamefont{and} \bibinfo{author}{\bibfnamefont{M.}~\bibnamefont{Troyer}},
  \bibinfo{journal}{Phys. Rev. A} \textbf{\bibinfo{volume}{92}},
  \bibinfo{pages}{042303} (\bibinfo{year}{2015}),
  \urlprefix\url{https://link.aps.org/doi/10.1103/PhysRevA.92.042303}.

\bibitem[{\citenamefont{Cade et~al.}(2020)\citenamefont{Cade, Mineh, Montanaro,
  and Stanisic}}]{cade_strategies_2020}
\bibinfo{author}{\bibfnamefont{C.}~\bibnamefont{Cade}},
  \bibinfo{author}{\bibfnamefont{L.}~\bibnamefont{Mineh}},
  \bibinfo{author}{\bibfnamefont{A.}~\bibnamefont{Montanaro}},
  \bibnamefont{and} \bibinfo{author}{\bibfnamefont{S.}~\bibnamefont{Stanisic}},
  \bibinfo{journal}{Physical Review B} \textbf{\bibinfo{volume}{102}}
  (\bibinfo{year}{2020}),
  \urlprefix\url{https://doi.org/10.1103%2Fphysrevb.102.235122}.

\bibitem[{\citenamefont{Wiersema et~al.}(2020)\citenamefont{Wiersema, Zhou,
  de~Sereville, Carrasquilla, Kim, and Yuen}}]{wiersema_exploring_2020}
\bibinfo{author}{\bibfnamefont{R.}~\bibnamefont{Wiersema}},
  \bibinfo{author}{\bibfnamefont{C.}~\bibnamefont{Zhou}},
  \bibinfo{author}{\bibfnamefont{Y.}~\bibnamefont{de~Sereville}},
  \bibinfo{author}{\bibfnamefont{J.~F.} \bibnamefont{Carrasquilla}},
  \bibinfo{author}{\bibfnamefont{Y.~B.} \bibnamefont{Kim}}, \bibnamefont{and}
  \bibinfo{author}{\bibfnamefont{H.}~\bibnamefont{Yuen}},
  \bibinfo{journal}{{PRX} Quantum} \textbf{\bibinfo{volume}{1}}
  (\bibinfo{year}{2020}),
  \urlprefix\url{https://doi.org/10.1103%2Fprxquantum.1.020319}.

\bibitem[{\citenamefont{Du et~al.}(2021)\citenamefont{Du, Qian, and
  Tao}}]{du_accelerating_2021}
\bibinfo{author}{\bibfnamefont{Y.}~\bibnamefont{Du}},
  \bibinfo{author}{\bibfnamefont{Y.}~\bibnamefont{Qian}}, \bibnamefont{and}
  \bibinfo{author}{\bibfnamefont{D.}~\bibnamefont{Tao}},
  \bibinfo{journal}{arXiv preprint arXiv:2106.12819}  (\bibinfo{year}{2021}),
  \urlprefix\url{https://arxiv.org/abs/2106.12819}.

\bibitem[{\citenamefont{Liao et~al.}(2021)\citenamefont{Liao, Ebler, Liu, and
  Dahlsten}}]{liao_quantum_2021}
\bibinfo{author}{\bibfnamefont{Y.}~\bibnamefont{Liao}},
  \bibinfo{author}{\bibfnamefont{D.}~\bibnamefont{Ebler}},
  \bibinfo{author}{\bibfnamefont{F.}~\bibnamefont{Liu}}, \bibnamefont{and}
  \bibinfo{author}{\bibfnamefont{O.}~\bibnamefont{Dahlsten}},
  \bibinfo{journal}{New Journal of Physics} \textbf{\bibinfo{volume}{23}},
  \bibinfo{pages}{063013} (\bibinfo{year}{2021}),
  \urlprefix\url{https://doi.org/10.1088/1367-2630/abc9ef}.

\bibitem[{\citenamefont{Pascanu et~al.}(2013)\citenamefont{Pascanu, Mikolov,
  and Bengio}}]{pascanu_difficulty_2013}
\bibinfo{author}{\bibfnamefont{R.}~\bibnamefont{Pascanu}},
  \bibinfo{author}{\bibfnamefont{T.}~\bibnamefont{Mikolov}}, \bibnamefont{and}
  \bibinfo{author}{\bibfnamefont{Y.}~\bibnamefont{Bengio}}, in
  \emph{\bibinfo{booktitle}{International conference on machine learning}}
  (\bibinfo{organization}{PMLR}, \bibinfo{year}{2013}), pp.
  \bibinfo{pages}{1310--1318}.

\bibitem[{\citenamefont{Shang and Wah}(1996)}]{shang_global_1996}
\bibinfo{author}{\bibfnamefont{Y.}~\bibnamefont{Shang}} \bibnamefont{and}
  \bibinfo{author}{\bibfnamefont{B.~W.} \bibnamefont{Wah}},
  \bibinfo{journal}{Computer} \textbf{\bibinfo{volume}{29}},
  \bibinfo{pages}{45} (\bibinfo{year}{1996}).

\bibitem[{\citenamefont{Peng et~al.}(2020)\citenamefont{Peng, Harrow, Ozols,
  and Wu}}]{peng_simulating_2020}
\bibinfo{author}{\bibfnamefont{T.}~\bibnamefont{Peng}},
  \bibinfo{author}{\bibfnamefont{A.~W.} \bibnamefont{Harrow}},
  \bibinfo{author}{\bibfnamefont{M.}~\bibnamefont{Ozols}}, \bibnamefont{and}
  \bibinfo{author}{\bibfnamefont{X.}~\bibnamefont{Wu}},
  \bibinfo{journal}{Physical Review Letters} \textbf{\bibinfo{volume}{125}},
  \bibinfo{pages}{150504} (\bibinfo{year}{2020}).

\bibitem[{\citenamefont{Marshall et~al.}(2022)\citenamefont{Marshall, Gyurik,
  and Dunjko}}]{marshall_high_2022}
\bibinfo{author}{\bibfnamefont{S.~C.} \bibnamefont{Marshall}},
  \bibinfo{author}{\bibfnamefont{C.}~\bibnamefont{Gyurik}}, \bibnamefont{and}
  \bibinfo{author}{\bibfnamefont{V.}~\bibnamefont{Dunjko}},
  \bibinfo{journal}{arXiv preprint arXiv:2203.13739}  (\bibinfo{year}{2022}).

\bibitem[{\citenamefont{Mitarai and
  Fujii}(2021{\natexlab{a}})}]{mitarai_constructing_2021}
\bibinfo{author}{\bibfnamefont{K.}~\bibnamefont{Mitarai}} \bibnamefont{and}
  \bibinfo{author}{\bibfnamefont{K.}~\bibnamefont{Fujii}},
  \bibinfo{journal}{New Journal of Physics} \textbf{\bibinfo{volume}{23}}
  (\bibinfo{year}{2021}{\natexlab{a}}).

\bibitem[{\citenamefont{Chen et~al.}(2018)\citenamefont{Chen, Zhou, Xue, Yang,
  Guo, and Guo}}]{chen_64qubit_2018}
\bibinfo{author}{\bibfnamefont{Z.-Y.} \bibnamefont{Chen}},
  \bibinfo{author}{\bibfnamefont{Q.}~\bibnamefont{Zhou}},
  \bibinfo{author}{\bibfnamefont{C.}~\bibnamefont{Xue}},
  \bibinfo{author}{\bibfnamefont{X.}~\bibnamefont{Yang}},
  \bibinfo{author}{\bibfnamefont{G.-C.} \bibnamefont{Guo}}, \bibnamefont{and}
  \bibinfo{author}{\bibfnamefont{G.-P.} \bibnamefont{Guo}},
  \bibinfo{journal}{Science Bulletin} \textbf{\bibinfo{volume}{63}},
  \bibinfo{pages}{964} (\bibinfo{year}{2018}), ISSN \bibinfo{issn}{2095-9273},
  \urlprefix\url{https://www.sciencedirect.com/science/article/pii/S2095927318302809}.

\bibitem[{\citenamefont{Eddins et~al.}(2022)\citenamefont{Eddins, Motta,
  Gujarati, Bravyi, Mezzacapo, Hadfield, and Sheldon}}]{eddins_doubling_2022}
\bibinfo{author}{\bibfnamefont{A.}~\bibnamefont{Eddins}},
  \bibinfo{author}{\bibfnamefont{M.}~\bibnamefont{Motta}},
  \bibinfo{author}{\bibfnamefont{T.~P.} \bibnamefont{Gujarati}},
  \bibinfo{author}{\bibfnamefont{S.}~\bibnamefont{Bravyi}},
  \bibinfo{author}{\bibfnamefont{A.}~\bibnamefont{Mezzacapo}},
  \bibinfo{author}{\bibfnamefont{C.}~\bibnamefont{Hadfield}}, \bibnamefont{and}
  \bibinfo{author}{\bibfnamefont{S.}~\bibnamefont{Sheldon}},
  \bibinfo{journal}{{PRX} Quantum} \textbf{\bibinfo{volume}{3}}
  (\bibinfo{year}{2022}),
  \urlprefix\url{https://doi.org/10.1103%2Fprxquantum.3.010309}.

\bibitem[{\citenamefont{Perlin et~al.}(2020)\citenamefont{Perlin, Saleem,
  Suchara, and Osborn}}]{perlin_quantum_2020}
\bibinfo{author}{\bibfnamefont{M.~A.} \bibnamefont{Perlin}},
  \bibinfo{author}{\bibfnamefont{Z.~H.} \bibnamefont{Saleem}},
  \bibinfo{author}{\bibfnamefont{M.}~\bibnamefont{Suchara}}, \bibnamefont{and}
  \bibinfo{author}{\bibfnamefont{J.~C.} \bibnamefont{Osborn}},
  \bibinfo{journal}{arXiv preprint arXiv:2005.12702}  (\bibinfo{year}{2020}),
  \urlprefix\url{https://arxiv.org/abs/2005.12702}.

\bibitem[{\citenamefont{Tang et~al.}(2021)\citenamefont{Tang, Tomesh, Suchara,
  Larson, and Martonosi}}]{tang_cutqc_2020}
\bibinfo{author}{\bibfnamefont{W.}~\bibnamefont{Tang}},
  \bibinfo{author}{\bibfnamefont{T.}~\bibnamefont{Tomesh}},
  \bibinfo{author}{\bibfnamefont{M.}~\bibnamefont{Suchara}},
  \bibinfo{author}{\bibfnamefont{J.}~\bibnamefont{Larson}}, \bibnamefont{and}
  \bibinfo{author}{\bibfnamefont{M.}~\bibnamefont{Martonosi}}, in
  \emph{\bibinfo{booktitle}{Proceedings of the 26th ACM International
  Conference on Architectural Support for Programming Languages and Operating
  Systems}} (\bibinfo{publisher}{Association for Computing Machinery},
  \bibinfo{address}{New York, NY, USA}, \bibinfo{year}{2021}), ASPLOS '21, p.
  \bibinfo{pages}{473–486}, ISBN \bibinfo{isbn}{9781450383172},
  \urlprefix\url{https://doi.org/10.1145/3445814.3446758}.

\bibitem[{\citenamefont{Saleem et~al.}(2021)\citenamefont{Saleem, Tomesh,
  Perlin, Gokhale, and Suchara}}]{saleem_divide_2021}
\bibinfo{author}{\bibfnamefont{Z.~H.} \bibnamefont{Saleem}},
  \bibinfo{author}{\bibfnamefont{T.}~\bibnamefont{Tomesh}},
  \bibinfo{author}{\bibfnamefont{M.~A.} \bibnamefont{Perlin}},
  \bibinfo{author}{\bibfnamefont{P.}~\bibnamefont{Gokhale}}, \bibnamefont{and}
  \bibinfo{author}{\bibfnamefont{M.}~\bibnamefont{Suchara}},
  \bibinfo{journal}{arXiv preprint arXiv:2107.07532}  (\bibinfo{year}{2021}),
  \urlprefix\url{https://arxiv.org/abs/2107.07532}.

\bibitem[{\citenamefont{Lowe et~al.}(2022)\citenamefont{Lowe, Medvidović,
  Hayes, O'Riordan, Bromley, Arrazola, and Killoran}}]{lowe_fast_2022}
\bibinfo{author}{\bibfnamefont{A.}~\bibnamefont{Lowe}},
  \bibinfo{author}{\bibfnamefont{M.}~\bibnamefont{Medvidović}},
  \bibinfo{author}{\bibfnamefont{A.}~\bibnamefont{Hayes}},
  \bibinfo{author}{\bibfnamefont{L.~J.} \bibnamefont{O'Riordan}},
  \bibinfo{author}{\bibfnamefont{T.~R.} \bibnamefont{Bromley}},
  \bibinfo{author}{\bibfnamefont{J.~M.} \bibnamefont{Arrazola}},
  \bibnamefont{and} \bibinfo{author}{\bibfnamefont{N.}~\bibnamefont{Killoran}},
  \bibinfo{journal}{arXiv preprint arXiv:2207.14734}  (\bibinfo{year}{2022}),
  \urlprefix\url{https://arxiv.org/abs/2207.14734}.

\bibitem[{\citenamefont{Piveteau and Sutter}(2022)}]{piveteau_circuit_2022}
\bibinfo{author}{\bibfnamefont{C.}~\bibnamefont{Piveteau}} \bibnamefont{and}
  \bibinfo{author}{\bibfnamefont{D.}~\bibnamefont{Sutter}},
  \bibinfo{journal}{arXiv preprint arXiv:2205.00016}  (\bibinfo{year}{2022}).

\bibitem[{\citenamefont{T{\"u}ys{\"u}z
  et~al.}(2022)\citenamefont{T{\"u}ys{\"u}z, Clemente, Crippa, Hartung,
  K{\"u}hn, and Jansen}}]{tuysuz_classical_2022}
\bibinfo{author}{\bibfnamefont{C.}~\bibnamefont{T{\"u}ys{\"u}z}},
  \bibinfo{author}{\bibfnamefont{G.}~\bibnamefont{Clemente}},
  \bibinfo{author}{\bibfnamefont{A.}~\bibnamefont{Crippa}},
  \bibinfo{author}{\bibfnamefont{T.}~\bibnamefont{Hartung}},
  \bibinfo{author}{\bibfnamefont{S.}~\bibnamefont{K{\"u}hn}}, \bibnamefont{and}
  \bibinfo{author}{\bibfnamefont{K.}~\bibnamefont{Jansen}},
  \bibinfo{journal}{arXiv preprint arXiv:2206.09641}  (\bibinfo{year}{2022}),
  \urlprefix\url{https://arxiv.org/abs/2206.09641}.

\bibitem[{\citenamefont{Mitarai and
  Fujii}(2021{\natexlab{b}})}]{mitarai_overhead_2021}
\bibinfo{author}{\bibfnamefont{K.}~\bibnamefont{Mitarai}} \bibnamefont{and}
  \bibinfo{author}{\bibfnamefont{K.}~\bibnamefont{Fujii}},
  \bibinfo{journal}{Quantum} \textbf{\bibinfo{volume}{5}}, \bibinfo{pages}{388}
  (\bibinfo{year}{2021}{\natexlab{b}}),
  \urlprefix\url{https://doi.org/10.22331%2Fq-2021-01-28-388}.

\bibitem[{\citenamefont{Bravyi et~al.}(2016)\citenamefont{Bravyi, Smith, and
  Smolin}}]{bravyi_trading_2016}
\bibinfo{author}{\bibfnamefont{S.}~\bibnamefont{Bravyi}},
  \bibinfo{author}{\bibfnamefont{G.}~\bibnamefont{Smith}}, \bibnamefont{and}
  \bibinfo{author}{\bibfnamefont{J.~A.} \bibnamefont{Smolin}},
  \bibinfo{journal}{Physical Review X} \textbf{\bibinfo{volume}{6}}
  (\bibinfo{year}{2016}).

\bibitem[{\citenamefont{Bachem et~al.}(2017)\citenamefont{Bachem, Lucic, and
  Krause}}]{bachem_practical_2017}
\bibinfo{author}{\bibfnamefont{O.}~\bibnamefont{Bachem}},
  \bibinfo{author}{\bibfnamefont{M.}~\bibnamefont{Lucic}}, \bibnamefont{and}
  \bibinfo{author}{\bibfnamefont{A.}~\bibnamefont{Krause}},
  \bibinfo{journal}{arXiv preprint arXiv:1703.06476}  (\bibinfo{year}{2017}).

\bibitem[{\citenamefont{Harrow}(2020)}]{harrow_small_2020}
\bibinfo{author}{\bibfnamefont{A.~W.} \bibnamefont{Harrow}},
  \bibinfo{journal}{arXiv preprint arXiv:2004.00026}  (\bibinfo{year}{2020}).

\bibitem[{\citenamefont{Tomesh et~al.}(2021)\citenamefont{Tomesh, Gokhale,
  Anschuetz, and Chong}}]{tomesh_coreset_2021}
\bibinfo{author}{\bibfnamefont{T.}~\bibnamefont{Tomesh}},
  \bibinfo{author}{\bibfnamefont{P.}~\bibnamefont{Gokhale}},
  \bibinfo{author}{\bibfnamefont{E.~R.} \bibnamefont{Anschuetz}},
  \bibnamefont{and} \bibinfo{author}{\bibfnamefont{F.~T.} \bibnamefont{Chong}},
  \bibinfo{journal}{Electronics} \textbf{\bibinfo{volume}{10}}
  (\bibinfo{year}{2021}), ISSN \bibinfo{issn}{2079-9292},
  \urlprefix\url{https://www.mdpi.com/2079-9292/10/14/1690}.

\bibitem[{\citenamefont{Self et~al.}(2021)\citenamefont{Self, Khosla, Smith,
  Sauvage, Haynes, Knolle, Mintert, and Kim}}]{self_variational_2021}
\bibinfo{author}{\bibfnamefont{C.~N.} \bibnamefont{Self}},
  \bibinfo{author}{\bibfnamefont{K.~E.} \bibnamefont{Khosla}},
  \bibinfo{author}{\bibfnamefont{A.~W.~R.} \bibnamefont{Smith}},
  \bibinfo{author}{\bibfnamefont{F.}~\bibnamefont{Sauvage}},
  \bibinfo{author}{\bibfnamefont{P.~D.} \bibnamefont{Haynes}},
  \bibinfo{author}{\bibfnamefont{J.}~\bibnamefont{Knolle}},
  \bibinfo{author}{\bibfnamefont{F.}~\bibnamefont{Mintert}}, \bibnamefont{and}
  \bibinfo{author}{\bibfnamefont{M.~S.} \bibnamefont{Kim}},
  \bibinfo{journal}{npj Quantum Information} \textbf{\bibinfo{volume}{7}}
  (\bibinfo{year}{2021}),
  \urlprefix\url{https://doi.org/10.1038%2Fs41534-021-00452-9}.

\bibitem[{\citenamefont{Chen et~al.}(2022)\citenamefont{Chen, Cotler, Huang,
  and Li}}]{chen_complexity_2022}
\bibinfo{author}{\bibfnamefont{S.}~\bibnamefont{Chen}},
  \bibinfo{author}{\bibfnamefont{J.}~\bibnamefont{Cotler}},
  \bibinfo{author}{\bibfnamefont{H.-Y.} \bibnamefont{Huang}}, \bibnamefont{and}
  \bibinfo{author}{\bibfnamefont{J.}~\bibnamefont{Li}}, \bibinfo{journal}{arXiv
  preprint arXiv:2210.07234}  (\bibinfo{year}{2022}),
  \urlprefix\url{https://arxiv.org/abs/2210.07234}.

\bibitem[{\citenamefont{Bravyi et~al.}(2018)\citenamefont{Bravyi, Gosset, and
  König}}]{bravyi_quantum_2018}
\bibinfo{author}{\bibfnamefont{S.}~\bibnamefont{Bravyi}},
  \bibinfo{author}{\bibfnamefont{D.}~\bibnamefont{Gosset}}, \bibnamefont{and}
  \bibinfo{author}{\bibfnamefont{R.}~\bibnamefont{König}},
  \bibinfo{journal}{Science} \textbf{\bibinfo{volume}{362}},
  \bibinfo{pages}{308} (\bibinfo{year}{2018}),
  \urlprefix\url{https://doi.org/10.1126%2Fscience.aar3106}.

\bibitem[{\citenamefont{Arute et~al.}(2019)\citenamefont{Arute, Arya, Babbush,
  Bacon, Bardin, Barends, Biswas, Boixo, Brandao, Buell
  et~al.}}]{arute_quantum_2019}
\bibinfo{author}{\bibfnamefont{F.}~\bibnamefont{Arute}},
  \bibinfo{author}{\bibfnamefont{K.}~\bibnamefont{Arya}},
  \bibinfo{author}{\bibfnamefont{R.}~\bibnamefont{Babbush}},
  \bibinfo{author}{\bibfnamefont{D.}~\bibnamefont{Bacon}},
  \bibinfo{author}{\bibfnamefont{J.~C.} \bibnamefont{Bardin}},
  \bibinfo{author}{\bibfnamefont{R.}~\bibnamefont{Barends}},
  \bibinfo{author}{\bibfnamefont{R.}~\bibnamefont{Biswas}},
  \bibinfo{author}{\bibfnamefont{S.}~\bibnamefont{Boixo}},
  \bibinfo{author}{\bibfnamefont{F.~G.} \bibnamefont{Brandao}},
  \bibinfo{author}{\bibfnamefont{D.~A.} \bibnamefont{Buell}},
  \bibnamefont{et~al.}, \bibinfo{journal}{Nature}
  \textbf{\bibinfo{volume}{574}}, \bibinfo{pages}{505} (\bibinfo{year}{2019}).

\bibitem[{\citenamefont{Abadi et~al.}(2016)\citenamefont{Abadi, Barham, Chen,
  Chen, Davis, Dean, Devin, Ghemawat, Irving, Isard
  et~al.}}]{abadi_tensorflow_2015}
\bibinfo{author}{\bibfnamefont{M.}~\bibnamefont{Abadi}},
  \bibinfo{author}{\bibfnamefont{P.}~\bibnamefont{Barham}},
  \bibinfo{author}{\bibfnamefont{J.}~\bibnamefont{Chen}},
  \bibinfo{author}{\bibfnamefont{Z.}~\bibnamefont{Chen}},
  \bibinfo{author}{\bibfnamefont{A.}~\bibnamefont{Davis}},
  \bibinfo{author}{\bibfnamefont{J.}~\bibnamefont{Dean}},
  \bibinfo{author}{\bibfnamefont{M.}~\bibnamefont{Devin}},
  \bibinfo{author}{\bibfnamefont{S.}~\bibnamefont{Ghemawat}},
  \bibinfo{author}{\bibfnamefont{G.}~\bibnamefont{Irving}},
  \bibinfo{author}{\bibfnamefont{M.}~\bibnamefont{Isard}},
  \bibnamefont{et~al.}, in \emph{\bibinfo{booktitle}{12th {USENIX} Symposium on
  Operating Systems Design and Implementation ({OSDI} 16)}}
  (\bibinfo{publisher}{{USENIX} Association}, \bibinfo{address}{Savannah, GA},
  \bibinfo{year}{2016}), pp. \bibinfo{pages}{265--283}.

\bibitem[{dis()}]{dist_tensorflow}
\emph{\bibinfo{title}{{Distributed training with TensorFlow}}},
  \bibinfo{note}{last Accessed 03.2022},
  \urlprefix\url{https://www.tensorflow.org/guide/distributed_training}.

\bibitem[{ncc()}]{nccl}
\emph{\bibinfo{title}{{NVIDIA Collective Communication Library} ({NCCL})}},
  \bibinfo{note}{https://developer.nvidia.com/nccl. Last Accessed 03.2022},
  \urlprefix\url{https://developer.nvidia.com/nccl}.

\bibitem[{\citenamefont{Chen et~al.}(2015)\citenamefont{Chen, Li, Li, Lin,
  Wang, Wang, Xiao, Xu, Zhang, and Zhang}}]{chen_mxnet_2015}
\bibinfo{author}{\bibfnamefont{T.}~\bibnamefont{Chen}},
  \bibinfo{author}{\bibfnamefont{M.}~\bibnamefont{Li}},
  \bibinfo{author}{\bibfnamefont{Y.}~\bibnamefont{Li}},
  \bibinfo{author}{\bibfnamefont{M.}~\bibnamefont{Lin}},
  \bibinfo{author}{\bibfnamefont{N.}~\bibnamefont{Wang}},
  \bibinfo{author}{\bibfnamefont{M.}~\bibnamefont{Wang}},
  \bibinfo{author}{\bibfnamefont{T.}~\bibnamefont{Xiao}},
  \bibinfo{author}{\bibfnamefont{B.}~\bibnamefont{Xu}},
  \bibinfo{author}{\bibfnamefont{C.}~\bibnamefont{Zhang}}, \bibnamefont{and}
  \bibinfo{author}{\bibfnamefont{Z.}~\bibnamefont{Zhang}},
  \bibinfo{journal}{arXiv preprint arXiv:1512.01274}  (\bibinfo{year}{2015}).

\bibitem[{\citenamefont{Paszke et~al.}(2017)\citenamefont{Paszke, Gross,
  Chintala, Chanan, Yang, DeVito, Lin, Desmaison, Antiga, and
  Lerer}}]{paszke_automatic_2017}
\bibinfo{author}{\bibfnamefont{A.}~\bibnamefont{Paszke}},
  \bibinfo{author}{\bibfnamefont{S.}~\bibnamefont{Gross}},
  \bibinfo{author}{\bibfnamefont{S.}~\bibnamefont{Chintala}},
  \bibinfo{author}{\bibfnamefont{G.}~\bibnamefont{Chanan}},
  \bibinfo{author}{\bibfnamefont{E.}~\bibnamefont{Yang}},
  \bibinfo{author}{\bibfnamefont{Z.}~\bibnamefont{DeVito}},
  \bibinfo{author}{\bibfnamefont{Z.}~\bibnamefont{Lin}},
  \bibinfo{author}{\bibfnamefont{A.}~\bibnamefont{Desmaison}},
  \bibinfo{author}{\bibfnamefont{L.}~\bibnamefont{Antiga}}, \bibnamefont{and}
  \bibinfo{author}{\bibfnamefont{A.}~\bibnamefont{Lerer}}, in
  \emph{\bibinfo{booktitle}{NIPS-W}} (\bibinfo{year}{2017}).

\bibitem[{\citenamefont{Seide and Agarwal}(2016)}]{seide_cntk_2016}
\bibinfo{author}{\bibfnamefont{F.}~\bibnamefont{Seide}} \bibnamefont{and}
  \bibinfo{author}{\bibfnamefont{A.}~\bibnamefont{Agarwal}}, in
  \emph{\bibinfo{booktitle}{Proceedings of the 22nd ACM SIGKDD international
  conference on knowledge discovery and data mining}} (\bibinfo{year}{2016}).

\bibitem[{\citenamefont{Broughton et~al.}(2020)\citenamefont{Broughton, Verdon,
  McCourt, Martinez, Yoo, Isakov, Massey, Halavati, Niu, Zlokapa
  et~al.}}]{broughton_tensorflow_2020}
\bibinfo{author}{\bibfnamefont{M.}~\bibnamefont{Broughton}},
  \bibinfo{author}{\bibfnamefont{G.}~\bibnamefont{Verdon}},
  \bibinfo{author}{\bibfnamefont{T.}~\bibnamefont{McCourt}},
  \bibinfo{author}{\bibfnamefont{A.~J.} \bibnamefont{Martinez}},
  \bibinfo{author}{\bibfnamefont{J.~H.} \bibnamefont{Yoo}},
  \bibinfo{author}{\bibfnamefont{S.~V.} \bibnamefont{Isakov}},
  \bibinfo{author}{\bibfnamefont{P.}~\bibnamefont{Massey}},
  \bibinfo{author}{\bibfnamefont{R.}~\bibnamefont{Halavati}},
  \bibinfo{author}{\bibfnamefont{M.~Y.} \bibnamefont{Niu}},
  \bibinfo{author}{\bibfnamefont{A.}~\bibnamefont{Zlokapa}},
  \bibnamefont{et~al.}, \bibinfo{journal}{arXiv preprint arXiv:2003.02989}
  (\bibinfo{year}{2020}).

\bibitem[{\citenamefont{Xing and Broughton}(2021)}]{xing_training_2021}
\bibinfo{author}{\bibfnamefont{C.}~\bibnamefont{Xing}} \bibnamefont{and}
  \bibinfo{author}{\bibfnamefont{M.}~\bibnamefont{Broughton}},
  \emph{\bibinfo{title}{Training with multiple workers using tensorflow
  quantum}} (\bibinfo{year}{2021}), \bibinfo{note}{last Accessed: 03 2022},
  \urlprefix\url{https://blog.tensorflow.org/2021/06/training-with-multiple-workers-using-tensorflow-quantum.html}.

\bibitem[{\citenamefont{Aleksandrowicz
  et~al.}(2019)\citenamefont{Aleksandrowicz, Alexander, Barkoutsos, Bello,
  Ben-Haim, Bucher, Cabrera-Hernández, Carballo-Franquis, Chen, Chen
  et~al.}}]{aleksandrowicz_qiskit_2019}
\bibinfo{author}{\bibfnamefont{G.}~\bibnamefont{Aleksandrowicz}},
  \bibinfo{author}{\bibfnamefont{T.}~\bibnamefont{Alexander}},
  \bibinfo{author}{\bibfnamefont{P.}~\bibnamefont{Barkoutsos}},
  \bibinfo{author}{\bibfnamefont{L.}~\bibnamefont{Bello}},
  \bibinfo{author}{\bibfnamefont{Y.}~\bibnamefont{Ben-Haim}},
  \bibinfo{author}{\bibfnamefont{D.}~\bibnamefont{Bucher}},
  \bibinfo{author}{\bibfnamefont{F.~J.} \bibnamefont{Cabrera-Hernández}},
  \bibinfo{author}{\bibfnamefont{J.}~\bibnamefont{Carballo-Franquis}},
  \bibinfo{author}{\bibfnamefont{A.}~\bibnamefont{Chen}},
  \bibinfo{author}{\bibfnamefont{C.-F.} \bibnamefont{Chen}},
  \bibnamefont{et~al.}, \emph{\bibinfo{title}{{Qiskit: An Open-source Framework
  for Quantum Computing}}} (\bibinfo{year}{2019}),
  \urlprefix\url{https://doi.org/10.5281/zenodo.2562111}.

\bibitem[{\citenamefont{Bergholm et~al.}(2018)\citenamefont{Bergholm, Izaac,
  Schuld, Gogolin, Ahmed, Ajith, Alam, Alonso-Linaje, AkashNarayanan, Asadi
  et~al.}}]{bergholm_pennylane_2018}
\bibinfo{author}{\bibfnamefont{V.}~\bibnamefont{Bergholm}},
  \bibinfo{author}{\bibfnamefont{J.}~\bibnamefont{Izaac}},
  \bibinfo{author}{\bibfnamefont{M.}~\bibnamefont{Schuld}},
  \bibinfo{author}{\bibfnamefont{C.}~\bibnamefont{Gogolin}},
  \bibinfo{author}{\bibfnamefont{S.}~\bibnamefont{Ahmed}},
  \bibinfo{author}{\bibfnamefont{V.}~\bibnamefont{Ajith}},
  \bibinfo{author}{\bibfnamefont{M.~S.} \bibnamefont{Alam}},
  \bibinfo{author}{\bibfnamefont{G.}~\bibnamefont{Alonso-Linaje}},
  \bibinfo{author}{\bibfnamefont{B.}~\bibnamefont{AkashNarayanan}},
  \bibinfo{author}{\bibfnamefont{A.}~\bibnamefont{Asadi}},
  \bibnamefont{et~al.}, \bibinfo{journal}{arXiv preprint arXiv:1811.04968}
  (\bibinfo{year}{2018}), \urlprefix\url{https://arxiv.org/abs/1811.04968}.

\bibitem[{\citenamefont{Fingerhuth et~al.}(2018)\citenamefont{Fingerhuth,
  Babej, and Wittek}}]{fingerhuth_open_2018}
\bibinfo{author}{\bibfnamefont{M.}~\bibnamefont{Fingerhuth}},
  \bibinfo{author}{\bibfnamefont{T.}~\bibnamefont{Babej}}, \bibnamefont{and}
  \bibinfo{author}{\bibfnamefont{P.}~\bibnamefont{Wittek}},
  \bibinfo{journal}{PLOS ONE} \textbf{\bibinfo{volume}{13}}, \bibinfo{pages}{1}
  (\bibinfo{year}{2018}),
  \urlprefix\url{https://doi.org/10.1371/journal.pone.0208561}.

\bibitem[{\citenamefont{Otterbach et~al.}(2017)\citenamefont{Otterbach,
  Manenti, Alidoust, Bestwick, Block, Bloom, Caldwell, Didier, Fried, Hong
  et~al.}}]{otterbach_unsupervised_2017}
\bibinfo{author}{\bibfnamefont{J.~S.} \bibnamefont{Otterbach}},
  \bibinfo{author}{\bibfnamefont{R.}~\bibnamefont{Manenti}},
  \bibinfo{author}{\bibfnamefont{N.}~\bibnamefont{Alidoust}},
  \bibinfo{author}{\bibfnamefont{A.}~\bibnamefont{Bestwick}},
  \bibinfo{author}{\bibfnamefont{M.}~\bibnamefont{Block}},
  \bibinfo{author}{\bibfnamefont{B.}~\bibnamefont{Bloom}},
  \bibinfo{author}{\bibfnamefont{S.}~\bibnamefont{Caldwell}},
  \bibinfo{author}{\bibfnamefont{N.}~\bibnamefont{Didier}},
  \bibinfo{author}{\bibfnamefont{E.~S.} \bibnamefont{Fried}},
  \bibinfo{author}{\bibfnamefont{S.}~\bibnamefont{Hong}}, \bibnamefont{et~al.},
  \bibinfo{journal}{arXiv preprint arXiv:1712.05771}  (\bibinfo{year}{2017}).

\bibitem[{\citenamefont{Rist{\`e} et~al.}(2017)\citenamefont{Rist{\`e},
  Da~Silva, Ryan, Cross, C{\'o}rcoles, Smolin, Gambetta, Chow, and
  Johnson}}]{riste_demonstration_2017}
\bibinfo{author}{\bibfnamefont{D.}~\bibnamefont{Rist{\`e}}},
  \bibinfo{author}{\bibfnamefont{M.~P.} \bibnamefont{Da~Silva}},
  \bibinfo{author}{\bibfnamefont{C.~A.} \bibnamefont{Ryan}},
  \bibinfo{author}{\bibfnamefont{A.~W.} \bibnamefont{Cross}},
  \bibinfo{author}{\bibfnamefont{A.~D.} \bibnamefont{C{\'o}rcoles}},
  \bibinfo{author}{\bibfnamefont{J.~A.} \bibnamefont{Smolin}},
  \bibinfo{author}{\bibfnamefont{J.~M.} \bibnamefont{Gambetta}},
  \bibinfo{author}{\bibfnamefont{J.~M.} \bibnamefont{Chow}}, \bibnamefont{and}
  \bibinfo{author}{\bibfnamefont{B.~R.} \bibnamefont{Johnson}},
  \bibinfo{journal}{npj Quantum Information} \textbf{\bibinfo{volume}{3}},
  \bibinfo{pages}{1} (\bibinfo{year}{2017}).

\bibitem[{\citenamefont{Grant et~al.}(2018)\citenamefont{Grant, Benedetti, Cao,
  Hallam, Lockhart, Stojevic, Green, and Severini}}]{grant_hierarchical_2018}
\bibinfo{author}{\bibfnamefont{E.}~\bibnamefont{Grant}},
  \bibinfo{author}{\bibfnamefont{M.}~\bibnamefont{Benedetti}},
  \bibinfo{author}{\bibfnamefont{S.}~\bibnamefont{Cao}},
  \bibinfo{author}{\bibfnamefont{A.}~\bibnamefont{Hallam}},
  \bibinfo{author}{\bibfnamefont{J.}~\bibnamefont{Lockhart}},
  \bibinfo{author}{\bibfnamefont{V.}~\bibnamefont{Stojevic}},
  \bibinfo{author}{\bibfnamefont{A.~G.} \bibnamefont{Green}}, \bibnamefont{and}
  \bibinfo{author}{\bibfnamefont{S.}~\bibnamefont{Severini}},
  \bibinfo{journal}{npj Quantum Information} \textbf{\bibinfo{volume}{4}},
  \bibinfo{pages}{1} (\bibinfo{year}{2018}).

\bibitem[{\citenamefont{Tacchino et~al.}(2019)\citenamefont{Tacchino,
  Macchiavello, Gerace, and Bajoni}}]{tacchino_artificial_2019}
\bibinfo{author}{\bibfnamefont{F.}~\bibnamefont{Tacchino}},
  \bibinfo{author}{\bibfnamefont{C.}~\bibnamefont{Macchiavello}},
  \bibinfo{author}{\bibfnamefont{D.}~\bibnamefont{Gerace}}, \bibnamefont{and}
  \bibinfo{author}{\bibfnamefont{D.}~\bibnamefont{Bajoni}},
  \bibinfo{journal}{npj Quantum Information} \textbf{\bibinfo{volume}{5}},
  \bibinfo{pages}{1} (\bibinfo{year}{2019}).

\bibitem[{\citenamefont{Benedetti
  et~al.}(2019{\natexlab{b}})\citenamefont{Benedetti, Garcia-Pintos, Perdomo,
  Leyton-Ortega, Nam, and Perdomo-Ortiz}}]{benedetti_generative_2019}
\bibinfo{author}{\bibfnamefont{M.}~\bibnamefont{Benedetti}},
  \bibinfo{author}{\bibfnamefont{D.}~\bibnamefont{Garcia-Pintos}},
  \bibinfo{author}{\bibfnamefont{O.}~\bibnamefont{Perdomo}},
  \bibinfo{author}{\bibfnamefont{V.}~\bibnamefont{Leyton-Ortega}},
  \bibinfo{author}{\bibfnamefont{Y.}~\bibnamefont{Nam}}, \bibnamefont{and}
  \bibinfo{author}{\bibfnamefont{A.}~\bibnamefont{Perdomo-Ortiz}},
  \bibinfo{journal}{npj Quantum Information} \textbf{\bibinfo{volume}{5}},
  \bibinfo{pages}{1} (\bibinfo{year}{2019}{\natexlab{b}}).

\bibitem[{\citenamefont{Coyle et~al.}(2020)\citenamefont{Coyle, Mills, Danos,
  and Kashefi}}]{coyle_born_2020}
\bibinfo{author}{\bibfnamefont{B.}~\bibnamefont{Coyle}},
  \bibinfo{author}{\bibfnamefont{D.}~\bibnamefont{Mills}},
  \bibinfo{author}{\bibfnamefont{V.}~\bibnamefont{Danos}}, \bibnamefont{and}
  \bibinfo{author}{\bibfnamefont{E.}~\bibnamefont{Kashefi}},
  \bibinfo{journal}{npj Quantum Information} \textbf{\bibinfo{volume}{6}},
  \bibinfo{pages}{1} (\bibinfo{year}{2020}).

\bibitem[{\citenamefont{Rocchetto et~al.}(2019)\citenamefont{Rocchetto,
  Aaronson, Severini, Carvacho, Poderini, Agresti, Bentivegna, and
  Sciarrino}}]{rocchetto_experimental_2019}
\bibinfo{author}{\bibfnamefont{A.}~\bibnamefont{Rocchetto}},
  \bibinfo{author}{\bibfnamefont{S.}~\bibnamefont{Aaronson}},
  \bibinfo{author}{\bibfnamefont{S.}~\bibnamefont{Severini}},
  \bibinfo{author}{\bibfnamefont{G.}~\bibnamefont{Carvacho}},
  \bibinfo{author}{\bibfnamefont{D.}~\bibnamefont{Poderini}},
  \bibinfo{author}{\bibfnamefont{I.}~\bibnamefont{Agresti}},
  \bibinfo{author}{\bibfnamefont{M.}~\bibnamefont{Bentivegna}},
  \bibnamefont{and}
  \bibinfo{author}{\bibfnamefont{F.}~\bibnamefont{Sciarrino}},
  \bibinfo{journal}{Science advances} \textbf{\bibinfo{volume}{5}},
  \bibinfo{pages}{eaau1946} (\bibinfo{year}{2019}).

\bibitem[{\citenamefont{Ding et~al.}(2019)\citenamefont{Ding, Lamata, Sanz,
  Chen, and Solano}}]{ding_experimental_2019}
\bibinfo{author}{\bibfnamefont{Y.}~\bibnamefont{Ding}},
  \bibinfo{author}{\bibfnamefont{L.}~\bibnamefont{Lamata}},
  \bibinfo{author}{\bibfnamefont{M.}~\bibnamefont{Sanz}},
  \bibinfo{author}{\bibfnamefont{X.}~\bibnamefont{Chen}}, \bibnamefont{and}
  \bibinfo{author}{\bibfnamefont{E.}~\bibnamefont{Solano}},
  \bibinfo{journal}{Advanced Quantum Technologies}
  \textbf{\bibinfo{volume}{2}}, \bibinfo{pages}{1800065}
  (\bibinfo{year}{2019}).

\bibitem[{\citenamefont{Parekh et~al.}(2021)\citenamefont{Parekh, Ricciardi,
  Darwish, and DiAdamo}}]{parekh_quantum_2021}
\bibinfo{author}{\bibfnamefont{R.}~\bibnamefont{Parekh}},
  \bibinfo{author}{\bibfnamefont{A.}~\bibnamefont{Ricciardi}},
  \bibinfo{author}{\bibfnamefont{A.}~\bibnamefont{Darwish}}, \bibnamefont{and}
  \bibinfo{author}{\bibfnamefont{S.}~\bibnamefont{DiAdamo}},
  \bibinfo{journal}{arXiv preprint arXiv:2106.06841}  (\bibinfo{year}{2021}).

\bibitem[{\citenamefont{Diadamo et~al.}(2021)\citenamefont{Diadamo, Notzel,
  Zanger, and Bese}}]{diadamo_qunetsim_2021}
\bibinfo{author}{\bibfnamefont{S.}~\bibnamefont{Diadamo}},
  \bibinfo{author}{\bibfnamefont{J.}~\bibnamefont{Notzel}},
  \bibinfo{author}{\bibfnamefont{B.}~\bibnamefont{Zanger}}, \bibnamefont{and}
  \bibinfo{author}{\bibfnamefont{M.~M.} \bibnamefont{Bese}},
  \bibinfo{journal}{{IEEE} Transactions on Quantum Engineering}
  \textbf{\bibinfo{volume}{2}}, \bibinfo{pages}{1} (\bibinfo{year}{2021}),
  \urlprefix\url{https://doi.org/10.1109%2Ftqe.2021.3092395}.

\bibitem[{\citenamefont{Cirac et~al.}(1999)\citenamefont{Cirac, Ekert, Huelga,
  and Macchiavello}}]{cirac_distributed_1999}
\bibinfo{author}{\bibfnamefont{J.~I.} \bibnamefont{Cirac}},
  \bibinfo{author}{\bibfnamefont{A.}~\bibnamefont{Ekert}},
  \bibinfo{author}{\bibfnamefont{S.~F.} \bibnamefont{Huelga}},
  \bibnamefont{and}
  \bibinfo{author}{\bibfnamefont{C.}~\bibnamefont{Macchiavello}},
  \bibinfo{journal}{Physical Review A} \textbf{\bibinfo{volume}{59}},
  \bibinfo{pages}{4249} (\bibinfo{year}{1999}).

\bibitem[{\citenamefont{Gyongyosi and
  Imre}(2019)}]{gyongyosi_entanglement_2019}
\bibinfo{author}{\bibfnamefont{L.}~\bibnamefont{Gyongyosi}} \bibnamefont{and}
  \bibinfo{author}{\bibfnamefont{S.}~\bibnamefont{Imre}},
  \bibinfo{journal}{Quantum Information Processing}
  \textbf{\bibinfo{volume}{18}}, \bibinfo{pages}{107} (\bibinfo{year}{2019}),
  ISSN \bibinfo{issn}{1573-1332}.

\bibitem[{\citenamefont{Streltsov et~al.}(2012)\citenamefont{Streltsov,
  Kampermann, and Bru{\ss}}}]{streltsov_quantum_2012}
\bibinfo{author}{\bibfnamefont{A.}~\bibnamefont{Streltsov}},
  \bibinfo{author}{\bibfnamefont{H.}~\bibnamefont{Kampermann}},
  \bibnamefont{and} \bibinfo{author}{\bibfnamefont{D.}~\bibnamefont{Bru{\ss}}},
  \bibinfo{journal}{Physical Review Letters} \textbf{\bibinfo{volume}{108}},
  \bibinfo{pages}{250501} (\bibinfo{year}{2012}).

\bibitem[{\citenamefont{Sharma et~al.}(2022)\citenamefont{Sharma, Cerezo,
  Holmes, Cincio, Sornborger, and Coles}}]{sharma_reformulation_2022}
\bibinfo{author}{\bibfnamefont{K.}~\bibnamefont{Sharma}},
  \bibinfo{author}{\bibfnamefont{M.}~\bibnamefont{Cerezo}},
  \bibinfo{author}{\bibfnamefont{Z.}~\bibnamefont{Holmes}},
  \bibinfo{author}{\bibfnamefont{L.}~\bibnamefont{Cincio}},
  \bibinfo{author}{\bibfnamefont{A.}~\bibnamefont{Sornborger}},
  \bibnamefont{and} \bibinfo{author}{\bibfnamefont{P.~J.} \bibnamefont{Coles}},
  \bibinfo{journal}{Physical Review Letters} \textbf{\bibinfo{volume}{128}}
  (\bibinfo{year}{2022}).

\bibitem[{\citenamefont{Raussendorf et~al.}(2003)\citenamefont{Raussendorf,
  Browne, and Briegel}}]{raussendorf_measurement_2003}
\bibinfo{author}{\bibfnamefont{R.}~\bibnamefont{Raussendorf}},
  \bibinfo{author}{\bibfnamefont{D.~E.} \bibnamefont{Browne}},
  \bibnamefont{and} \bibinfo{author}{\bibfnamefont{H.~J.}
  \bibnamefont{Briegel}}, \bibinfo{journal}{Physical review A}
  \textbf{\bibinfo{volume}{68}} (\bibinfo{year}{2003}).

\bibitem[{\citenamefont{McMahan et~al.}(2017)\citenamefont{McMahan, Moore,
  Ramage, Hampson, and Arcas}}]{mcmahan_communication_2017}
\bibinfo{author}{\bibfnamefont{B.}~\bibnamefont{McMahan}},
  \bibinfo{author}{\bibfnamefont{E.}~\bibnamefont{Moore}},
  \bibinfo{author}{\bibfnamefont{D.}~\bibnamefont{Ramage}},
  \bibinfo{author}{\bibfnamefont{S.}~\bibnamefont{Hampson}}, \bibnamefont{and}
  \bibinfo{author}{\bibfnamefont{B.~A.~y.} \bibnamefont{Arcas}}, in
  \emph{\bibinfo{booktitle}{Proceedings of the 20th International Conference on
  Artificial Intelligence and Statistics}}, edited by
  \bibinfo{editor}{\bibfnamefont{A.}~\bibnamefont{Singh}} \bibnamefont{and}
  \bibinfo{editor}{\bibfnamefont{J.}~\bibnamefont{Zhu}}
  (\bibinfo{publisher}{PMLR}, \bibinfo{year}{2017}), vol.~\bibinfo{volume}{54}
  of \emph{\bibinfo{series}{Proceedings of Machine Learning Research}}, pp.
  \bibinfo{pages}{1273--1282}.

\bibitem[{\citenamefont{Chehimi and Saad}(2021)}]{chehimi_quantum_2021}
\bibinfo{author}{\bibfnamefont{M.}~\bibnamefont{Chehimi}} \bibnamefont{and}
  \bibinfo{author}{\bibfnamefont{W.}~\bibnamefont{Saad}},
  \bibinfo{journal}{arXiv preprint arXiv:2106.00005}  (\bibinfo{year}{2021}).

\bibitem[{\citenamefont{Chen and Yoo}(2021)}]{chen_federated_2021}
\bibinfo{author}{\bibfnamefont{S.~Y.-C.} \bibnamefont{Chen}} \bibnamefont{and}
  \bibinfo{author}{\bibfnamefont{S.}~\bibnamefont{Yoo}},
  \bibinfo{journal}{Entropy} \textbf{\bibinfo{volume}{23}},
  \bibinfo{pages}{460} (\bibinfo{year}{2021}).

\bibitem[{\citenamefont{Morello}(2018)}]{morello_what_2018}
\bibinfo{author}{\bibfnamefont{A.}~\bibnamefont{Morello}},
  \bibinfo{journal}{Quantum Science and Technology}
  \textbf{\bibinfo{volume}{3}}, \bibinfo{pages}{030201} (\bibinfo{year}{2018}),
  \urlprefix\url{https://doi.org/10.1088/2058-9565/aac869}.

\bibitem[{\citenamefont{Perdomo-Ortiz et~al.}(2018)\citenamefont{Perdomo-Ortiz,
  Benedetti, Realpe-G{\'{o}}mez, and Biswas}}]{perdomo_opportunities_2018}
\bibinfo{author}{\bibfnamefont{A.}~\bibnamefont{Perdomo-Ortiz}},
  \bibinfo{author}{\bibfnamefont{M.}~\bibnamefont{Benedetti}},
  \bibinfo{author}{\bibfnamefont{J.}~\bibnamefont{Realpe-G{\'{o}}mez}},
  \bibnamefont{and} \bibinfo{author}{\bibfnamefont{R.}~\bibnamefont{Biswas}},
  \bibinfo{journal}{Quantum Science and Technology}
  \textbf{\bibinfo{volume}{3}}, \bibinfo{pages}{030502} (\bibinfo{year}{2018}),
  \urlprefix\url{https://doi.org/10.1088/2058-9565/aab859}.

\end{thebibliography}

\appendix

\end{document}